\documentclass{article}

\usepackage{microtype}
\usepackage{graphicx}
\usepackage{subcaption}
\usepackage{booktabs} 

\usepackage{hyperref}

\usepackage[preprint]{icml2026}

\usepackage{amsmath}
\usepackage{amssymb}
\usepackage{mathtools}
\usepackage{amsthm}
\usepackage[utf8]{inputenc} 
\usepackage[T1]{fontenc}    
\usepackage{hyperref}       
\usepackage{url}            
\usepackage{amsfonts}       
\usepackage{nicefrac}       
\usepackage{xcolor}         
\usepackage{graphicx}
\usepackage{placeins}
\usepackage[table]{xcolor}
\usepackage{tabularx}
\usepackage{pdflscape}
\usepackage{colortbl}

\theoremstyle{plain}

\theoremstyle{definition}

\theoremstyle{remark}

\usepackage[textsize=tiny]{todonotes}

\icmltitlerunning{MAEB: Massive Audio Embedding Benchmark}

\begin{document}

\twocolumn[
  \icmltitle{MAEB: Massive Audio Embedding Benchmark}



  \icmlsetsymbol{equal}{*}

  \begin{icmlauthorlist}
    \icmlauthor{Adnan El Assadi}{carleton}
    \icmlauthor{Isaac Chung}{zd}
    \icmlauthor{Chenghao Xiao}{du}
    \icmlauthor{Roman Solomatin}{mirai,salute}
    \icmlauthor{Animesh Jha}{stanf}
    \icmlauthor{Rahul Chand}{stanf}
    \icmlauthor{Silky Singh}{stanf}
    \icmlauthor{Kaitlyn Wang}{stanf}
    \icmlauthor{Ali Sartaz Khan}{stanf}
    \icmlauthor{Marc Moussa Nasser}{stanf}
    \icmlauthor{Sufen Fong}{stanf}
    \icmlauthor{Pengfei He}{stanf}
    \icmlauthor{Alan Xiao}{stanf}
    \icmlauthor{Ayush Sunil Munot}{iitkgp}
    \icmlauthor{Aditya Shrivastava}{c1}
    \icmlauthor{Artem Gazizov}{harv}
    \icmlauthor{Niklas Muennighoff}{stanf}
    \icmlauthor{Kenneth Enevoldsen}{au}

  \end{icmlauthorlist}

  
  \icmlaffiliation{carleton}{Carleton University}
  \icmlaffiliation{zd}{Zendesk}
  \icmlaffiliation{du}{Durham University}
  \icmlaffiliation{salute}{SaluteDevices}
  \icmlaffiliation{mirai}{MIRAI}
  \icmlaffiliation{stanf}{Stanford University}
  \icmlaffiliation{au}{Aarhus University}
  \icmlaffiliation{iitkgp}{Indian Institute of Technology, Kharagpur}
  
\icmlaffiliation{harv}{Harvard University}
\icmlaffiliation{c1}{Capital One}

  \icmlcorrespondingauthor{Adnan El Assadi}{adnanelassadi@cmail.carleton.ca}

  \icmlkeywords{audio, embedding, benchmark, multilingual}

  \vskip 0.3in
]



\printAffiliationsAndNotice{}  

\begin{abstract}

We introduce the \textbf{M}assive \textbf{A}udio \textbf{E}mbedding \textbf{B}enchmark (MAEB), a large-scale benchmark covering 30 tasks across speech, music, environmental sounds, and cross-modal audio-text reasoning in 100+ languages. We evaluate 50+ models and find that no single model dominates across all tasks: contrastive audio-text models excel at environmental sound classification (e.g., ESC50) but score near random on multilingual speech tasks (e.g., SIB-FLEURS), while speech-pretrained models show the opposite pattern. Clustering remains challenging for all models, with even the best-performing model achieving only modest results. We observe that models excelling on acoustic understanding often perform poorly on linguistic tasks, and vice versa. We also show that the performance of audio encoders on MAEB correlates highly with their performance when used in audio large language models.
MAEB is derived from MAEB+, a collection of 98 tasks. MAEB is designed to maintain task diversity while reducing evaluation cost, and it integrates into the MTEB ecosystem for unified evaluation across text, image, and audio modalities. We release MAEB and all 98 tasks along with
code and a leaderboard at \url{https://github.com/embeddings-benchmark/mteb}.

\end{abstract}

\begin{figure*}
    \centering
    \includegraphics[width=\linewidth]{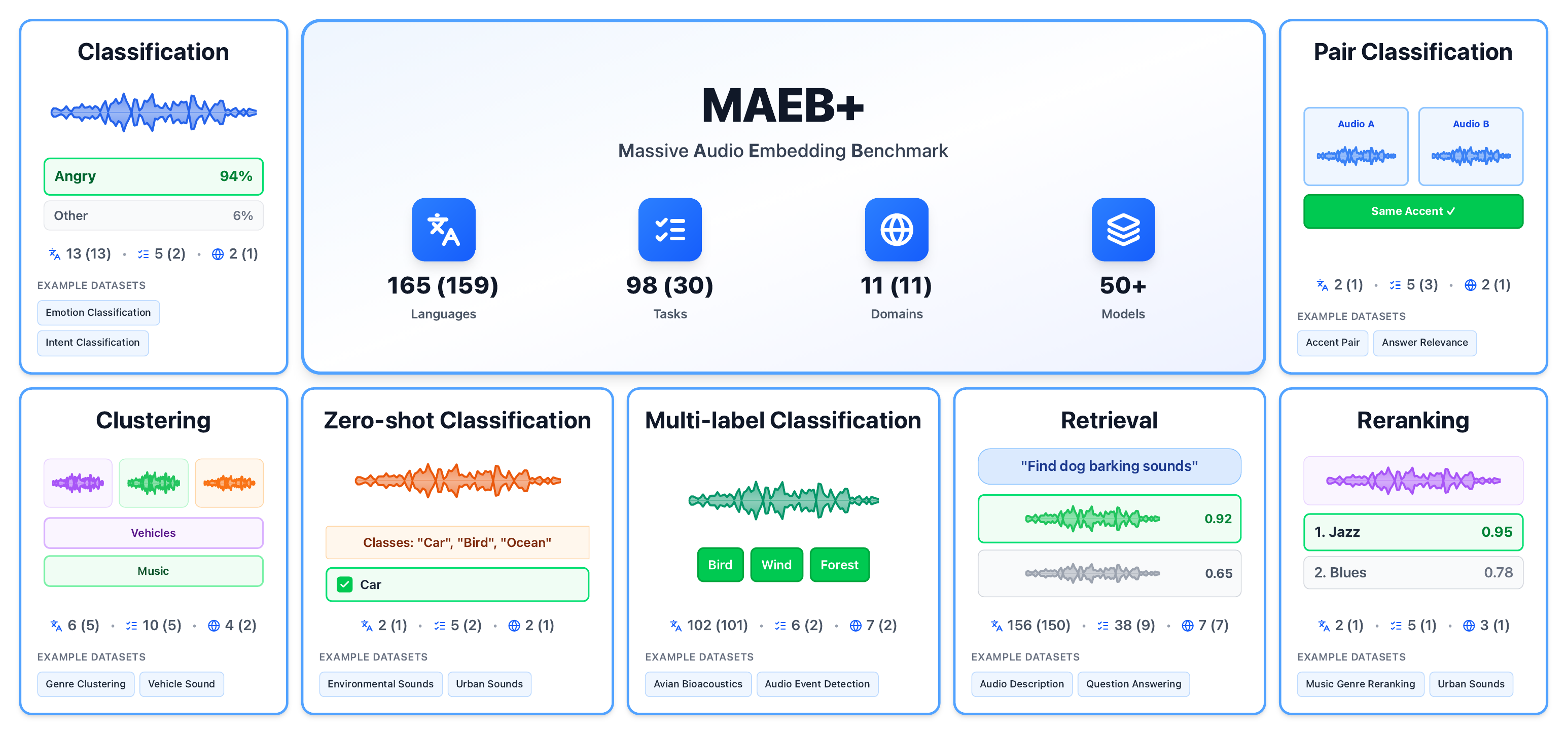}
    \caption{ overview of task types and example subtypes in MAEB+. Values in parentheses denote numbers for MAEB.}
    \label{fig:overview}
\end{figure*}

\section{Introduction}

Audio and speech representations support diverse applications such as voice assistants and music recommendation systems. However, evaluation protocols for audio embedding models vary significantly, spanning speech recognition, zero-shot classification, and audio-text retrieval. Existing audio benchmarks often focus on specific tasks (e.g., vocal sound classification \cite{Gong_2022}) or narrow domains (e.g., environmental sounds \cite{piczak2015dataset}) while often ignoring others, limiting insight into how well embeddings transfer across different applications. Without a unified evaluation framework, the field remains fragmented, making it difficult to compare models or track meaningful progress across the full landscape of audio tasks. Additionally, the absence of integrated development and maintenance infrastructure has led to stagnation in existing benchmarks, with many becoming outdated as the field rapidly evolves.  

We introduce the \textbf{M}assive \textbf{A}udio \textbf{E}mbedding \textbf{B}enchmark (MAEB) to provide a unified, comprehensive evaluation protocol to spur the field's advancement toward universal audio embedding models. Building on the success of MTEB \cite{muennighoff2022mteb}, MMTEB \cite{enevoldsen2025mmteb}, and MIEB \cite{xiao2025mieb}, which have unified and expanded evaluation of embedding models for text and image through continual development and community maintenance, we extend this proven framework to the audio domain. %

MAEB spans 30 audio tasks grouped into 7 categories. Aligning with MTEB's approach, we include Classification, Zero-shot Classification, Clustering, Pair Classification, Retrieval, and Reranking tasks adapted for audio data. Notably, we consider audio-specific aspects such as multilingual audio understanding, long-form audio processing, and cross-modal audio-text tasks that have been largely absent from prior audio benchmarks. Beyond traditional speech recognition tasks, we emphasize comprehensive audio understanding capabilities through: 1) Diverse acoustic domains, including speech, music, environmental sounds, and bioacoustics; 2) Cross-modal abilities, particularly in zero-shot settings leveraging text descriptions; 3) Complex recognition tasks requiring fine-grained audio understanding; 4) Multilingual audio processing across various languages and dialects.


To ensure efficient evaluation and broader adoption, MAEB allows for evaluation of a small audio-only model in ~2 GPU hours while not compromising on coverage. We also provide MAEB(audio), a 19-task audio-only subset for evaluating audio-only models, and MAEB+, our full unfiltered collection of 98 tasks. Additionally, we provide a modular architecture that simplifies the addition of new audio models and datasets, ensuring that MAEB can evolve with the rapidly advancing field of audio representation learning.

Our evaluation of 53 models reveals that no single model dominates across all audio domains; each excels in specific areas while underperforming in others. Preliminary evidence from four Audio LLMs suggests that MAEB encoder quality may correlate with downstream Audio LLM performance ($R^2$~=~0.86, $n$~=~4; see \autoref{fig:maeb_correlation}), validating the benchmark's relevance for multimodal audio understanding.

To summarize, MAEB makes the following key contributions: 
\begin{enumerate}
    \item  We provide the first comprehensive benchmark for audio embeddings that spans multiple domains, languages, and task types,
    \item We establish baseline evaluations using a representative set of 53 models, revealing strengths and weaknesses across different audio understanding capabilities,
    \item We identify critical areas where current models struggle, particularly in multilingual contexts and cross-modal understanding, providing clear directions for future research,
    \item We create a flexible, extensible framework that enables the audio research community to standardize evaluation practices and track progress more effectively.
\end{enumerate}

\section{MAEB}

MAEB is fully integrated into the MTEB ecosystem~\citep{muennighoff2022mteb}, extending its unified evaluation framework to the audio modality alongside text~\citep{enevoldsen2025mmteb} and image~\citep{xiao2025mieb} embeddings. This integration provides several advantages: (1) \textit{tried-and-tested implementations} with standardized metrics and evaluation protocols validated across thousands of submissions; (2) \textit{extensibility} through a minimal interface that allows adding new models or tasks with minimal code changes; (3) \textit{reproducibility} via versioned code and artifacts, with results stored in a public repository; and (4) \textit{long-term maintenance} and community-driven development~\citep{chung2025maintainingmteblongterm}. MAEB seeks to broadly evaluate \textit{embedding quality} for downstream tasks--it does not assess transcription, generation, or other capabilities outside the scope of representation learning.

\subsection{Benchmark Construction}
\label{sec:benchmark-construction}

\paragraph{Dataset Selection}
We curate datasets according to four guiding principles: (1) \textit{domain diversity} across speech, music, environmental sounds, and bioacoustics; (2) \textit{task diversity} spanning classification, clustering, pair classification, retrieval, and reranking; (3) \textit{linguistic diversity} across languages and dialects; and (4) \textit{quality and accessibility}, prioritizing datasets with established usage, clear licensing, and public availability.

\paragraph{Task Selection}
Evaluating models across our full dataset collection, MAEB+, would be prohibitively expensive for most groups. Following MMTEB and MIEB, which demonstrated that principled filtering maintains high rank correlation with exhaustive evaluation, we construct MAEB using five selection criteria: (1) \textit{Validity}: For directional tasks (e.g., retrieval), we prioritize the more semantically valid direction (e.g., text-to-audio over audio-to-text when text queries better reflect realistic use cases); (2) \textit{Unique coverage}: Tasks providing exclusive coverage of a domain or capability are retained regardless of other factors (e.g., the only bioacoustics clustering task); (3) \textit{Linguistic breadth}: Among comparable tasks, we retain those covering more languages; (4) \textit{Redundancy removal}: We compute pairwise correlation matrices across model rankings and remove tasks with Spearman $\rho > 0.8$ to a retained task, keeping the task with broader coverage or lower runtime; (5) \textit{Runtime efficiency}: Among otherwise equivalent tasks, we select those with lower computational cost.

As an intermediate step in task selection, we create MAEB(extended) with 89 tasks by applying initial validity and unique coverage filters to MAEB+. From this intermediate collection, we apply redundancy removal and runtime efficiency criteria to produce the final MAEB (30 tasks). \autoref{tab:benchmark-runtime} compares GPU runtime between MAEB and MAEB(extended) across representative models, showing a 2.2--3.3$\times$ speedup depending on model type. MAEB maintains strong correlation with MAEB(extended) in terms of model scores (Pearson $r$=0.981) and model ranking (Spearman $\rho$=0.912), indicating that it preserves relative model performance while substantially reducing evaluation time.

\begin{table}[h]
    \centering
    \caption{Benchmark runtime comparison (GPU hours) between MAEB and MAEB(extended). Runtime measured on a single NVIDIA A100 GPU.}
    \label{tab:benchmark-runtime}
    \resizebox{\columnwidth}{!}{
    \begin{tabular}{llccc}
    \toprule
    \textbf{Model} & \textbf{Params} & \textbf{MAEB} & \textbf{Extended} & \textbf{Speedup} \\
    \midrule
    YAMNet & 3.7M & 2.01 & 6.02 & 3.0$\times$ \\
    wav2vec2-xls-r-2b & 2B & 26.93 & 45.62 & 1.7$\times$ \\
    larger\_clap\_general & 630M & 11.52 & 32.23 & 2.8$\times$ \\
    CLAP-htsat-fused & 194M & 13.03 & 35.35 & 2.7$\times$ \\
    \bottomrule
    \end{tabular}
    }
\end{table}


For comprehensive evaluation, we release the full unfiltered collection as MAEB+. See the full dataset list in \autoref{sec:overview}. 

\paragraph{Benchmark Ranking} 
Following the same protocol in MMTEB \cite{enevoldsen2025mmteb}, we compute model ranks using a Borda count \cite{colombo2022what} by treating each task as a preference voter over models. While the Borda count has several advantages over the mean (including scale invariance and robustness to outliers), it is not a continuous measure; thus, we provide both the Borda rank and the mean in the leaderboard.



\subsection{Tasks and Evaluation}
We follow a similar approach to MMTEB and MIEB to extend tasks to the audio domain. 

\paragraph{Classification} A logistic regression is trained on audio embeddings to predict labels~\citep{alain2018understandingintermediatelayersusing,radford2021learning}. We use few-shot linear probing~\citep{muennighoff2022mteb,cherti2023reproducible} with 8 examples per class, balancing evaluation quality with computational efficiency.

\paragraph{Zero-shot Classification} Audio embeddings are directly matched to class labels converted to text prompts (e.g., ``This is a sound of dog bark'') without training a classifier. We measure accuracy following~\citet{radford2021learning}.

\paragraph{Clustering} We use MiniBatchKMeans (with k set to the number of true labels) and V-measure~\citep{rosenberg-hirschberg-2007-v} as the main metric to evaluate whether embeddings group meaningfully according to semantic categories.

\paragraph{Retrieval} Retrieval evaluates finding relevant documents from a corpus given a query, including uni-modal (audio-to-audio) and cross-modal (text-to-audio, audio-to-text) scenarios. Documents are ranked by cosine similarity, with CV Recall@5 (cross-validation recall at 5) as the main metric.

\paragraph{Pair Classification} Given two audio inputs, the task is to predict whether they are similar according to a criterion (e.g., same speaker, same sound class). Similarity is computed between embeddings, and average precision based on cosine similarity serves as the main metric.

\paragraph{Reranking} Unlike retrieval over full corpora, reranking evaluates ranking quality on pre-selected candidate sets containing relevant documents and hard negatives. This tests fine-grained discrimination, with MAP@1000 (mean average precision at 1000) as the main metric.

\section{Experimental Settings}

\subsection{Models}

We seek to evaluate the broad category of audio embedding models, and select 50+ audio encoders representing four broad development categories.

\textbf{Audio Encoders} includes models trained specifically on audio through various methods. Self-supervised speech models learn contextualized representations through masked prediction and clustering objectives, including Wav2Vec2/XLS-R \cite{baevski2020wav2vec,babu2021xlsr}, WavLM \cite{Chen_2022}, HuBERT \cite{hsu2021hubert}, Data2Vec \cite{baevski2022data2vecgeneralframeworkselfsupervised}, UniSpeech \cite{wang2021unispeechunifiedspeechrepresentation}, SEW-D \cite{wu2021performanceefficiencytradeoffsunsupervisedpretraining}, and MCTCT \cite{lugosch2022pseudolabelingmassivelymultilingualspeech}. Transformer-based models apply vision transformer architectures to audio spectrograms, including AST \cite{gong2021astaudiospectrogramtransformer}. CNN-based models employ convolutional architectures trained on large-scale audio datasets, including CNN14 \cite{kong2020pannslargescalepretrainedaudio}, YAMNet \cite{audioset}, and VGGish \cite{hershey2017cnn}. Neural codec models provide audio compression through learned representations, including Encodec \cite{défossez2022highfidelityneuralaudio}.

\textbf{Sequence-to-Sequence Models} includes models trained for a sequence-to-sequence objective, e.g., for speech recognition and translation. This category includes Whisper \cite{radford2022whisper}, MMS \cite{pratap2023scalingspeechtechnology1000}, SeamlessM4T \cite{communication2023seamlessmultilingualexpressivestreaming}, and SpeechT5 ASR \cite{ao-etal-2022-speecht5}.

\textbf{Contrastive Alignment Models} includes models that learn joint audio-text embedding spaces through a contrastive alignment objective, including CLAP \cite{wu2024largescalecontrastivelanguageaudiopretraining}, MS-CLAP \cite{elizalde2023msclap}, Wav2CLIP \cite{wu2022wav2clip}, MuQ-MuLan \cite{zhu2025muqselfsupervisedmusicrepresentation}, and SpeechT5 Multimodal \cite{ao-etal-2022-speecht5}.

\textbf{Large Audio-Language Models} are models derived from generative multimodal LLMs, which are then adapted for embeddings, e.g., by utilizing their hidden states or through contrastive refinement. These include Qwen2-Audio \cite{chu2024qwen2audio} and LCO-Embedding \cite{xiaoscaling}.

Note that the categories are not perfect; for instance, LCO-Embedding \cite{xiaoscaling} and Wav2Vec2/XLS-R \cite{baevski2020wav2vec,babu2021xlsr} both utilize a contrastive loss during training. Please refer to \autoref{appdx:models} for all model details.

\subsection{Implementation Details}

All models implement consistent preprocessing with audio truncated to a maximum of 30 seconds, or shorter where required by model architecture or memory constraints. Audio is resampled to model-specific sampling rates (16kHz for speech models, 48kHz for CLAP and MS-CLAP variants, 24kHz for MuQ-MuLan and Encodec) and converted to mono when required. 

For embedding extraction, we use model-native approaches: transformer models employ mean pooling over temporal dimensions, CNN models use global average pooling, and specialized architectures follow their intended pooling strategies. Contrastive models (CLAP, MS-CLAP, Wav2CLIP, MuQ-MuLan) use their audio encoder branches with L2 normalization for retrieval compatibility. Large audio-language models extract embeddings from the final hidden layer using last-token pooling.

\section{Results}

\autoref{tab:maeb-performance} presents the top 30 models on the MAEB benchmark. The table includes both MAEB rank (over all 30 tasks) and Audio-only rank (over the 19 audio-only subset tasks) to highlight how models perform differently across task types.
LCO-Embedding-Omni-7B ranks first overall by Borda count, achieving the highest average scores (52.2\% overall, 50.3\% cross-modal retrieval, 64.5\% zero-shot) across all categories. Qwen2-Audio-7B ranks second overall by Borda count (overall average 33.7\%) but ranks first on audio-only tasks by Borda count (50.8\% average) and excels in reranking (80.8\%) and clustering (12.7\%). Whisper-medium achieves third place overall by Borda count (overall average 46.7\%) with strong audio-only performance (48.2\%) but cannot perform cross-modal tasks. CLAP variants (larger\_clap\_general at 4th, larger\_clap\_music\_and\_speech at 6th) demonstrate balanced cross-modal capabilities. We provide detailed per-task results for each category in \autoref{appdx:task_results}.

Figure~\ref{fig:radar} visualizes the performance of leading models on 94 tasks in MAEB+ across 5 acoustic domains (see Appendix~\ref{app:radar_methodology} for task details). For each domain, we select the model achieving the highest average score across all task types. We observe distinct specialization patterns: LCO-Embedding-Omni-7B leads in the Speech domain with an aggregate score of 68.2, driven by strong speech-text alignment, while the Audio Spectrogram Transformer (AST) dominates the Music (71.6), Environmental (63.8), and Bioacoustics (45.2) domains, likely benefiting from its AudioSet pre-training on diverse non-speech events. Qwen2-Audio establishes itself as the leader in Emotion recognition (44.7), demonstrating the advantages of multimodal instruction-tuning for paralinguistic understanding. The disjointed, non-overlapping shapes confirm that no single encoder achieves universal performance across all acoustic domains, the dashed target of 80 remains unmet in every category. This validates our finding that specialized models excel in their respective domains but fail to generalize broadly across the full acoustic spectrum.

\begin{figure}[t]
    \centering
    \includegraphics[width=0.90\linewidth]{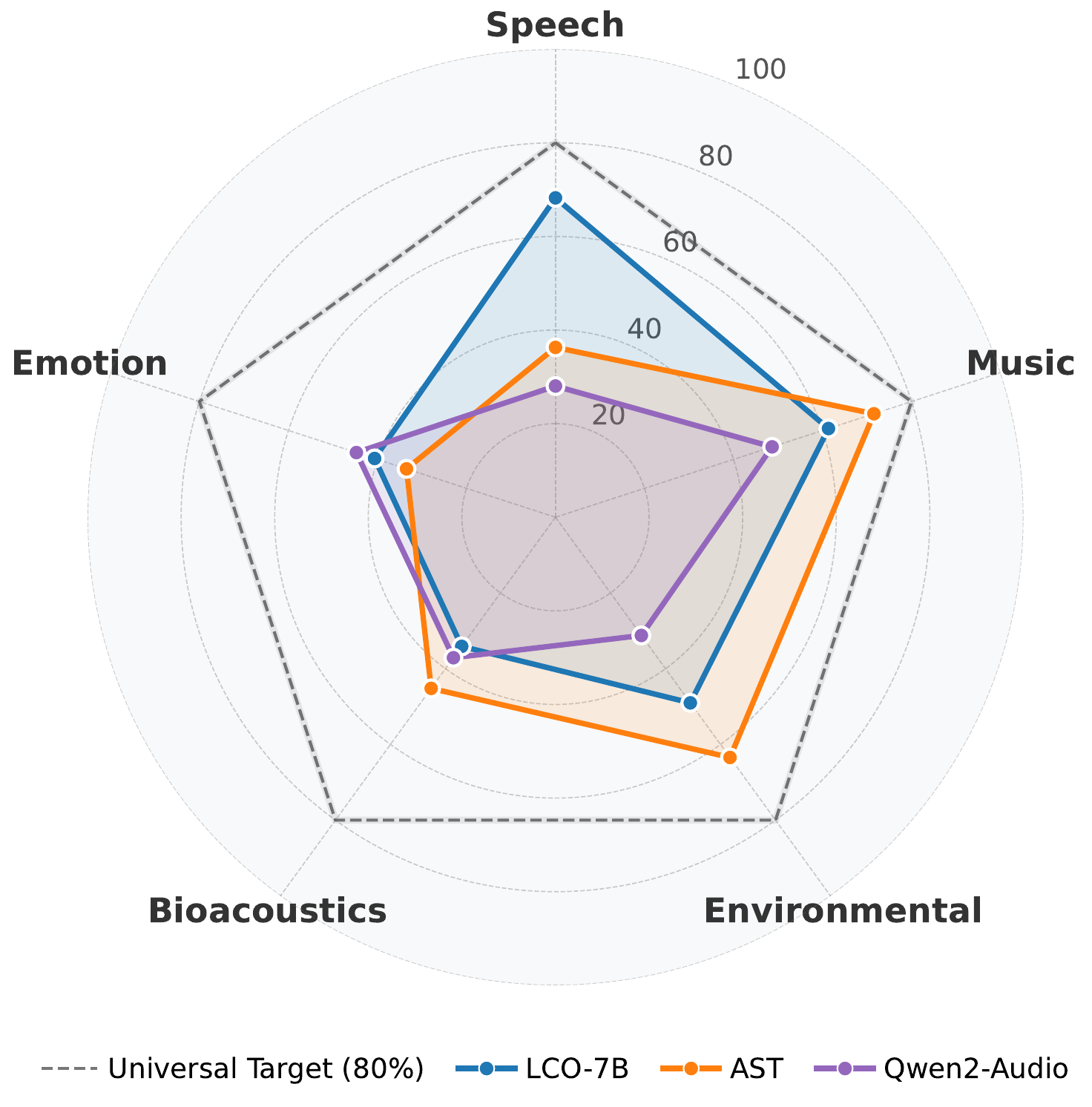}
    \caption{Domain-level performance on 94 tasks in MAEB+. Radial plot shows the top-performing model for each of the five acoustic domains: Speech (44 tasks), Music (13), Environmental (29), Bioacoustics (2), and Emotion (6).  The dashed line represents an 80 target for universal performance, which remains unmet. Scores are averaged across all available task types (classification, clustering, retrieval, reranking). See Appendix~\ref{app:radar_methodology} for methodology.}

    \label{fig:radar}
\end{figure}

\begin{table*}[!th]
    \centering
    \caption{
    Top 30 models on the MAEB benchmark (30 tasks spanning audio-only and audio-text evaluation). Results are ranked using Borda count. The ``Audio'' column shows the model's rank on MAEB(audio-only) for reference. We provide averages across all tasks, and per task category. ``Eng.'' shows the average for English-only tasks, ``Multi.'' shows the average excluding tasks with no linguistic content (zxx), and ``Aud.'' shows the average for audio-only tasks. Task categories are abbreviated as: Classification (Clf), Multilabel Classification (M.Clf), Pair Classification (PC), Reranking (Rrnk), Clustering (Clust), Audio Retrieval (A. Rtrvl), Cross-modal Retrieval (X. Rtrvl), Zero-shot Classification (Zero Clf.). We highlight the best score in \textbf{bold} and the best score with each model category using a grey cell.
    }
    \label{tab:maeb-performance}
    \resizebox{\textwidth}{!}{
    \setlength{\tabcolsep}{4pt}
    {\footnotesize
    \begin{tabular}{lcc|ccccc|cccccccc}
    \toprule
    & \multicolumn{2}{c}{\textbf{Rank} ($\downarrow$)} & \multicolumn{5}{c}{\textbf{Average}} & \multicolumn{8}{c}{\textbf{Average per Category}} \\
    \cmidrule(r){2-3} \cmidrule{4-8} \cmidrule(l){9-16}
    \textbf{Model} & MAEB & Audio & All & Cat. & Eng. & Multi. & Aud. & Clf & M.Clf & PC & Rrnk & Clust & A. Rtrvl & X. Rtrvl & Zero Clf. \\
    \midrule
\multicolumn{16}{c}{\vspace{2mm} \normalsize \texttt{MAEB}} \\
\textcolor{gray}{Number of datasets} & & & \textcolor{gray}{(30)} & \textcolor{gray}{(30)} & \textcolor{gray}{(15)} & \textcolor{gray}{(23)} & \textcolor{gray}{(19)} & \textcolor{gray}{(10)} & \textcolor{gray}{(2)} & \textcolor{gray}{(3)} & \textcolor{gray}{(1)} & \textcolor{gray}{(3)} & \textcolor{gray}{(1)} & \textcolor{gray}{(8)} & \textcolor{gray}{(2)} \\
\midrule
\textbf{Large audio-language models}\\
\midrule
LCO-Embedding-Omni-7B & \cellcolor{gray!20}1 & 5 & \cellcolor{gray!20}\textbf{52.2} & \cellcolor{gray!20}\textbf{55.6} & \cellcolor{gray!20}\textbf{50.9} & \cellcolor{gray!20}\textbf{53.6} & \cellcolor{gray!20}\textbf{52.2} & 58.0 & \cellcolor{gray!20}\textbf{45.7} & \cellcolor{gray!20}\textbf{67.3} & 78.7 & 1.7 & \cellcolor{gray!20}78.2 & \cellcolor{gray!20}\textbf{50.3} & \cellcolor{gray!20}\textbf{64.5} \\
Qwen2-Audio-7B & 2 & \cellcolor{gray!20}1 & 33.7 & 34.0 & 30.1 & 27.6 & 50.8 & \cellcolor{gray!20}\textbf{62.7} & 10.7 & 56.9 & \cellcolor{gray!20}80.8 & \cellcolor{gray!20}12.7 & 33.9 & 1.6 & 12.4 \\
LCO-Embedding-Omni-3B & 5 & 11 & 50.7 & 52.7 & 49.0 & 52.0 & 50.0 & 56.4 & 41.6 & 66.7 & 75.4 & 1.3 & 67.7 & 50.3 & 62.2 \\
\midrule
\textbf{Contrastive Alignment Models}\\
\midrule
larger\_clap\_general & \cellcolor{gray!20}4 & \cellcolor{gray!20}3 & \cellcolor{gray!20}32.2 & 37.1 & \cellcolor{gray!20}29.8 & \cellcolor{gray!20}28.3 & \cellcolor{gray!20}45.1 & \cellcolor{gray!20}51.7 & 2.3 & 51.9 & 66.8 & 6.6 & 93.2 & 9.8 & \cellcolor{gray!20}14.9 \\
larger\_clap\_music\_and\_speech & 6 & 4 & 31.9 & 37.0 & 29.7 & 28.1 & \cellcolor{gray!20}45.1 & 51.3 & 2.7 & 52.1 & 65.6 & 7.7 & 94.3 & 9.3 & 13.2 \\
clap-htsat-unfused & 7 & 9 & 30.0 & 35.9 & 29.1 & 25.9 & 42.4 & 45.2 & 1.8 & 52.6 & 66.5 & 12.5 & 88.8 & 8.8 & 11.3 \\
clap-htsat-fused & 10 & 14 & 30.7 & 36.2 & 29.0 & 27.3 & 43.2 & 44.5 & 4.0 & 52.0 & 61.3 & \cellcolor{gray!20}\textbf{22.7} & 82.8 & 9.2 & 13.2 \\
msclap-2023 & 12 & 12 & 31.1 & \cellcolor{gray!20}38.0 & 28.7 & 26.7 & 43.7 & 45.0 & 5.8 & \cellcolor{gray!20}53.6 & 75.4 & 15.2 & 87.3 & 9.4 & 12.6 \\
wav2clip & 14 & 13 & 25.5 & 32.7 & 23.2 & 21.5 & 38.8 & 39.4 & \cellcolor{gray!20}13.0 & \cellcolor{gray!20}53.6 & 68.9 & 6.0 & 68.9 & 1.0 & 10.8 \\
MuQ-MuLan-large & 16 & 16 & 27.0 & 37.7 & 22.2 & 22.3 & 40.9 & 40.7 & 10.3 & 51.9 & \cellcolor{gray!20}\textbf{85.4} & 4.3 & \cellcolor{gray!20}\textbf{95.2} & 1.1 & 12.6 \\
msclap-2022 & 19 & 28 & 29.8 & 36.1 & 29.7 & 27.3 & 39.9 & 38.3 & 7.6 & 51.7 & 62.9 & 19.9 & 82.4 & \cellcolor{gray!20}13.7 & 12.1 \\
\midrule
\textbf{Sequence-to-sequence Models}\\
\midrule
whisper-medium & \cellcolor{gray!20}3 & \cellcolor{gray!20}2 & \cellcolor{gray!20}46.7 & \cellcolor{gray!20}46.0 & \cellcolor{gray!20}41.7 & \cellcolor{gray!20}44.2 & \cellcolor{gray!20}48.2 & \cellcolor{gray!20}57.5 & \cellcolor{gray!20}22.3 & 53.9 & \cellcolor{gray!20}67.6 & 5.0 & \cellcolor{gray!20}69.5 & - & - \\
whisper-base & 8 & 6 & 42.7 & 41.9 & 38.7 & 39.6 & 44.4 & 53.0 & 11.7 & 52.1 & 65.0 & 5.0 & 64.5 & - & - \\
whisper-small & 9 & 7 & 43.2 & 42.6 & 38.8 & 40.5 & 44.8 & 53.4 & 15.5 & 52.6 & 64.2 & 3.9 & 66.2 & - & - \\
whisper-large-v3 & 11 & 8 & 42.1 & 42.8 & 37.3 & 40.0 & 43.8 & 50.7 & 17.1 & 52.5 & 63.9 & 3.4 & 69.1 & - & - \\
whisper-tiny & 13 & 10 & 42.1 & 41.8 & 37.0 & 39.0 & 44.0 & 51.0 & 14.9 & 51.5 & 63.4 & \cellcolor{gray!20}7.4 & 62.7 & - & - \\
speecht5\_multimodal & 22 & 37 & 25.8 & 29.6 & 23.2 & 23.5 & 38.4 & 42.9 & 5.9 & \cellcolor{gray!20}57.9 & 56.5 & 1.1 & 55.6 & \cellcolor{gray!20}1.3 & \cellcolor{gray!20}15.9 \\
mms-1b-l1107 & 25 & 27 & 38.6 & 37.0 & 32.5 & 37.4 & 40.5 & 48.1 & 12.4 & 51.5 & 58.8 & 1.0 & 50.3 & - & - \\
mms-1b-all & 29 & 29 & 38.8 & 37.5 & 33.3 & 38.0 & 40.6 & 47.4 & 14.9 & 52.8 & 59.5 & 1.6 & 48.8 & - & - \\
\midrule
\textbf{Audio Encoders}\\
\midrule
ast-finetuned-audioset-10-10-0.4593 & \cellcolor{gray!20}15 & \cellcolor{gray!20}15 & \cellcolor{gray!20}44.2 & \cellcolor{gray!20}50.1 & \cellcolor{gray!20}40.4 & \cellcolor{gray!20}36.8 & \cellcolor{gray!20}44.5 & \cellcolor{gray!20}48.9 & \cellcolor{gray!20}26.1 & 51.2 & 77.6 & 6.9 & \cellcolor{gray!20}90.2 & - & - \\
vggish & 17 & 17 & 39.1 & 45.8 & 38.0 & 34.9 & 40.9 & 41.8 & 9.7 & 52.8 & 78.7 & \cellcolor{gray!20}7.8 & 83.8 & - & - \\
wavlm-large & 18 & 18 & 37.9 & 41.1 & 35.4 & 36.6 & 39.7 & 43.9 & 7.1 & 52.3 & 68.8 & 2.4 & 71.8 & - & - \\
hubert-base-ls960 & 20 & 19 & 37.5 & 40.5 & 36.7 & 35.6 & 39.3 & 43.2 & 8.3 & 51.9 & 66.3 & 2.7 & 70.7 & - & - \\
yamnet & 21 & 20 & 38.0 & 44.9 & 37.1 & 32.6 & 39.0 & 40.1 & 16.6 & \cellcolor{gray!20}54.6 & \cellcolor{gray!20}81.7 & 1.6 & 74.5 & - & - \\
wav2vec2-lv-60-espeak-cv-ft & 23 & 22 & 38.5 & 35.8 & 34.9 & 36.6 & 40.4 & 48.6 & 8.2 & 53.7 & 55.6 & 1.6 & 46.9 & - & - \\
wav2vec2-xls-r-2b & 24 & 23 & 38.7 & 37.5 & 35.8 & 34.1 & 40.5 & 48.4 & 7.8 & 50.8 & 62.9 & 1.4 & 53.7 & - & - \\
cnn14-esc50 & 30 & 21 & 33.2 & 38.4 & 34.0 & 31.8 & 35.0 & 33.5 & 9.4 & 54.2 & 53.8 & 7.4 & 72.3 & - & - \\
    \bottomrule
    \end{tabular}
    }
    } 
\end{table*}

\subsection{Key Findings on Model Performance}

Our comprehensive evaluation over MAEB reveals four critical weaknesses in current audio representations, each suggesting specific directions for future model development.

\paragraph{(a) No universal audio model exists.} Speech-trained models (Wav2Vec2, Whisper) underperform on music tasks, while music-focused models (CLAP variants) struggle with speech understanding, confirming that no single architecture achieves universal audio representation. As shown in \autoref{tab:maeb-performance}, Whisper-medium achieves strong classification performance (51.7\%) but struggles with clustering (5.0\%), whereas CLAP variants show more balanced performance across categories but lower peak scores on speech-specific tasks.

Models pretrained on massively multilingual automatic speech recognition data (SeamlessM4T, MMS) substantially outperform other approaches on multilingual classification---SeamlessM4T-v2-large achieves the best performance on 10 of 12 languages in MInDS-14 (\autoref{tab:minds14_results}). Yet this strength does not transfer to music or environmental sound tasks. Conversely, audio-text models like CLAP variants, despite their strength on environmental audio, score below 15\% across all languages on MInDS-14, near random chance for intent classification.

While LCO-Embedding-Omni-7B and Qwen2-Audio-7B both rank at the top and leverage similar training approaches, they obtain drastically different scores on cross-modal retrieval tasks (50.3\% and 1.6\%, respectively). This highlights that scale and multimodal pretraining do not guarantee balanced performance. This indicates that training paradigm, data curation, and architectural choices matter more than parameter count for general audio embedding quality, echoing findings from text embedding research.

\textit{Direction:} The specialization gap calls for domain-agnostic architectures that generalize across speech, music, and environmental sound without sacrificing domain-specific capabilities. Future work should explore unified training objectives and architectural innovations that maintain strong performance across the full acoustic spectrum.

\paragraph{(b) Multilingual audio understanding remains unsolved.} Despite evaluation across 200+ languages via SIB-FLEURS~\cite{adelani2023sib200} (94 languages), CommonVoice~\cite{ardila2019common} (43 languages), MInDS-14~\cite{DBLP:journals/corr/abs-2104-08524} (14 languages), VoxPopuli~\cite{wang-etal-2021-voxpopuli} (5 languages), and FLEURS~\cite{schmidt2025fleursslumassivelymultilingualbenchmark} (102 languages), models demonstrate a strong bias toward high-resource languages with severely degraded performance on African, Indigenous, and minority languages. On SIB-FLEURS classification (\autoref{tab:sibfleurs_part1}), high-resource European languages achieve 40--60\% accuracy while low-resource languages like Umbundu, Yoruba, and Xhosa remain below 20\% even for the best models.

This disparity becomes catastrophic for cross-modal tasks. While audio-to-audio retrieval maintains reasonable performance across languages (50--99\% on JamAlt, \autoref{tab:jamalt_a2a_part1}), cross-modal audio-text retrieval collapses in multilingual settings. On FLEURS retrieval across 102 languages (Tables~\ref{tab:fleurs_a2t_part1}--\ref{tab:fleurs_t2a_part3}), even the best CLAP models achieve below 3\% for most language pairs, with audio-to-text and text-to-audio retrieval scores often below 1\%. Current audio-text alignment approaches, trained predominantly on English data, fail completely to generalize to multilingual scenarios---a critical gap for global audio retrieval applications.

\textit{Direction:} We recommend extending contrastive audio-text pretraining to multilingual corpora and implementing cross-lingual transfer learning to leverage high-resource language knowledge for the 100+ languages where current models achieve near-random performance.

\paragraph{(c) Acoustic versus linguistic representations trade off.} Multilingual evaluation reveals fundamental trade-offs between acoustic and linguistic representations that current architectures cannot reconcile. On VoxPopuli tasks (\autoref{tab:voxpopuli_results}), CLAP-htsat-unfused achieves 94.4\% on gender identification but only 30.0\% on language identification, while Whisper-medium shows the inverse pattern (59.2\% vs 99.4\%). This suggests that models optimized for acoustic properties (timbre, speaker characteristics) develop fundamentally different representations than those optimized for linguistic content.

This trade-off extends to audio-text alignment more broadly. The performance gap between audio-only and audio-text tasks is substantial: as shown in \autoref{tab:maeb-performance}, AST achieves 44.2\% overall but cannot perform cross-modal tasks (showing ``-'' for Retrieval and Zero-shot Classification), while CLAP variants achieve around 30-32\% overall despite enabling cross-modal tasks. Within audio-text tasks, most models show weak retrieval performance (CLAP variants around 8-14\%), though LCO-Embedding-Omni-7B achieves 50.3\% cross-modal retrieval and 64.5\% zero-shot classification, demonstrating that stronger cross-modal alignment is possible with appropriate training. Models struggle especially with complex audio scenes and abstract musical concepts, suggesting current training objectives fail to capture deeper semantic relationships beyond surface-level correspondences.

\textit{Direction:} Future architectures should explore disentangled representations or multi-task learning approaches that capture both acoustic properties (speaker, timbre) and linguistic content simultaneously, enabling models to perform well on both gender identification and language identification without sacrificing one for the other.

\paragraph{(d) Clustering exposes fundamental representation gaps.} Clustering tasks prove universally challenging across all evaluated models, revealing a consistent weakness in semantic structure. Even the best-performing model on clustering (clap-htsat-fused) achieves only 22.7\%, while top-ranked models show inconsistent clustering performance: Qwen2-Audio-7B (2nd overall) scores 12.7\%, LCO-Embedding-Omni-7B (1st overall, highest average scores) achieves only 1.7\%, and whisper-medium (3rd overall) reaches just 5.0\%. This disconnect between supervised and unsupervised task performance suggests that current audio embeddings lack the semantic organization necessary for grouping related audio without explicit labels---a fundamental limitation for applications requiring audio organization, discovery, or similarity-based retrieval at scale.

\textit{Direction:} Incorporating clustering-aware losses or contrastive objectives that explicitly encourage semantically coherent embedding neighborhoods could address this gap, enabling applications that require audio organization without explicit labels.

\subsection{Correlation with Audio LLM Performance}
To assess whether MAEB scores translate to real-world multimodal capabilities, we examine the relationship between encoder quality and Audio LLM performance on the MMAU benchmark \citep{sakshi2024mmaumassivemultitaskaudio}. 
MMAU evaluates multimodal audio understanding through expert-annotated questions organized into three domains: Speech, Music, and Sound. 
To ensure a direct comparison, we compute the encoder's embedding quality using a subset of 26 classification tasks from MAEB+ selected to align with these three domains (see \autoref{app:correlation_tasks} for the full task list).

\begin{figure}[t]
    \centering
    \includegraphics[width=0.9\linewidth]{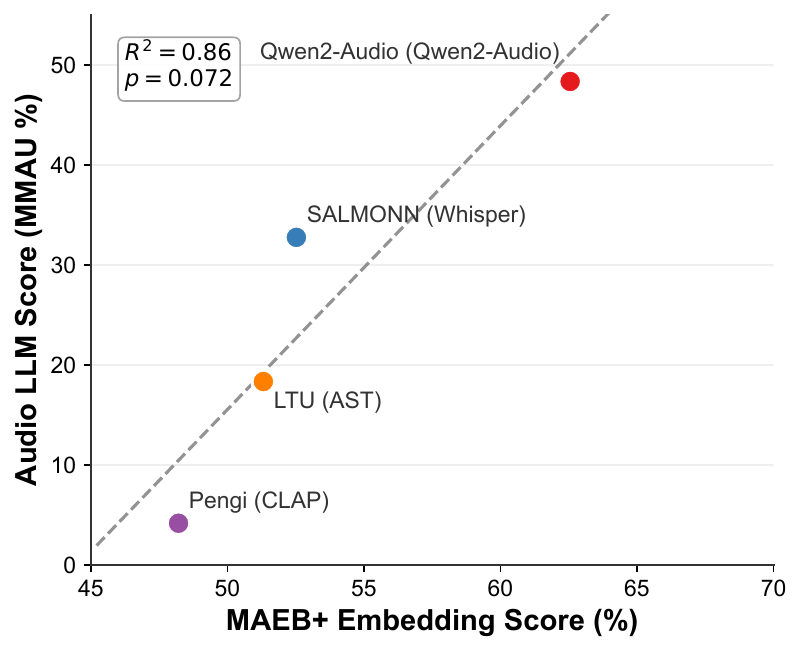}
    \caption{
    \textbf{MAEB+ embedding quality correlates with Audio LLM performance.}
    MMAU evaluates Audio LLMs across Speech, Music, and Sound, the same domains 
    covered by MAEB+. Each point plots an Audio LLM's overall MMAU score (y-axis, 
    averaged across domains) against its encoder's MAEB+ score (x-axis, 
    computed from 26 classification tasks aligned with MMAU domains). Preliminary correlation (R²=0.86, p=0.072, n=4) suggests a positive relationship between embedding quality and downstream reasoning, though the small sample size and statistical marginality warrant caution in interpreting this relationship.}
    \label{fig:maeb_correlation}
\end{figure}

We compare four Audio LLMs that use different encoder architectures: Qwen2-Audio
(Qwen2-Audio encoder), SALMONN (Whisper), LTU (AST), and Pengi (CLAP). \autoref{fig:maeb_correlation} shows a preliminary positive correlation across four models. Given the strong correlation between MAEB and MAEB(extended) established in \autoref{sec:benchmark-construction}, this result suggests that the efficient MAEB benchmark serves as a reliable predictive signal for downstream Audio LLM performance.

\section{Limitations}

\paragraph{Technical Constraints} While our evaluation includes 50+ models spanning multiple architectures, this represents only a subset of available models. Audio length management poses challenges: models with native limits below 30 seconds retain those settings, while others are limited to 30 seconds for memory management, restricting applicability to long-form content like podcasts or lectures. While future standardization around pre-processing pipelines could streamline evaluation, our approach currently reflects the diverse sampling rate requirements inherent to different audio domains rather than a benchmark limitation. Large-scale models (Whisper-large-v3: 
1.55B parameters, Wav2Vec2-XLS-R-2B: 2B parameters) require substantial 
computational resources, limiting accessibility.

\begin{figure*}[t]
    \centering
    \includegraphics[width=\linewidth]{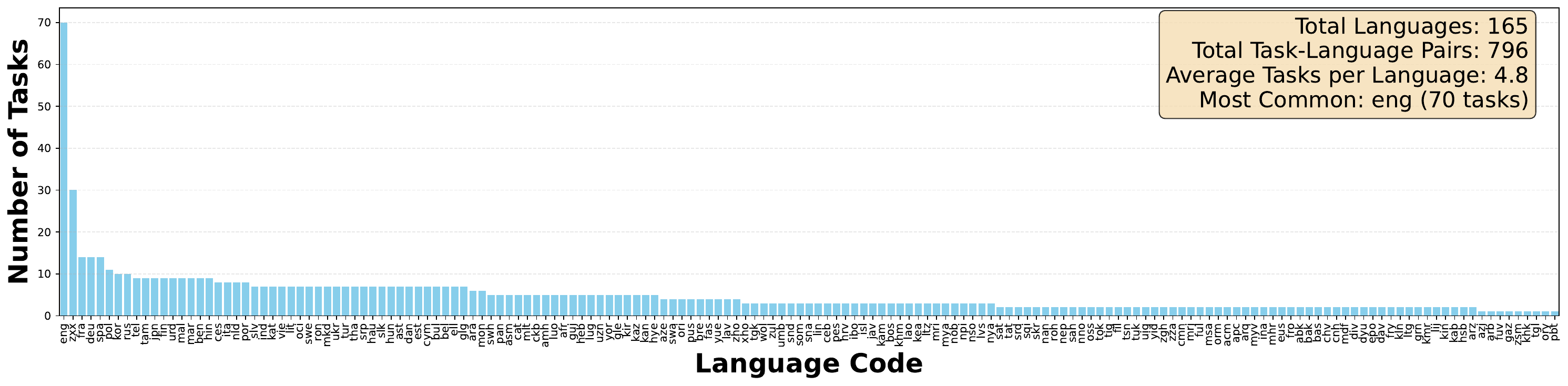}
    \caption{Language distribution in the MAEB+ collection. English dominates with 70 tasks. We use \texttt{zxx} (No Linguistic Content) to tag datasets with no languages present. }
    \label{fig:maeb_language_distribution}
\end{figure*}

\paragraph{Dataset Coverage Limitations} The benchmark exhibits several coverage gaps. Domain representation skews toward Western musical traditions and standard speech patterns. Language coverage, while spanning 100+ languages, remains limited for many underrepresented language families, with some languages appearing in only a single datasets, preventing comprehensive cross-task evaluation. The language distribution of MAEB is shown in \autoref{fig:maeb_language_distribution}. 

\begin{table*}[t]
    \centering
    \caption{MAEB+ Audio-Only Tasks Overview. Tasks are grouped by type and show MAEB benchmark membership, dataset size, total audio duration, language coverage, domains, and main evaluation metric. * denotes values from huge datasets. }
    \resizebox{\linewidth}{!}{
    \begin{tabular}{llcccclc}
    \toprule
    \textbf{Dataset} & \textbf{Citation} & \textbf{MAEB} & \textbf{N. Samples} & \textbf{Total Duration(s)} & \textbf{N. Langs} & \textbf{Domains} & \textbf{Main Metric} \\
    \midrule
    \multicolumn{8}{l}{\textit{Any2AnyRetrieval}} \\
    \hline
    JamAltArtistA2ARetrieval & \cite{cifka-2024-jam-alt} & \checkmark & 6.7k & 22992 & 4 & Music & ndcg\_at\_10 \\
    \hline
    \multicolumn{8}{l}{\textit{Classification}} \\
    \hline
    AmbientAcousticContext & \cite{10.1145/3379503.3403535} &  & 1k & 1046 & 1 & Spoken, Speech & accuracy \\
    BeijingOpera & \cite{6853981} & \checkmark & 236 & 393 & 1 & Music & accuracy \\
    BirdCLEF & \cite{birdclef2025} & \checkmark & 1k & 33602 & 1 & Spoken, Speech, Bioacoustics & accuracy \\
    CREMA\_D & \cite{Cao2014-ih} & \checkmark & 7.4k & 18924 & 1 & Emotion & accuracy \\
    CommonLanguageAgeDetection & \cite{ganesh_sinisetty_2021_5036977} & \checkmark & 2k & 8685 & 1 & Spoken, Scene, Speech & accuracy \\
    CommonLanguageGenderDetection & \cite{ganesh_sinisetty_2021_5036977} &  & 2k & 8777 & 1 & Spoken, Scene, Speech & accuracy \\
    CommonLanguageLanguageDetection & \cite{ganesh_sinisetty_2021_5036977} &  & 2k & 8637 & 1 & Spoken, Scene, Speech & accuracy \\
    ESC50 & \cite{piczak2015dataset} &  & 2k & 10000 & 1 & Spoken & accuracy \\
    FSDD & \cite{zohar2018free} &  & 300 & 129 & 1 & Music & accuracy \\
    GTZANGenre & \cite{1021072} & \checkmark & 1k & 30024 & 1 & Music & accuracy \\
    GunshotTriangulation & \cite{raponi2021soundgunsdigitalforensics} &  & 88 & 132 & 1 &  & accuracy \\
    IEMOCAPEmotion & \cite{busso2008iemocap} &  & 10k & 44775 & 1 & Spoken, Emotion & accuracy \\
    IEMOCAPGender & \cite{busso2008iemocap} & \checkmark & 10k & 44775 & 1 & Spoken, Speech & accuracy \\
    LibriCount & \cite{Stoter_2018} &  & 5.7k & 28600 & 1 & Speech & accuracy \\
    MInDS14 & \cite{DBLP:journals/corr/abs-2104-08524} & \checkmark & 7k & 78225 & 12 & Speech, Spoken & accuracy \\
    MridinghamStroke & \cite{6637633} &  & 7k & 2462 & 1 & Music & accuracy \\
    MridinghamTonic & \cite{6637633} & \checkmark & 7k & 2462 & 1 & Music & accuracy \\
    NSynth & \cite{engel2017neuralaudiosynthesismusical} &  & 3k & 12008 & 1 & Music & accuracy \\
    SpeechCommands & \cite{DBLP:journals/corr/abs-1804-03209} &  & 4.9k & 4890 & 1 & Speech & accuracy \\
    SpokeNEnglish & \cite{groh2024spoken100crosslingualbenchmarkingdataset} &  & 3.2k & 2829 & 1 & Spoken & accuracy \\
    SpokenQAForIC & \cite{shon2023sluephase2benchmarksuite} &  & 6.1k & 12967 & 1 & Spoken & accuracy \\
    TUTAcousticScenes & \cite{Mesaros2018_DCASE} &  & 2k & 20000 & 1 & AudioScene & accuracy \\
    UrbanSound8k & \cite{Salamon:UrbanSound:ACMMM:14} &  & 8.7k & 31501 & 1 & AudioScene & accuracy \\
    VocalSound & \cite{Gong_2022} &  & 3.6k & 14934 & 1 & Spoken & accuracy \\
    VoxCelebSA & \cite{shon2022sluenewbenchmarktasks} & \checkmark & 3.4k & 27337 & 1 & Spoken & accuracy \\
    VoxLingua107\_Top10 & \cite{valk2020voxlingua107datasetspokenlanguage} &  & 972 & 9634 & 1 & Speech & accuracy \\
    VoxPopuliAccentID & \cite{wang-etal-2021-voxpopuli} &  & 2k & 22381 & 1 & Spoken, Speech & accuracy \\
    VoxPopuliGenderID & \cite{wang-etal-2021-voxpopuli} &  & 500 & 5122 & 5 & Spoken, Speech & accuracy \\
    VoxPopuliLanguageID & \cite{wang-etal-2021-voxpopuli} & \checkmark & 500 & 5122 & 5 & Spoken, Speech & accuracy \\
    \hline
    \multicolumn{8}{l}{\textit{Clustering}} \\
    \hline
    AmbientAcousticContextClustering & \cite{10.1145/3379503.3403535} &  & 1k & 1046 & 1 & Spoken, Speech & v\_measure \\
    CREMA\_DClustering & \cite{Cao2014-ih} & \checkmark & 2k & 5246 & 1 & Speech & v\_measure \\
    ESC50Clustering & \cite{piczak2015dataset} &  & 2k & 10000 & 1 & Spoken, Speech & v\_measure \\
    GTZANGenreClustering & \cite{1021072} &  & 1k & 30024 & 1 & Music & v\_measure \\
    MusicGenreClustering & \cite{homburg2005benchmark} &  & 1.9k & 18965 & 1 & Music & v\_measure \\
    VehicleSoundClustering & \cite{bazilinskyy2018auditory} & \checkmark & 1.7k & 6819 & 1 & Scene & v\_measure \\
    VoiceGenderClustering & \cite{Chung18b} &  & 2k & 14559 & 1 & Spoken & v\_measure \\
    VoxCelebClustering & \cite{shon2022sluenewbenchmarktasks} &  & 2k & 16124 & 1 & Spoken, Speech & v\_measure \\
    VoxPopuliAccentClustering & \cite{wang-etal-2021-voxpopuli} &  & 2k & 23097 & 1 & Spoken, Speech & v\_measure \\
    VoxPopuliGenderClustering & \cite{wang-etal-2021-voxpopuli} & \checkmark & 500 & 5122 & 5 & Spoken, Speech & v\_measure \\
    \hline
    \multicolumn{8}{l}{\textit{MultilabelClassification}} \\
    \hline
    AudioSet & \cite{audioset} &  & * & * & 1 & Web, Music, Speech... & lrap \\
    AudioSetMini & \cite{audioset} &  & 2.2k & 21316 & 1 & Web, Music, Speech... & lrap \\
    BirdSet & \cite{rauch2024birdsetlargescaledatasetaudio} &  & * & * & 1 & Spoken, Speech, Bioacoustics & accuracy \\
    FSD2019Kaggle & \cite{fonseca2021fsd50k} & \checkmark & 9k & 92834 & 1 & Web & accuracy \\
    FSD50K & \cite{fonseca2021fsd50k} &  & 2k & 21157 & 1 & Web & accuracy \\
    SIBFLEURS & \cite{schmidt2025fleursslumassivelymultilingualbenchmark} & \checkmark & 11.4k & 152396 & 101 & Encyclopaedic & accuracy \\
    \hline
    \multicolumn{8}{l}{\textit{PairClassification}} \\
    \hline
    CREMADPairClassification & \cite{Cao2014-ih} & \checkmark & 7.4k & 37858 & 1 & Spoken & max\_ap \\
    ESC50PairClassification & \cite{piczak2015dataset} &  & 2k & 20000 & 1 & Encyclopaedic & max\_ap \\
    NMSQAPairClassification & \cite{lin2022dualdiscretespokenunit} & \checkmark & 171 & 3245 & 1 & Spoken & max\_ap \\
    VocalSoundPairClassification & \cite{Gong_2022} &  & 720 & 6010 & 1 & Spoken & max\_ap \\
    VoxPopuliAccentPairClassification & \cite{wang-etal-2021-voxpopuli} & \checkmark & 7.4k & 169638 & 1 & Spoken & max\_ap \\
    \hline
    \multicolumn{8}{l}{\textit{Reranking}} \\
    \hline
    ESC50AudioReranking & \cite{piczak2015dataset} &  & 4.4k & 22000 & 1 & AudioScene & map\_at\_1000 \\
    FSDnoisy18kAudioReranking & \cite{fonseca2019fsdnoisy18k} &  & 4.2k & 21924 & 1 & AudioScene & map\_at\_1000 \\
    GTZANAudioReranking & \cite{1021072} & \checkmark & 1.4k & 42033 & 1 & Music & map\_at\_1000 \\
    UrbanSound8KAudioReranking & \cite{Salamon:UrbanSound:ACMMM:14} &  & 5.2k & 17904 & 1 & Spoken & map\_at\_1000 \\
    VocalSoundAudioReranking & \cite{Gong_2022} &  & 4.2k & 17371 & 1 & Spoken & map\_at\_1000 \\
    \bottomrule
    \end{tabular}
    }
    \label{tab:maeb_audio_overview}
\end{table*}

\begin{table*}[t]
    \centering
    \caption{MAEB+ Audio-Text Cross-Modal Tasks Overview. Tasks include zero-shot classification and bidirectional retrieval between audio and text modalities, with dataset size, total audio duration, and main evaluation metric. * denotes values from huge datasets. }
    \resizebox{\linewidth}{!}{
    \begin{tabular}{llccccllc}
    \toprule
    \textbf{Dataset} & \textbf{Citation} & \textbf{MAEB} & \textbf{N. Samples} & \textbf{Total Secs} & \textbf{N. Langs} & \textbf{Modality} & \textbf{Domains} & \textbf{Main Metric} \\
    \midrule
    \multicolumn{9}{l}{\textit{Audio-to-Text Retrieval}} \\
    \hline
    AudioCapsA2TRetrieval & \cite{kim2019audiocaps} &  & 5.3k & 8708 & 2 & a2t & Encyclopaedic, Written & cv\_recall\_at\_5 \\
    AudioSetStrongA2TRetrieval & \cite{hershey2021benefittemporallystronglabelsaudio} &  & 1k & 5065 & 1 & a2t & AudioScene & cv\_recall\_at\_5 \\
    CMUArcticA2TRetrieval & \cite{cmu-lti-03-177} &  & 2.6k & 4134 & 1 & a2t & Spoken & cv\_recall\_at\_5 \\
    ClothoA2TRetrieval & \cite{drossos2019clothoaudiocaptioningdataset} &  & 6.6k & 23636 & 1 & a2t & Encyclopaedic, Written & cv\_recall\_at\_5 \\
    CommonVoiceMini17A2TRetrieval & \cite{ardila2019common} &  & 46.8k & 120220 & 50 & a2t & Spoken & cv\_recall\_at\_5 \\
    CommonVoiceMini21A2TRetrieval & \cite{ardila2019common} &  & 58.5k & 149040 & 114 & a2t & Spoken & cv\_recall\_at\_5 \\
    EmoVDBA2TRetrieval & \cite{adigwe2018emotional} &  & 2.9k & 7231 & 1 & a2t & Spoken & cv\_recall\_at\_5 \\
    FleursA2TRetrieval & \cite{conneau2023fleurs} &  & 155620 & 1018098  & 102 & a2t & Spoken & cv\_recall\_at\_5 \\
    GigaSpeechA2TRetrieval & \cite{GigaSpeech2021} &  & 13.5k & 44982 & 1 & a2t & Spoken & cv\_recall\_at\_5 \\
    GoogleSVQA2TRetrieval & \cite{heigold2025massive} &  & 342.9k & 879901 & 20 & a2t & Spoken & cv\_recall\_at\_5 \\
    HiFiTTSA2TRetrieval & \cite{bakhturina2021hi} &  & 600 & 1280 & 1 & a2t & Spoken & cv\_recall\_at\_5 \\
    JLCorpusA2TRetrieval & \cite{james2018open} &  & 2.5k & 5083 & 1 & a2t & Spoken & cv\_recall\_at\_5 \\
    JamAltLyricA2TRetrieval & \cite{cifka-2024-jam-alt} & \checkmark & 6.7k & 11496 & 4 & a2t & Music & ndcg\_at\_10 \\
    LibriTTSA2TRetrieval & \cite{zen2019librittscorpusderivedlibrispeech} &  & 9.4k & 30433 & 1 & a2t & Spoken & cv\_recall\_at\_5 \\
    MACSA2TRetrieval & \cite{martinmorato2021groundtruthreliabilitymultiannotator} &  & 786 & 3930 & 1 & a2t & AudioScene & cv\_recall\_at\_5 \\
    MusicCapsA2TRetrieval & \cite{agostinelli2023musiclmgeneratingmusictext} &  & 8.6k & 42844 & 1 & a2t & Music & cv\_recall\_at\_5 \\
    SoundDescsA2TRetrieval & \cite{Koepke2022} &  & * & * & 1 & a2t & Encyclopaedic, Written & cv\_recall\_at\_5 \\
    UrbanSound8KA2TRetrieval & \cite{Salamon:UrbanSound:ACMMM:14} &  & 10.2k & 18334 & 1 & a2t & AudioScene & cv\_recall\_at\_5 \\
    \hline
    \multicolumn{9}{l}{\textit{Text-to-Audio Retrieval}} \\
    \hline
    AudioCapsT2ARetrieval & \cite{kim2019audiocaps} &  & 5.3k & 8708 & 2 & t2a & Encyclopaedic, Written & cv\_recall\_at\_5 \\
    AudioSetStrongT2ARetrieval & \cite{hershey2021benefittemporallystronglabelsaudio} &  & 1k & 5065 & 1 & t2a & AudioScene & cv\_recall\_at\_5 \\
    CMUArcticT2ARetrieval & \cite{cmu-lti-03-177} &  & 2.6k & 4134 & 1 & t2a & Spoken & cv\_recall\_at\_5 \\
    ClothoT2ARetrieval & \cite{drossos2019clothoaudiocaptioningdataset} & \checkmark & 6.6k & 23636 & 1 & t2a & Encyclopaedic, Written & cv\_recall\_at\_5 \\
    CommonVoiceMini17T2ARetrieval & \cite{ardila2019common} &  & 46.8k & 120220 & 50 & t2a & Spoken & cv\_recall\_at\_5 \\
    CommonVoiceMini21T2ARetrieval & \cite{ardila2019common} & \checkmark & 58.5k & 149040 & 114 & t2a & Spoken & cv\_recall\_at\_5 \\
    EmoVDBT2ARetrieval & \cite{adigwe2018emotional} &  & 2.9k & 7231 & 1 & t2a & Spoken & cv\_recall\_at\_5 \\
    FleursT2ARetrieval & \cite{conneau2023fleurs} & \checkmark & 155620 & 1018098 & 102 & t2a & Spoken & cv\_recall\_at\_5 \\
    GigaSpeechT2ARetrieval & \cite{GigaSpeech2021} & \checkmark & 13.5k & 44982 & 1 & t2a & Spoken & cv\_recall\_at\_5 \\
    GoogleSVQT2ARetrieval & \cite{heigold2025massive} &  & 342.9k & 879901 & 20 & t2a & Spoken & cv\_recall\_at\_5 \\
    HiFiTTST2ARetrieval & \cite{bakhturina2021hi} &  & 600 & 1280 & 1 & t2a & Spoken & cv\_recall\_at\_5 \\
    JLCorpusT2ARetrieval & \cite{james2018open} &  & 2.5k & 5083 & 1 & t2a & Spoken & cv\_recall\_at\_5 \\
    JamAltLyricT2ARetrieval & \cite{cifka-2024-jam-alt} &  & 6.7k & 11496 & 4 & t2a & Music & ndcg\_at\_10 \\
    LibriTTST2ARetrieval & \cite{zen2019librittscorpusderivedlibrispeech} &  & 9.4k & 30433 & 1 & t2a & Spoken & cv\_recall\_at\_5 \\
    MACST2ARetrieval & \cite{martinmorato2021groundtruthreliabilitymultiannotator} & \checkmark & 786 & 3930 & 1 & t2a & AudioScene & cv\_recall\_at\_5 \\
    MusicCapsT2ARetrieval & \cite{agostinelli2023musiclmgeneratingmusictext} &  & 8.6k & 42844 & 1 & t2a & Music & cv\_recall\_at\_5 \\
    SoundDescsT2ARetrieval & \cite{Koepke2022} &  & * & * & 1 & t2a & Encyclopaedic, Written & cv\_recall\_at\_5 \\
    SpokenSQuADT2ARetrieval & \cite{li2018spokensquad} & \checkmark & 600 & 3557 & 1 & t2a & Academic, Encyclopaedic, Non-fiction & cv\_recall\_at\_5 \\
    UrbanSound8KT2ARetrieval & \cite{Salamon:UrbanSound:ACMMM:14} & \checkmark & 10.2k & 18334 & 1 & t2a & AudioScene & cv\_recall\_at\_5 \\
    \hline
    \multicolumn{9}{l}{\textit{Zero-shot Classification}} \\
    \hline
    ESC50\_Zeroshot & \cite{piczak2015dataset} &  & 2k & 10000 & 1 & a2t & Spoken & accuracy \\
    RavdessZeroshot & \cite{10.1371/journal.pone.0196391} & \checkmark & 1.4k & 5329 & 1 & a2t & Spoken & accuracy \\
    SpeechCommandsZeroshotv0.01 & \cite{DBLP:journals/corr/abs-1804-03209} &  & 2.6k & 2567 & 1 & a2t & Spoken & accuracy \\
    SpeechCommandsZeroshotv0.02 & \cite{DBLP:journals/corr/abs-1804-03209} & \checkmark & 4.1k & 4074 & 1 & a2t & Spoken & accuracy \\
    UrbanSound8kZeroshot & \cite{Salamon:UrbanSound:ACMMM:14} &  & 2k & 7378 & 1 & a2t & AudioScene & accuracy \\
    \bottomrule
    \end{tabular}
    }
    \label{tab:maeb_audio_text_overview}
\end{table*}

Task coverage across 30 tasks in MAEB (98 in MAEB+) and 7 categories still lacks certain capabilities including audio generation quality assessment and real-time processing evaluation. Ecological validity is limited as many tasks use clean, studio-recorded audio that does not reflect real-world conditions with noise, reverberation, and compression artifacts.

\section{Related Work}

 \textbf{Text Embedding Benchmarks} Large, standardized benchmarks have been critical for driving progress in representation learning. For text, MTEB provides a comprehensive evaluation suite spanning 8 task families across 58 datasets and 112 languages, enabling systematic assessment of generalization beyond task-specific setups~\cite{muennighoff2022mteb}. Recent expansions toward massive multilingual and multimodal evaluations such as MMTEB for multilingual text embeddings and MIEB for image embeddings reinforce the value of broad, regularly maintained leaderboards with consistent protocols~\cite{enevoldsen2025mmteb,xiao2025mieb}. These efforts motivate analogous, up-to-date benchmarking for audio embeddings.

\textbf{Audio Representation Benchmarks} HEAR~\cite{turian2022hear} represents one of the first attempts to evaluate general-purpose audio embeddings across diverse domains such as speech recognition, music tagging, and environmental sound classification. Evaluating 29 models on 19 downstream tasks, HEAR primarily tests pretrained features with simple classifiers like multilayer perceptrons (MLPs), leaving room for exploration with more complex architectures.

Despite this progress, comprehensive evaluation of audio embeddings remains limited. Task coverage is narrow, focusing primarily on classification while neglecting systematic evaluation across fundamental applications such as retrieval, and clustering. Similarly, zero-shot performance testing remains fragmented with prior work exploring approaches such as using textual label embeddings, sentence descriptions, or even image embeddings of sound classes~\cite{xie2021zero, mercea2022zero}, but these efforts are isolated and not integrated into comprehensive evaluation frameworks. Large-scale multilingual support also remains an outstanding issue despite the importance of supporting diverse languages and accents~\cite{xu2024multilingual}.
Maintenance and reproducibility pose ongoing challenges, with outdated datasets and inconsistent evaluation protocols hindering fair model comparison of current models. MAEB addresses these limitations by building into an existing and maintained framework for evaluating embeddings, drawing on lessons from MTEB while adapting to the unique challenges of audio representation learning.
Separately, AudioBench~\cite{wang2024audiobench} and MMAU~\cite{sakshi2024mmaumassivemultitaskaudio} focus on evaluating AudioLLMs rather than embedding models. AudioBench evaluates instruction-following capabilities across eight tasks using 26 datasets, while MMAU introduces multimodal benchmarks requiring reasoning across speech, sound, and music domains.

\section{Conclusion}
We introduce the Massive Audio Embedding Benchmark (MAEB), comprising 30 tasks across 100+ languages with baselines from 50+ models.

Our evaluation reveals critical gaps in current audio representations. No single model achieves universal performance: LCO-Embedding-Omni-7B ranks first overall, achieving the strongest cross-modal retrieval (50.3\%) and zero-shot classification (64.5\%) averages in our MAEB evaluation. Qwen2-Audio-7B ranks second overall and ranks first on audio-only tasks, excelling particularly in reranking (80.8\%) and clustering (12.7\%). Speech-pretrained models (e.g., Whisper) perform strongly on audio-only tasks but cannot support cross-modal evaluation, while contrastive audio-text models (e.g., CLAP variants) provide cross-modal capabilities but remain weak on multilingual speech tasks.

Clustering proves universally challenging (best model: 22.7\%), exposing fundamental limitations in semantic structure. We observe stark trade-offs between acoustic and linguistic features, with models excelling at gender identification struggling on language identification and vice versa. Cross-modal multilingual retrieval reveals a stark capability gap: LCO models achieve 50\%+ accuracy across 100+ languages, while most other models (CLAP, Whisper, ASR encoders) remain below 2\%, highlighting the critical role of speech-text alignment for this task. Preliminary analysis across four Audio LLMs suggests a positive relationship between MAEB encoder quality and downstream performance, validating the benchmark's relevance for multimodal audio understanding.

MAEB integrates into the MTEB ecosystem, enabling unified evaluation across text, image, and audio modalities. We release code, tasks, and leaderboards to support community-driven progress toward robust, multilingual audio representations.

\section*{Impact Statement}

Large benchmarks create barriers for low-resource communities and incur high environmental costs. We have reduced large datasets to reasonable sizes and include kilogram $CO_2$ measures per task, allowing users to assess environmental benchmarking costs.

\bibliography{icml2026}
\bibliographystyle{icml2026}

\newpage
\appendix
\onecolumn

\section{Tasks overview}
\label{sec:overview}
\begin{figure*}[t]
    \centering
    \includegraphics[width=\linewidth]{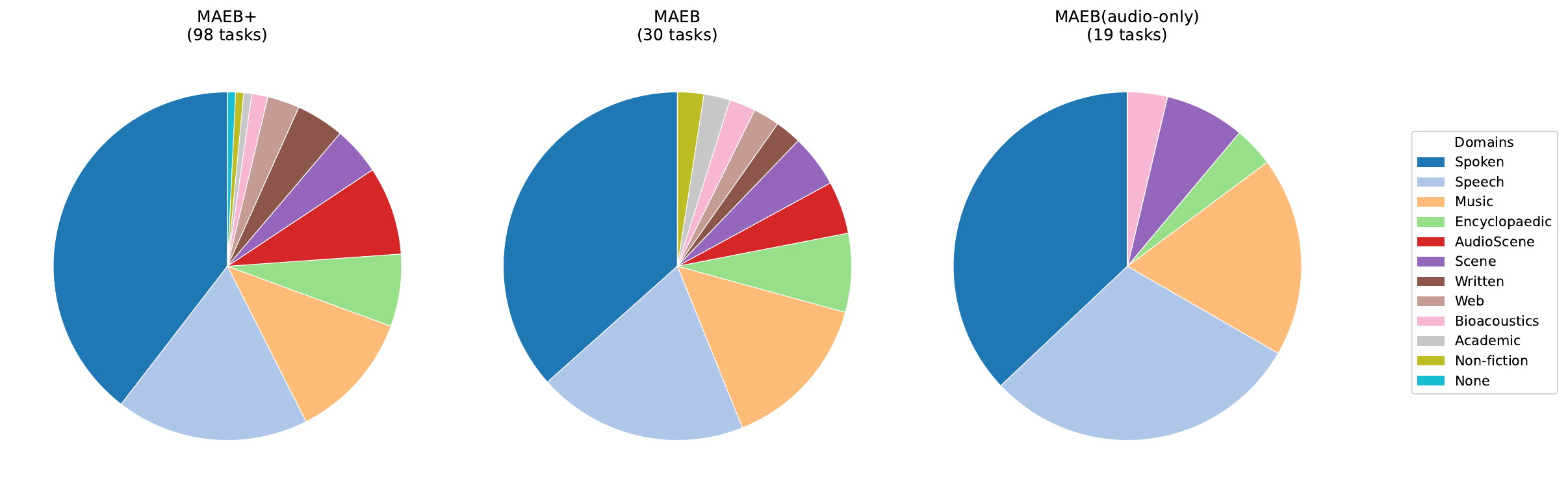}
    \caption{Domain distributions in the MAEB+ collection, MAEB, and MAEB(audio-only). }
    \label{fig:maeb_domain_distribution}
\end{figure*}

This appendix provides detailed information on all tasks within MAEB, including size, language, metrics, and other relevant details in \autoref{tab:maeb_audio_overview} and \autoref{tab:maeb_audio_text_overview}. The domain distribution of MAEB is shown in \autoref{fig:maeb_domain_distribution}. 

\section{Overview of Models}
\label{appdx:models}

All models used in the evaluations are listed in \autoref{tab: list of models}. 

\subsection{Audio Encoders}

\textbf{Transformer-based Models:} AST (Audio Spectrogram Transformer) \cite{gong2021astaudiospectrogramtransformer} applies vision transformer architecture to mel-spectrograms. For retrieval evaluation, we extract the pooler output embedding (768-dim), which corresponds to the [CLS] token representation that captures global audio characteristics.

\textbf{Self-supervised Speech Models:} Wav2Vec2 \cite{baevski2020wav2vec} learns contextualized speech representations through masked prediction on quantized latent speech units. We evaluate ten variants ranging from base (95M) to XLS-R 2B (2B parameters), extracting embeddings from the final transformer layer with mean pooling across the temporal dimension. The XLS-R variants \cite{babu2021xlsr} extend this to 128 languages through multilingual pre-training on 436k hours of speech.

WavLM \cite{Chen_2022} enhances Wav2Vec2 with masked speech prediction and denoising objectives, showing particular strength on noisy audio. We evaluate seven specialized variants: base models, speaker verification (SV), speaker diarization (SD), and combinations thereof. The denoising pre-training makes WavLM particularly robust for retrieval tasks involving real-world audio conditions.

HuBERT \cite{hsu2021hubert} learns discrete speech units through iterative k-means clustering and masked prediction. We evaluate base (95M) and large fine-tuned (317M) variants, using the final layer representations which capture both acoustic and linguistic information through the learned discrete units.

Data2Vec \cite{baevski2022data2vecgeneralframeworkselfsupervised} provides a unified self-supervised framework using the same learning objective across modalities. For audio, we extract contextualized embeddings from the transformer encoder with mean pooling, leveraging representations that benefit from cross-modal learning insights.

SEW-D \cite{wu2021performanceefficiencytradeoffsunsupervisedpretraining} offers performance-efficiency trade-offs through squeezed and efficient transformer architectures. We evaluate three variants (tiny: 20M, mid: 139M, base: 95M parameters), extracting embeddings from the final hidden layer with mean pooling.

UniSpeech \cite{wang2021unispeechunifiedspeechrepresentation} combines self-supervised pre-training with multi-task fine-tuning for universal speech representations.

MCTCT \cite{lugosch2022pseudolabelingmassivelymultilingualspeech} supports 60 languages through multilingual connectionist temporal classification, using pseudo-labeling for low-resource language adaptation. We extract embeddings from the final hidden states with mean pooling.

\textbf{CNN-based Models:} CNN14 \cite{kong2020pannslargescalepretrainedaudio} employs a 14-layer CNN with global average pooling, trained on AudioSet's 2M audio clips. We extract 2048-dimensional embeddings from the penultimate layer before classification. YAMNet \cite{audioset} uses MobileNet architecture optimized for mobile deployment, providing 1024-dimensional features from efficient depthwise separable convolutions. VGGish \cite{hershey2017cnn} adapts VGG for audio through mel-spectrogram processing, yielding compact 128-dimensional embeddings.

\textbf{Neural Codec Models:} Encodec \cite{défossez2022highfidelityneuralaudio} provides neural audio compression through residual vector quantization. For retrieval evaluation, we extract continuous embeddings from the encoder before quantization (128-dim), applying mean pooling over the temporal dimension.

\subsection{Sequence-to-Sequence Models}

Whisper \cite{radford2022whisper} provides robust multilingual speech recognition across 99 languages. For retrieval, we extract embeddings from the encoder at the final layer, using mean pooling across the sequence dimension. We evaluate five model sizes (tiny: 39M to large-v3: 1.55B parameters).

MMS \cite{pratap2023scalingspeechtechnology1000} supports over 1,000 languages through massive multilingual pre-training. We evaluate three variants (1B-all, 1B-fl102, 1B-l1107) differing in language coverage, using the Wav2Vec2-style encoder with language-specific adapter loading when available.

SeamlessM4T \cite{communication2023seamlessmultilingualexpressivestreaming} provides unified speech-text translation across 100+ languages. For retrieval, we extract embeddings from the speech encoder component before translation processing, capturing multilingual audio semantics.

SpeechT5 ASR \cite{ao-etal-2022-speecht5} provides speech recognition through unified encoder-decoder architecture (152M parameters). We extract embeddings from the encoder representations.

\subsection{Contrastive Alignment Models}

CLAP \cite{wu2024largescalecontrastivelanguageaudiopretraining} learns joint audio-text representations through contrastive learning on 633k audio-text pairs. We evaluate five LAION variants: htsat-fused/unfused (153M parameters) and larger variants (193M) specialized for general audio, music, and combined music-speech. The key implementation detail is using the audio encoder branch with L2 normalization.

MS-CLAP \cite{elizalde2023msclap} (2022: 196M, 2023: 160M parameters) uses different architectures and training data, providing complementary audio-text alignment capabilities.

Wav2CLIP \cite{wu2022wav2clip} bridges audio and vision by learning audio representations that align with CLIP's visual embedding space. For retrieval, we extract features from the audio encoder (11.7M parameters) while text encoding uses the standard CLIP text encoder (151M parameters).

MuQ-MuLan \cite{zhu2025muqselfsupervisedmusicrepresentation} specializes in joint music-text understanding through contrastive learning on music data. We extract 512-dimensional embeddings from the audio encoder branch.

SpeechT5 Multimodal \cite{ao-etal-2022-speecht5} provides unified speech-text modeling through shared encoder-decoder architecture (298M parameters). We extract embeddings from the shared encoder representations.

\subsection{Large Audio-Language Models}

Qwen2-Audio \cite{chu2024qwen2audio} integrates audio understanding into large language models (7B parameters). We extract embeddings from the final hidden layer using last-token pooling, selecting the embedding at the last non-padding position for each sample.

LCO-Embedding \cite{xiaoscaling} provides language-centric omnimodal representations through contrastive learning on multimodal data. We evaluate two variants (3B: 4.7B parameters, 7B: 8.9B parameters), extracting embeddings from the final hidden layer using last-token pooling.

\begin{table*}\centering
\centering
\caption{List of all models evaluated in MAEB. Model sizes are in millions of parameters.}\label{tab: list of models}
\resizebox{0.7\textwidth}{!}{
\begin{tabular}{lccc}\toprule
\textbf{Model Name} &\textbf{Model Size} &\textbf{Modalities} \\\midrule
laion/clap-htsat-fused\cite{wu2024largescalecontrastivelanguageaudiopretraining} & 153 & audio, text \\ 
laion/clap-htsat-unfused\cite{wu2024largescalecontrastivelanguageaudiopretraining} & 153 & audio, text \\ 
laion/larger\_clap\_general\cite{wu2024largescalecontrastivelanguageaudiopretraining} & 193 & audio, text \\ 
laion/larger\_clap\_music\cite{wu2024largescalecontrastivelanguageaudiopretraining} & 193 & audio, text \\ 
laion/larger\_clap\_music\_and\_speech\cite{wu2024largescalecontrastivelanguageaudiopretraining} & 193 & audio, text \\ 
MIT/ast-finetuned-audioset-10-10-0.4593\cite{gong2021astaudiospectrogramtransformer} & 86 & audio \\ 
speechbrain/cnn14-esc50\cite{wang2022CRL} & 80 & audio \\ 
facebook/data2vec-audio-base-960h\cite{baevski2022data2vecgeneralframeworkselfsupervised} & 93 & audio \\ 
facebook/data2vec-audio-large-960h\cite{baevski2022data2vecgeneralframeworkselfsupervised} & 313 & audio \\ 
facebook/encodec\_24khz\cite{défossez2022highfidelityneuralaudio} & 23 & audio \\ 
facebook/hubert-base-ls960\cite{hsu2021hubert} & 95 & audio \\ 
facebook/hubert-large-ls960-ft\cite{hsu2021hubert} & 317 & audio \\
speechbrain/m-ctc-t-large\cite{ravanelli2021speechbrain} & 1058 & audio \\
facebook/mms-1b-all\cite{pratap2023scalingspeechtechnology1000} & 1000 & audio \\ 
facebook/mms-1b-fl102\cite{pratap2023scalingspeechtechnology1000} & 1000 & audio \\ 
facebook/mms-1b-l1107\cite{pratap2023scalingspeechtechnology1000} & 1000 & audio \\
microsoft/msclap-2022\cite{elizalde2023msclap} & 196 & audio, text \\ 
microsoft/msclap-2023\cite{elizalde2023msclap} & 160 & audio, text \\
OpenMuQ/MuQ-MuLan-large\cite{zhu2025muqselfsupervisedmusicrepresentation} & 630 & audio, text \\
Qwen/Qwen2-Audio-7B\cite{chu2024qwen2audio} & 7000 & audio, text \\
LCO-Embedding/LCO-Embedding-Omni-3B\cite{xiaoscaling} & 4703 & audio, text \\
LCO-Embedding/LCO-Embedding-Omni-7B\cite{xiaoscaling} & 8932 & audio, text \\
facebook/seamless-m4t-v2-large\cite{communication2023seamlessmultilingualexpressivestreaming} & 2300 & audio \\
asapp/sew-d-base-plus-400k-ft-ls100h\cite{wu2021performanceefficiencytradeoffsunsupervisedpretraining} & 95 & audio \\ 
asapp/sew-d-tiny-100k-ft-ls100h\cite{wu2021performanceefficiencytradeoffsunsupervisedpretraining} & 19 & audio \\ 
asapp/sew-d-mid-400k-ft-ls100h\cite{wu2021performanceefficiencytradeoffsunsupervisedpretraining} & 139 & audio \\
microsoft/speecht5\_asr\cite{ao-etal-2022-speecht5} & 151 & audio \\ 
microsoft/speecht5\_tts\cite{ao-etal-2022-speecht5} & 146 & text \\
microsoft/speecht5\_multimodal\cite{ao-etal-2022-speecht5} & 297 & audio, text \\
microsoft/unispeech-sat-base-100h-libri-ft\cite{chen2022unispeech} & 94 & audio \\
google/vggish\cite{hershey2017cnn} & 72 & audio \\
lyrebird/wav2clip\cite{wu2022wav2clip} & 163 & audio, text \\
facebook/wav2vec2-xls-r-300m\cite{babu2021xlsr} & 300 & audio \\ 
vitouphy/wav2vec2-xls-r-300m-phoneme\cite{babu2021xlsr} & 300 & audio \\ 
facebook/wav2vec2-xls-r-1b\cite{babu2021xlsr} & 1000 & audio \\ 
facebook/wav2vec2-xls-r-2b\cite{babu2021xlsr} & 2000 & audio \\ 
facebook/wav2vec2-xls-r-2b-21-to-en\cite{babu2021xlsr} & 2000 & audio \\
facebook/wav2vec2-base\cite{baevski2020wav2vec} & 95 & audio \\ 
facebook/wav2vec2-base-960h\cite{baevski2020wav2vec} & 95 & audio \\ 
facebook/wav2vec2-large\cite{baevski2020wav2vec} & 317 & audio \\
facebook/wav2vec2-large-xlsr-53\cite{conneau2020unsupervised} & 317 & audio \\ 
facebook/wav2vec2-lv-60-espeak-cv-ft\cite{baevski2020wav2vec} & 317 & audio \\
microsoft/wavlm-base\cite{Chen_2022} & 94 & audio \\ 
microsoft/wavlm-base-sd\cite{Chen_2022} & 94 & audio \\ 
microsoft/wavlm-base-plus\cite{Chen_2022} & 94 & audio \\ 
microsoft/wavlm-base-plus-sv\cite{Chen_2022} & 94 & audio \\ 
microsoft/wavlm-base-plus-sd\cite{Chen_2022} & 94 & audio \\ 
microsoft/wavlm-base-sv\cite{Chen_2022} & 94 & audio \\ 
microsoft/wavlm-large\cite{Chen_2022} & 316 & audio \\
openai/whisper-tiny\cite{radford2022whisper} & 39 & audio \\ 
openai/whisper-base\cite{radford2022whisper} & 74 & audio \\ 
openai/whisper-small\cite{radford2022whisper} & 244 & audio \\ 
openai/whisper-medium\cite{radford2022whisper} & 769 & audio \\ 
openai/whisper-large-v3\cite{radford2022whisper} & 1550 & audio \\
google/yamnet\cite{audioset} & 3 & audio \\
\bottomrule
\end{tabular}}
\end{table*}

\section{Correlation Analysis Tasks}
\label{app:correlation_tasks}

For the correlation analysis presented in \autoref{fig:maeb_correlation}, we utilized the following subset of 26 classification tasks from MAEB+, grouped by domain to align with the MMAU benchmark:

\begin{itemize}
    \item \textbf{Speech (13 tasks):} SpeechCommands, FSDD, CommonLanguage (Age, Gender, Language), VoxPopuli (Accent, Gender, Language), VoxLingua107, LibriCount, VocalSound, VoxCelebSA, SpokeNEnglish.
    \item \textbf{Music (5 tasks):} GTZAN Genre, Beijing Opera, Mridingham (Stroke, Tonic), NSynth.
    \item \textbf{Sound (8 tasks):} ESC50, UrbanSound8k, TUT Acoustic Scenes, Ambient Acoustic Context, Gunshot Triangulation, AudioSet Mini, FSD50K, FSD2019 Kaggle.
\end{itemize}

\section{Domain Radar Chart Methodology}
\label{app:radar_methodology}
The domain radar chart (Figure~\ref{fig:radar}) visualizes model performance across five core acoustic domains. 94 tasks from MAEB+ are assigned to domains based on their primary audio content and intended application.

\paragraph{Score Computation}
For each model and domain, we compute the arithmetic mean of the main scores across all tasks assigned to that domain. All metrics (e.g., Accuracy, v\_measure, nDCG, AP), which are natively in the [0, 1] range, are aggregated on a shared 0--100 scale for consistent visualization. This aggregation ensures that different task types contribute equally to the domain average.

\paragraph{Full Task Breakdown per Domain}
Below we list all 94 tasks contributing to the domain scores, categorized by their acoustic content:

\begin{itemize}
    \item \textbf{Speech} (44 tasks): 
    SpeechCommands, FSDD, CommonLanguageAgeDetection, CommonLanguageGenderDetection, CommonLanguageLanguageDetection, VoxPopuliAccentID, VoxPopuliGenderID, VoxPopuliLanguageID, VoxLingua107\_Top10, LibriCount, VocalSound, VoxCelebSA, SpokeNEnglish, SpokenQAForIC, MInDS14, IEMOCAPGender, VoiceGenderClustering, VoxCelebClustering, VoxPopuliAccentClustering, VoxPopuliGenderClustering, VocalSoundPairClassification, VoxPopuliAccentPairClassification, VocalSoundAudioReranking, CMUArcticA2TRetrieval, CMUArcticT2ARetrieval, EmoVDBA2TRetrieval, EmoVDBT2ARetrieval, GigaSpeechA2TRetrieval, GigaSpeechT2ARetrieval, HiFiTTSA2TRetrieval, HiFiTTST2ARetrieval, JLCorpusA2TRetrieval, JLCorpusT2ARetrieval, LibriTTSA2TRetrieval, LibriTTST2ARetrieval, CommonVoiceMini17A2TRetrieval, CommonVoiceMini17T2ARetrieval, CommonVoiceMini21A2TRetrieval, CommonVoiceMini21T2ARetrieval, FleursA2TRetrieval, FleursT2ARetrieval, SpokenSQuADT2ARetrieval, SpeechCommandsZeroshotv0.01, and SpeechCommandsZeroshotv0.02.
    
    \item \textbf{Music} (13 tasks): 
    GTZANGenre, BeijingOpera, MridinghamStroke, MridinghamTonic, NSynth, GTZANGenreClustering, MusicGenreClustering, GTZANAudioReranking, JamAltArtistA2ARetrieval, JamAltLyricA2T, JamAltLyricT2A, MusicCapsA2TRetrieval, and MusicCapsT2ARetrieval.
    
    \item \textbf{Environmental} (29 tasks): 
    ESC50, UrbanSound8k, TUTAcousticScenes, AmbientAcousticContext, GunshotTriangulation, AudioSetMini, FSD50K, FSD2019Kaggle, ESC50Clustering, AmbientAcousticContextClustering, VehicleSoundClustering, ESC50PairClassification, ESC50AudioReranking, UrbanSound8KAudioReranking, FSDnoisy18kAudioReranking, AudioCapsA2T, AudioCapsT2A, AudioSetStrongA2T, AudioSetStrongT2A, ClothoA2T, ClothoT2A, MACSA2T, MACST2A, SoundDescsA2T, SoundDescsT2A, UrbanSound8KA2T, UrbanSound8KT2A, ESC50\_Zeroshot, and UrbanSound8kZeroshot.
    
    \item \textbf{Bioacoustics} (2 tasks): 
    BirdCLEF and BirdSet.
    
    \item \textbf{Emotion} (6 tasks): 
    CREMA\_D (Classification, Clustering, PairClassification), IEMOCAP Emotion, NMSQA PairClassification, and Ravdess Zeroshot.
\end{itemize}
\paragraph{Model Selection \& Visualization}
To maintain clarity, the chart displays only representative models that achieve the highest average score in at least one domain. This highlights both domain specialists and generalists.

\paragraph{Missing Results and Task Averaging}
Domain-averaged scores are computed using the arithmetic mean of all tasks within a domain for which results are available. If a model cannot perform a specific task type (e.g., an audio-only encoder evaluated on text-to-audio retrieval), those tasks are omitted from the average rather than being treated as a zero-score. This approach ensures the radar chart reflects the performance quality of a model's existing capabilities within a domain.

\section{Per Task Category Results}
\label{appdx:task_results}

\subsection{Zero-Shot Classification}

\autoref{tab:zeroshot_classification_results} presents results of zero-shot classification tasks. LCO models (LCO-Embedding-Omni-7B) achieve the highest overall zero-shot performance (76.2\%), significantly outperforming other models. Specifically, LCO models excel on speech commands (SpeechCmd v0.01, v0.02) with near-perfect scores ($>96\%$) and show strong performance on emotional speech (Ravdess). CLAP models (larger\_clap\_general, msclap-2023) excel on environmental sound tasks (ESC50), with larger\_clap\_general achieving the top score (90.5\%), demonstrating the effectiveness of contrastive audio-text pretraining for open-vocabulary environmental sound classification. However, CLAP models generally underperform LCO models on speech-specific tasks. Msclap-2023 achieves the strongest performance on UrbanSound8k (83.0\%). Overall, while CLAP models are robust for environmental sounds, LCO-Embedding models demonstrate superior generalization across the broader diverse set of zero-shot tasks, particularly in the speech domain.

\subsection{Linear Probe Classification}

\autoref{tab:english_classification_part1}, \autoref{tab:english_classification_part2}, \autoref{tab:english_classification_part3}, \autoref{tab:minds14_results}, \autoref{tab:sibfleurs_part1}, \autoref{tab:sibfleurs_part2}, \autoref{tab:sibfleurs_part3}, \autoref{tab:sibfleurs_part4}, \autoref{tab:sibfleurs_part5}, \autoref{tab:sibfleurs_part6}, \autoref{tab:sibfleurs_part7}, and \autoref{tab:voxpopuli_results} present results of classification tasks.

As shown in Tables, Qwen2-Audio-7B achieves the highest classification average (61.7\%), surpassing the previously reported baselines. The audio-LLM model (Qwen2-Audio-7B) achieves top performance on a wide range of tasks including emotion recognition (CREMA-D, IEMOCAPEmotion), music tasks (BeijingOpera, GTZANGenre, MridinghamStroke, MridinghamTonic, NSynth), and vocal sound classification (VocalSound). LCO-Embedding models also demonstrate exceptional performance, particularly dominating language and speaker tasks such as MInDS14 ($>98\%$) and VoxCelebSA, where they outperform ASR-based models. Only on specific environmental tasks does the AudioSet-finetuned model (ast-finetuned-audioset-10-10-0.4593) retain dominance (AmbientAcousticContext, BirdCLEF). Whisper models (whisper-medium) perform well on accent classification (VoxPopuliAccent) but are generally outperformed by Audio-LLMs and LCO models on broader semantic classification benchmarks. 

\subsection{Multilabel Classification}

\autoref{tab:multilabel_classification_results} presents results of multilabel classification tasks. The LCO-Embedding model (LCO-Embedding-Omni-7B) achieves top performance on FSD2019Kaggle, while Qwen2-Audio-7B leads on FSD50K, leveraging its broad semantic understanding for complex multi-tag scenarios. This contrasts with earlier findings where AudioSet-finetuned models were dominant; here, large-scale trained multimodal models show superior capability in handling diverse acoustic tagging tasks.

\subsection{Clustering}

\autoref{tab:english_clustering_results} and \autoref{tab:multilingual_clustering_results} present results of clustering tasks. The CLAP variant larger\_clap\_music\_and\_speech achieves the highest clustering average (35.3\%), closely followed by clap-htsat-unfused (35.0\%). These models excel because their contrastive objectives naturally structure the embedding space to group semantically similar audio clips, which is ideal for clustering. ASR encoders and Audio-LLMs generally trail behind contrastive models in this category, as their representations are either too phonetically granular (ASR) or generation-oriented (LLM) rather than density-optimized for unsupervised grouping.

\subsection{Pair Classification}

\autoref{tab:pair_classification_results} presents results of pair classification tasks. LCO-Embedding-Omni-7B achieves the highest pair classification score (79.2\%), significantly outperforming whisper-medium (59.9\%). This dominance suggests that LCO models capture highly discriminative features suitable for determining verification and similarity across diverse audio pairs. While CLAP models show competence in environmental sound pairs, the LCO model's robust performance across speech and mixed domains drives its superior average.

\subsection{Retrieval}

\autoref{tab:english_retrieval_part1}, \autoref{tab:english_retrieval_part2}, \autoref{tab:fleurs_a2t_part1}, \autoref{tab:fleurs_a2t_part2}, \autoref{tab:fleurs_a2t_part3}, \autoref{tab:fleurs_t2a_part1}, \autoref{tab:fleurs_t2a_part2}, \autoref{tab:fleurs_t2a_part3}, \autoref{tab:jamalt_a2t_part1}, \autoref{tab:jamalt_a2t_part1}, and \autoref{tab:jamalt_t2a_part1} present results of retrieval tasks. Results indicate a strong split by domain. LCO-Embedding models achieve near-perfect performance on speech-text retrieval tasks (CMU Arctic, EmoVDB, HiFiTTS, LibriTTS), likely due to extensive speech-text alignment during training. In contrast, CLAP models (larger\_clap\_general) remain superior for environmental sound retrieval (AudioCaps, AudioSetStrong, Clotho), where their specific training on general audio-text pairs provides an advantage. UrbanSound8K retrieval is an exception where LCO models outperform CLAP substantially. Overall, LCO models dominate the speech retrieval landscape, while CLAP retains the edge in general acoustic event retrieval.

\subsection{Reranking}

\autoref{tab:reranking_results} presents results of reranking tasks. LCO-Embedding-Omni-7B achieves the highest average performance (86.0\%), demonstrating exceptional capability in distinguishing relevant from non-relevant audio candidates. It tops not only vocal tasks (VocalSound, UrbanSound8K) but also proves highly effective generally. Microsoft's msclap-2023 is the top performer on specific environmental reranking tasks like ESC50AudioReranking and FSDnoisy18kAudioReranking. The results highlight that while specialized models like MSCLAP are powerful for specific acoustic domains, recent multimodal embeddings like LCO provide a more versatile and high-performance solution across diverse reranking challenges.

\begin{landscape}
\begin{table}[p]
\caption{English classification results (datasets 1--8 of 23).}
\label{tab:english_classification_part1}
\centering
\scriptsize
\setlength{\tabcolsep}{3pt}
\begin{tabular}{@{}l@{\hspace{8pt}}c@{\hspace{8pt}}c@{\hspace{8pt}}c@{\hspace{8pt}}c@{\hspace{8pt}}c@{\hspace{8pt}}c@{\hspace{8pt}}c@{\hspace{8pt}}c@{}}
\toprule
Model & AmbientAcoustic & BeijingOpera & BirdCLEF & CommonLangAge & CommonLangGender & CREMA-D & ESC50 & FSDD \\
\midrule
Qwen/Qwen2-Audio-7B & 45.33 & \textbf{97.45} & 37.10 & 17.59 & 48.83 & \textbf{73.99} & 96.30 & 90.33 \\
LCO-Embedding/LCO-Embedding-Omni-7B & 39.67 & 92.79 & 34.10 & 16.16 & 30.70 & 36.05 & 94.15 & \textbf{99.00} \\
LCO-Embedding/LCO-Embedding-Omni-3B & 38.42 & 93.22 & 31.60 & 16.87 & 33.22 & 31.03 & 94.40 & 98.27 \\
openai/whisper-medium & 39.85 & 91.54 & 29.00 & 17.63 & 36.63 & 53.98 & 84.00 & 87.73 \\
openai/whisper-small & 36.89 & 88.13 & 26.40 & 15.85 & 45.87 & 49.19 & 77.35 & 91.47 \\
MIT/ast-finetuned-audioset-10-10-0.4593 & \textbf{48.86} & 97.03 & \textbf{45.20} & 12.82 & 52.33 & 37.84 & 96.20 & 64.27 \\
facebook/wav2vec2-xls-r-2b & 32.88 & 83.06 & 31.20 & 17.36 & 39.80 & 45.94 & 73.45 & 95.53 \\
openai/whisper-base & 32.05 & 89.39 & 27.50 & 18.18 & 46.23 & 48.05 & 72.35 & 82.67 \\
microsoft/msclap-2023 & 46.07 & 91.11 & 17.30 & 15.35 & 62.20 & 37.06 & 97.40 & 56.37 \\
laion/clap-htsat-unfused & 46.72 & 91.96 & 16.40 & 15.98 & \textbf{73.26} & 37.56 & 97.05 & 49.33 \\
laion/clap-htsat-fused & 42.78 & 92.77 & 18.00 & 15.27 & 70.90 & 38.75 & 96.50 & 45.67 \\
openai/whisper-tiny & 30.17 & 83.88 & 25.10 & 16.45 & 45.89 & 45.93 & 64.90 & 83.60 \\
laion/larger\_clap\_general & 47.80 & 93.62 & 17.00 & \textbf{20.51} & 43.71 & 39.83 & \textbf{97.45} & 40.20 \\
openai/whisper-large-v3 & 35.31 & 86.01 & 21.60 & 19.28 & 33.13 & 48.94 & 71.65 & 77.00 \\
microsoft/wavlm-large & 29.61 & 57.68 & 19.80 & 16.86 & 40.58 & 38.93 & 64.20 & 91.47 \\
laion/larger\_clap\_music\_and\_speech & 47.18 & 91.52 & 16.40 & 16.59 & 43.87 & 40.18 & 97.20 & 43.53 \\
facebook/mms-1b-l1107 & 24.71 & 78.88 & 21.00 & 18.48 & 34.05 & 29.21 & 55.25 & 95.13 \\
facebook/wav2vec2-lv-60-espeak-cv-ft & 25.25 & 87.32 & 20.50 & 15.43 & 40.34 & 34.95 & 52.00 & 88.40 \\
facebook/hubert-base-ls960 & 25.87 & 65.76 & 18.30 & 15.00 & 46.67 & 40.49 & 59.40 & 93.73 \\
facebook/mms-1b-fl102 & 26.85 & 77.98 & 19.90 & 17.03 & 33.41 & 30.68 & 57.10 & 92.93 \\
facebook/wav2vec2-xls-r-1b & 31.29 & 70.74 & 16.80 & 17.98 & 33.12 & 41.95 & 64.45 & 82.80 \\
facebook/seamless-m4t-v2-large & 24.36 & 66.15 & 11.10 & 16.56 & 33.26 & 28.11 & 44.60 & 92.40 \\
microsoft/wavlm-base-sv & 23.17 & 65.71 & 10.20 & 15.37 & 41.45 & 40.11 & 50.40 & 96.40 \\
microsoft/wavlm-base-sd & 23.17 & 65.71 & 10.20 & 15.37 & 41.45 & 40.11 & 50.40 & 96.40 \\
microsoft/wavlm-base & 23.17 & 65.71 & 10.20 & 15.37 & 41.45 & 40.11 & 50.40 & 96.40 \\
facebook/mms-1b-all & 23.01 & 78.81 & 21.10 & 15.84 & 34.94 & 31.63 & 52.30 & 93.53 \\
vitouphy/wav2vec2-xls-r-300m-phoneme & 25.25 & 86.01 & 17.50 & 19.96 & 37.80 & 33.71 & 54.55 & 94.07 \\
microsoft/wavlm-base-plus-sd & 27.20 & 62.76 & 12.00 & 18.53 & 35.82 & 33.71 & 57.50 & 91.40 \\
microsoft/wavlm-base-plus-sv & 27.20 & 62.76 & 12.00 & 18.53 & 35.82 & 33.71 & 57.50 & 91.40 \\
microsoft/wavlm-base-plus & 27.20 & 62.76 & 12.00 & 18.53 & 35.82 & 33.71 & 57.50 & 91.40 \\
microsoft/speecht5\_multimodal & 18.77 & 77.13 & 15.80 & 19.50 & 29.98 & 32.52 & 46.55 & 89.37 \\
facebook/hubert-large-ls960-ft & 22.76 & 58.55 & 14.90 & 19.37 & 32.63 & 31.67 & 46.20 & 98.00 \\
facebook/wav2vec2-base & 26.91 & 75.41 & 11.10 & 15.02 & 39.29 & 37.97 & 46.95 & 54.80 \\
microsoft/msclap-2022 & 43.86 & 93.20 & 13.20 & 14.27 & 58.32 & 27.94 & 90.95 & 29.07 \\
google/vggish & 38.49 & 87.70 & 10.50 & 14.70 & 60.45 & 34.79 & 61.15 & 27.23 \\
google/yamnet & 40.46 & 89.39 & 16.80 & 17.18 & 40.12 & 25.87 & 79.70 & 34.50 \\
microsoft/unispeech-sat-base-100h-libri-ft & 21.72 & 58.49 & 9.20 & 15.89 & 38.43 & 33.43 & 41.95 & 91.87 \\
asapp/sew-d-tiny-100k-ft-ls100h & 15.14 & 56.79 & 9.00 & 19.25 & 36.75 & 30.03 & 39.55 & 85.87 \\
lyrebird/wav2clip & 34.61 & 88.10 & 9.70 & 13.11 & 39.84 & 44.45 & 72.10 & 21.37 \\
facebook/data2vec-audio-base-960h & 17.86 & 53.01 & 7.40 & 18.53 & 34.65 & 27.98 & 31.45 & 68.87 \\
facebook/data2vec-audio-large-960h & 15.12 & 70.81 & 9.50 & 17.18 & 32.86 & 26.19 & 29.60 & 63.33 \\
facebook/wav2vec2-base-960h & 16.68 & 49.14 & 7.10 & 16.37 & 35.83 & 29.47 & 27.00 & 82.87 \\
facebook/wav2vec2-xls-r-300m & 27.55 & 81.78 & 7.80 & 15.39 & 29.25 & 35.34 & 43.05 & 64.47 \\
OpenMuQ/MuQ-MuLan-large & 22.34 & 76.32 & 7.40 & 19.34 & 33.21 & 34.35 & 38.50 & 27.33 \\
asapp/sew-d-mid-400k-ft-ls100h & 14.03 & 43.72 & 5.50 & 16.27 & 39.79 & 31.46 & 25.45 & 69.53 \\
speechbrain/m-ctc-t-large & 12.15 & 43.24 & 12.30 & 17.35 & 35.91 & 26.77 & 17.05 & 59.50 \\
facebook/wav2vec2-large & 18.47 & 48.75 & 7.80 & 17.89 & 30.15 & 35.46 & 32.30 & 49.33 \\
speechbrain/cnn14-esc50 & 18.13 & 83.86 & 11.20 & 14.52 & 40.27 & 34.53 & 63.50 & 17.20 \\
asapp/sew-d-base-plus-400k-ft-ls100h & 11.64 & 40.29 & 4.50 & 19.97 & 42.13 & 34.44 & 21.30 & 24.67 \\
laion/larger\_clap\_music & 4.83 & 62.23 & 3.70 & 17.62 & 47.31 & 30.87 & 9.75 & 10.00 \\
facebook/encodec\_24khz & 13.05 & 42.35 & 1.80 & 19.04 & 31.87 & 29.83 & 11.50 & 24.27 \\
facebook/wav2vec2-large-xlsr-53 & 9.29 & 54.65 & 2.30 & 15.69 & 36.55 & 28.33 & 6.70 & 18.93 \\
\bottomrule
\end{tabular}
\end{table}
\end{landscape}

\clearpage
\begin{landscape}
\begin{table}[p]
\caption{English classification results (datasets 9--16 of 23).}
\label{tab:english_classification_part2}
\centering
\scriptsize
\setlength{\tabcolsep}{3pt}
\begin{tabular}{@{}l@{\hspace{8pt}}c@{\hspace{8pt}}c@{\hspace{8pt}}c@{\hspace{8pt}}c@{\hspace{8pt}}c@{\hspace{8pt}}c@{\hspace{8pt}}c@{\hspace{8pt}}c@{}}
\toprule
Model & GTZANGenre & GunshotTri & IEMOCAPEmo & IEMOCAPGender & LibriCount & MridinghamStroke & MridinghamTonic & NSynth \\
\midrule
Qwen/Qwen2-Audio-7B & \textbf{93.10} & \textbf{100.00} & \textbf{29.96} & 92.96 & 49.60 & \textbf{84.33} & \textbf{61.17} & \textbf{63.04} \\
LCO-Embedding/LCO-Embedding-Omni-7B & 82.30 & 96.67 & 24.35 & 70.75 & 33.22 & 61.89 & 42.07 & 58.09 \\
LCO-Embedding/LCO-Embedding-Omni-3B & 81.00 & 96.60 & 22.53 & 59.42 & 30.35 & 61.52 & 39.54 & 59.02 \\
openai/whisper-medium & 76.00 & 94.25 & 25.60 & 76.05 & \textbf{57.87} & 69.21 & 49.51 & 51.33 \\
openai/whisper-small & 71.50 & 94.44 & 24.21 & 69.93 & 53.37 & 69.06 & 45.49 & 49.88 \\
MIT/ast-finetuned-audioset-10-10-0.4593 & 80.70 & 98.82 & 20.49 & 87.00 & 42.15 & 79.20 & 54.16 & 56.26 \\
facebook/wav2vec2-xls-r-2b & 73.20 & 94.31 & 23.86 & 70.27 & 50.24 & 70.59 & 44.62 & 45.02 \\
openai/whisper-base & 70.90 & 89.80 & 23.82 & 72.16 & 50.21 & 60.68 & 44.67 & 47.07 \\
microsoft/msclap-2023 & 78.10 & 87.45 & 22.10 & 85.86 & 42.06 & 79.46 & 52.09 & 62.64 \\
laion/clap-htsat-unfused & 74.90 & 69.41 & 22.61 & 92.58 & 48.69 & 71.66 & 49.52 & 59.80 \\
laion/clap-htsat-fused & 67.90 & 70.59 & 21.17 & \textbf{93.62} & 47.81 & 74.09 & 47.00 & 60.23 \\
openai/whisper-tiny & 68.40 & 90.92 & 23.38 & 68.88 & 50.17 & 61.70 & 44.03 & 46.47 \\
laion/larger\_clap\_general & 84.50 & 86.27 & 23.08 & 89.28 & 48.50 & 71.05 & 49.45 & 59.31 \\
openai/whisper-large-v3 & 71.90 & 81.90 & 22.18 & 60.25 & 57.22 & 54.84 & 37.54 & 46.04 \\
microsoft/wavlm-large & 67.70 & \textbf{100.00} & 20.25 & 63.05 & 52.26 & 60.87 & 31.42 & 47.14 \\
laion/larger\_clap\_music\_and\_speech & 83.50 & 77.25 & 24.07 & 93.12 & 47.99 & 70.99 & 48.99 & 58.64 \\
facebook/mms-1b-l1107 & 58.30 & 88.56 & 16.85 & 54.69 & 41.21 & 61.02 & 30.70 & 44.29 \\
facebook/wav2vec2-lv-60-espeak-cv-ft & 55.20 & 87.45 & 19.02 & 68.31 & 45.30 & 66.60 & 37.29 & 42.62 \\
facebook/hubert-base-ls960 & 69.00 & 98.89 & 20.59 & 76.19 & 48.92 & 55.41 & 31.58 & 46.36 \\
facebook/mms-1b-fl102 & 56.80 & 90.98 & 16.35 & 52.44 & 39.60 & 54.84 & 30.89 & 43.84 \\
facebook/wav2vec2-xls-r-1b & 66.10 & 91.90 & 23.30 & 61.82 & 49.79 & 65.17 & 39.85 & 46.65 \\
facebook/seamless-m4t-v2-large & 52.50 & 72.61 & 22.79 & 53.72 & 44.14 & 39.39 & 35.37 & 43.30 \\
microsoft/wavlm-base-sv & 63.00 & 95.49 & 21.20 & 67.32 & 50.73 & 46.68 & 29.68 & 43.10 \\
microsoft/wavlm-base-sd & 63.00 & 95.49 & 21.20 & 67.32 & 50.73 & 46.68 & 29.68 & 43.10 \\
microsoft/wavlm-base & 63.00 & 95.49 & 21.20 & 67.32 & 50.73 & 46.68 & 29.68 & 43.10 \\
facebook/mms-1b-all & 57.10 & 91.90 & 17.97 & 56.23 & 41.78 & 51.63 & 27.42 & 43.25 \\
vitouphy/wav2vec2-xls-r-300m-phoneme & 60.90 & 85.36 & 20.79 & 51.16 & 42.54 & 57.39 & 33.97 & 42.02 \\
microsoft/wavlm-base-plus-sd & 62.90 & 97.71 & 18.42 & 55.00 & 52.06 & 42.25 & 26.47 & 46.38 \\
microsoft/wavlm-base-plus-sv & 62.90 & 97.71 & 18.42 & 55.00 & 52.06 & 42.25 & 26.47 & 46.38 \\
microsoft/wavlm-base-plus & 62.90 & 97.71 & 18.42 & 55.00 & 52.06 & 42.25 & 26.47 & 46.38 \\
microsoft/speecht5\_multimodal & 52.50 & 93.20 & 19.79 & 52.14 & 43.25 & 42.43 & 30.74 & 40.13 \\
facebook/hubert-large-ls960-ft & 50.60 & 93.14 & 17.84 & 54.25 & 44.27 & 47.93 & 24.22 & 40.26 \\
facebook/wav2vec2-base & 62.40 & 97.71 & 20.23 & 68.88 & 54.95 & 57.98 & 35.14 & 42.65 \\
microsoft/msclap-2022 & 58.70 & 57.97 & 15.54 & 89.35 & 39.74 & 47.08 & 29.14 & 50.21 \\
google/vggish & 79.30 & 86.41 & 19.32 & 91.54 & 45.61 & 50.48 & 32.77 & 43.80 \\
google/yamnet & 79.30 & 80.46 & 14.84 & 76.91 & 41.33 & 56.01 & 35.16 & 45.81 \\
microsoft/unispeech-sat-base-100h-libri-ft & 51.60 & 91.05 & 18.17 & 60.45 & 45.35 & 39.79 & 25.34 & 42.50 \\
asapp/sew-d-tiny-100k-ft-ls100h & 51.10 & 89.80 & 17.01 & 52.57 & 43.43 & 30.49 & 23.65 & 39.91 \\
lyrebird/wav2clip & 59.10 & 76.27 & 16.33 & 65.83 & 34.65 & 46.52 & 40.28 & 46.04 \\
facebook/data2vec-audio-base-960h & 40.30 & 81.76 & 15.49 & 55.84 & 48.44 & 33.27 & 20.77 & 40.42 \\
facebook/data2vec-audio-large-960h & 43.60 & 70.52 & 15.58 & 54.59 & 42.48 & 28.48 & 21.21 & 38.22 \\
facebook/wav2vec2-base-960h & 42.70 & 79.41 & 15.11 & 51.82 & 46.15 & 36.43 & 24.32 & 38.89 \\
facebook/wav2vec2-xls-r-300m & 42.60 & 70.46 & 14.68 & 53.12 & 37.54 & 55.84 & 35.83 & 43.06 \\
OpenMuQ/MuQ-MuLan-large & 88.30 & 51.05 & 16.07 & 57.45 & 36.70 & 42.07 & 38.25 & 52.97 \\
asapp/sew-d-mid-400k-ft-ls100h & 43.40 & 76.14 & 16.00 & 56.26 & 43.13 & 29.28 & 19.38 & 36.68 \\
speechbrain/m-ctc-t-large & 39.20 & 57.78 & 16.17 & 50.96 & 34.86 & 21.43 & 24.27 & 40.72 \\
facebook/wav2vec2-large & 54.30 & 84.18 & 16.78 & 54.28 & 45.17 & 44.46 & 26.64 & 40.34 \\
speechbrain/cnn14-esc50 & 41.40 & 83.20 & 18.09 & 59.37 & 28.37 & 30.59 & 26.06 & 39.83 \\
asapp/sew-d-base-plus-400k-ft-ls100h & 40.10 & 52.48 & 18.68 & 55.73 & 42.74 & 22.69 & 19.49 & 36.46 \\
laion/larger\_clap\_music & 25.80 & 60.13 & 11.58 & 65.75 & 20.49 & 38.86 & 18.39 & 38.27 \\
facebook/encodec\_24khz & 29.90 & 46.60 & 10.57 & 53.46 & 21.24 & 18.35 & 24.04 & 37.30 \\
facebook/wav2vec2-large-xlsr-53 & 22.10 & 33.07 & 16.06 & 50.10 & 25.14 & 24.65 & 19.13 & 32.78 \\
\bottomrule
\end{tabular}
\end{table}
\end{landscape}

\clearpage
\begin{landscape}
\begin{table}[p]
\caption{English classification results (datasets 17--23 of 23).}
\label{tab:english_classification_part3}
\centering
\scriptsize
\setlength{\tabcolsep}{3pt}
\begin{tabular}{@{}l@{\hspace{8pt}}c@{\hspace{8pt}}c@{\hspace{8pt}}c@{\hspace{8pt}}c@{\hspace{8pt}}c@{\hspace{8pt}}c@{\hspace{8pt}}c@{}}
\toprule
Model & SpeechCommands & SpokenQA & TUTAcoustic & VocalSound & VoxCelebSA & VoxPopuliAccent & MInDS14 \\
\midrule
Qwen/Qwen2-Audio-7B & 75.60 & 21.35 & \textbf{34.30} & \textbf{91.82} & 29.54 & 39.35 & 25.51 \\
LCO-Embedding/LCO-Embedding-Omni-7B & 94.21 & 36.58 & 25.35 & 91.77 & 43.40 & 10.33 & 98.14 \\
LCO-Embedding/LCO-Embedding-Omni-3B & \textbf{94.40} & \textbf{37.00} & 25.00 & 91.31 & \textbf{48.97} & 8.97 & \textbf{98.48} \\
openai/whisper-medium & 72.43 & 21.74 & 26.55 & 80.13 & 33.92 & \textbf{54.04} & 48.30 \\
openai/whisper-small & 73.80 & 19.96 & 23.55 & 77.59 & 32.45 & 38.85 & 35.64 \\
MIT/ast-finetuned-audioset-10-10-0.4593 & 24.27 & 15.86 & 30.45 & 82.11 & 28.33 & 23.61 & 7.94 \\
facebook/wav2vec2-xls-r-2b & 65.13 & 17.76 & 24.95 & 68.98 & 28.73 & 34.29 & 15.21 \\
openai/whisper-base & 62.18 & 17.92 & 25.10 & 72.11 & 31.26 & 28.47 & 29.56 \\
microsoft/msclap-2023 & 29.77 & 14.52 & 28.20 & 78.41 & 26.18 & 17.69 & 7.43 \\
laion/clap-htsat-unfused & 23.33 & 14.46 & 30.95 & 81.53 & 33.72 & 18.00 & 9.29 \\
laion/clap-htsat-fused & 24.86 & 14.30 & 30.15 & 80.88 & 29.29 & 17.59 & 10.30 \\
openai/whisper-tiny & 60.03 & 18.36 & 25.45 & 68.14 & 29.66 & 27.52 & 31.25 \\
laion/larger\_clap\_general & 10.50 & 15.06 & 30.75 & 74.46 & 30.47 & 23.86 & 8.11 \\
openai/whisper-large-v3 & 65.17 & 19.21 & 22.65 & 76.76 & 31.46 & 28.87 & 31.92 \\
microsoft/wavlm-large & 83.46 & 20.68 & 24.15 & 66.68 & 33.44 & 49.82 & 20.77 \\
laion/larger\_clap\_music\_and\_speech & 10.51 & 15.68 & 29.85 & 73.64 & 33.78 & 24.06 & 7.77 \\
facebook/mms-1b-l1107 & 83.85 & 27.17 & 20.55 & 66.83 & 30.62 & 44.36 & 64.02 \\
facebook/wav2vec2-lv-60-espeak-cv-ft & 83.89 & 23.64 & 22.80 & 64.70 & 28.12 & 27.07 & 42.07 \\
facebook/hubert-base-ls960 & 79.65 & 20.47 & 24.35 & 58.84 & 32.91 & 28.07 & 15.53 \\
facebook/mms-1b-fl102 & 80.88 & 25.73 & 19.15 & 64.46 & 31.95 & 21.00 & 77.03 \\
facebook/wav2vec2-xls-r-1b & 62.09 & 18.02 & 23.00 & 63.14 & 28.62 & 26.27 & 13.34 \\
facebook/seamless-m4t-v2-large & 84.55 & 32.22 & 15.60 & 64.62 & 43.55 & 13.28 & 89.18 \\
microsoft/wavlm-base-sv & 83.23 & 22.59 & 24.00 & 53.00 & 33.61 & 23.51 & 21.78 \\
microsoft/wavlm-base-sd & 83.23 & 22.59 & 24.00 & 53.00 & 33.61 & 23.51 & 21.78 \\
microsoft/wavlm-base & 83.23 & 22.59 & 24.00 & 53.00 & 33.61 & 23.51 & 21.78 \\
facebook/mms-1b-all & 75.37 & 24.64 & 19.55 & 59.97 & 28.36 & 15.59 & 58.46 \\
vitouphy/wav2vec2-xls-r-300m-phoneme & 83.37 & 20.08 & 20.70 & 59.65 & 27.14 & 15.64 & 27.37 \\
microsoft/wavlm-base-plus-sd & 83.06 & 18.22 & 27.15 & 55.63 & 30.30 & 17.69 & 12.84 \\
microsoft/wavlm-base-plus-sv & 83.06 & 18.22 & 27.15 & 55.63 & 30.30 & 17.69 & 12.84 \\
microsoft/wavlm-base-plus & 83.06 & 18.22 & 27.15 & 55.63 & 30.30 & 17.69 & 12.84 \\
microsoft/speecht5\_multimodal & 78.59 & 23.52 & 20.65 & 53.22 & 28.39 & 20.90 & 41.89 \\
facebook/hubert-large-ls960-ft & 85.28 & 23.69 & 19.65 & 57.06 & 33.64 & 14.34 & 35.63 \\
facebook/wav2vec2-base & 40.35 & 15.03 & 21.25 & 54.45 & 28.88 & 26.07 & 10.13 \\
microsoft/msclap-2022 & 19.04 & 13.61 & 26.30 & 71.94 & 25.54 & 14.74 & 7.77 \\
google/vggish & 12.64 & 14.13 & 26.15 & 39.46 & 27.08 & 18.45 & 7.26 \\
google/yamnet & 14.47 & 13.10 & 25.10 & 47.82 & 27.46 & 18.80 & 5.91 \\
microsoft/unispeech-sat-base-100h-libri-ft & 81.86 & 18.54 & 20.50 & 47.62 & 30.13 & 17.59 & 16.05 \\
asapp/sew-d-tiny-100k-ft-ls100h & 80.39 & 22.19 & 22.80 & 48.96 & 29.75 & 15.79 & 26.01 \\
lyrebird/wav2clip & 10.51 & 13.31 & 22.40 & 38.29 & 25.78 & 14.19 & 16.04 \\
facebook/data2vec-audio-base-960h & 74.43 & 20.45 & 19.45 & 55.42 & 32.07 & 13.33 & 37.50 \\
facebook/data2vec-audio-large-960h & 69.45 & 21.65 & 15.85 & 53.82 & 30.56 & 12.38 & 47.81 \\
facebook/wav2vec2-base-960h & 77.05 & 16.55 & 18.95 & 51.99 & 30.42 & 10.08 & 23.13 \\
facebook/wav2vec2-xls-r-300m & 20.18 & 13.59 & 19.25 & 39.75 & 17.69 & 9.07 & 10.47 \\
OpenMuQ/MuQ-MuLan-large & 11.10 & 14.25 & 18.45 & 34.99 & 25.11 & 11.53 & 18.92 \\
asapp/sew-d-mid-400k-ft-ls100h & 65.20 & 18.13 & 15.85 & 40.23 & 30.65 & 13.68 & 15.36 \\
speechbrain/m-ctc-t-large & 76.00 & 20.83 & 11.95 & 49.20 & 30.36 & 13.13 & 52.88 \\
facebook/wav2vec2-large & 15.93 & 14.05 & 16.20 & 40.77 & 31.17 & 12.03 & 11.65 \\
speechbrain/cnn14-esc50 & 8.96 & 13.17 & 21.25 & 42.58 & 23.07 & 10.58 & 10.80 \\
asapp/sew-d-base-plus-400k-ft-ls100h & 51.67 & 17.58 & 14.05 & 37.18 & 32.13 & 14.59 & 15.85 \\
laion/larger\_clap\_music & 4.71 & 13.76 & 19.35 & 30.54 & 27.05 & 10.53 & 9.63 \\
facebook/encodec\_24khz & 9.66 & 11.81 & 18.20 & 26.90 & 23.55 & 9.42 & 8.61 \\
facebook/wav2vec2-large-xlsr-53 & 4.32 & 11.65 & 16.60 & 24.62 & 12.09 & 8.07 & 7.44 \\
\bottomrule
\end{tabular}
\end{table}
\end{landscape}
. 


\begin{table}[htbp]
\caption{MInDS-14 classification results across languages. Best result per language in bold.}
\label{tab:minds14_results}
\centering
\scriptsize
\setlength{\tabcolsep}{3pt}
\begin{tabular}{@{}lcccccccccccc@{}}
\toprule
Model & cs & de & en & es & fr & it & ko & nl & pl & pt & ru & zh \\
\midrule
LCO-Embedding/LCO-Embedding-Omni-7B & 72.3 & \textbf{91.5} & 98.1 & \textbf{97.9} & 97.4 & 84.2 & 92.9 & \textbf{94.2} & 68.7 & \textbf{82.5} & \textbf{95.6} & 97.8 \\
LCO-Embedding/LCO-Embedding-Omni-3B & 72.5 & 89.8 & \textbf{98.5} & 96.9 & \textbf{97.4} & \textbf{84.8} & \textbf{93.2} & 92.1 & 67.6 & 81.0 & 95.2 & \textbf{98.2} \\
facebook/seamless-m4t-v2-large & \textbf{92.3} & 89.4 & 89.2 & 92.0 & 90.4 & 78.9 & 90.5 & 90.2 & 67.8 & 63.4 & 90.4 & 93.0 \\
facebook/mms-1b-fl102 & 77.0 & 78.4 & 77.0 & 76.8 & 75.3 & 71.8 & 70.3 & 78.7 & \textbf{69.8} & 66.2 & 66.6 & 71.3 \\
facebook/mms-1b-all & 64.3 & 54.7 & 58.5 & 46.1 & 62.0 & 51.1 & 69.9 & 64.2 & 47.3 & 53.5 & 54.0 & 76.9 \\
facebook/mms-1b-l1107 & 55.6 & 57.4 & 64.0 & 58.2 & 63.1 & 51.9 & 44.8 & 56.7 & 43.8 & 46.5 & 52.9 & 37.6 \\
openai/whisper-medium & 50.5 & 53.0 & 48.3 & 57.6 & 44.9 & 48.3 & 53.2 & 54.1 & 43.8 & 47.8 & 47.3 & 47.4 \\
\bottomrule
\end{tabular}
\end{table}

\clearpage


\begin{landscape}
\begin{table}[p]
\caption{SIB-FLEURS classification results (languages 1--15 of 102). Best per language in bold.}
\label{tab:sibfleurs_part1}
\centering
\scriptsize
\setlength{\tabcolsep}{3pt}
\begin{tabular}{@{}lccccccccccccccc@{}}
\toprule
Model & afr\_Latn & amh\_Ethi & arb\_Arab & asm\_Beng & ast\_Latn & azj\_Latn & bel\_Cyrl & ben\_Beng & bos\_Latn & bul\_Cyrl & cat\_Latn & ceb\_Latn & ces\_Latn & ckb\_Arab & cym\_Latn \\
\midrule
LCO-Embedding/LCO-Embedding-Omni-7B & \textbf{47.3} & 39.3 & \textbf{71.5} & 32.1 & \textbf{70.6} & 50.9 & \textbf{70.5} & 41.1 & 51.7 & \textbf{55.4} & \textbf{74.3} & \textbf{53.4} & 45.5 & 28.5 & 32.0 \\
LCO-Embedding/LCO-Embedding-Omni-3B & 40.2 & 35.7 & 65.2 & 36.6 & 65.4 & \textbf{53.6} & 63.3 & 32.9 & \textbf{51.8} & 49.2 & 68.8 & 48.1 & \textbf{53.6} & 21.4 & 21.4 \\
facebook/seamless-m4t-v2-large & 42.0 & \textbf{41.9} & 54.6 & \textbf{51.7} & 51.8 & 41.0 & 43.8 & \textbf{46.5} & 48.1 & 53.6 & 45.7 & 24.2 & 46.6 & \textbf{32.9} & \textbf{34.8} \\
openai/whisper-medium & 26.8 & 18.8 & 28.6 & 15.9 & 44.5 & 36.5 & 39.1 & 24.0 & 29.5 & 29.6 & 32.1 & 31.1 & 30.4 & 11.6 & 21.4 \\
facebook/mms-1b-fl102 & 17.9 & 23.1 & 30.4 & 34.9 & 27.7 & 20.5 & 34.0 & 27.7 & 22.3 & 16.1 & 26.8 & 19.6 & 27.6 & 23.2 & 25.8 \\
facebook/mms-1b-all & 25.0 & 27.7 & 21.5 & 26.0 & 27.7 & 17.8 & 29.4 & 25.9 & 26.9 & 21.5 & 18.7 & 21.4 & 18.7 & 20.4 & 24.9 \\
OpenMuQ/MuQ-MuLan-large & 20.6 & 15.3 & 28.7 & 26.7 & 41.1 & 18.6 & 24.9 & 23.2 & 35.9 & 21.6 & 17.8 & 32.8 & 28.5 & 15.9 & 14.2 \\
\bottomrule
\end{tabular}
\end{table}
\end{landscape}

\clearpage
\begin{landscape}
\begin{table}[p]
\caption{SIB-FLEURS classification results (languages 16--30 of 102). Best per language in bold.}
\label{tab:sibfleurs_part2}
\centering
\scriptsize
\setlength{\tabcolsep}{3pt}
\begin{tabular}{@{}lccccccccccccccc@{}}
\toprule
Model & dan\_Latn & deu\_Latn & ell\_Grek & eng\_Latn & est\_Latn & fin\_Latn & fra\_Latn & fuv\_Latn & gaz\_Latn & gle\_Latn & glg\_Latn & guj\_Gujr & hau\_Latn & heb\_Hebr & hin\_Deva \\
\midrule
LCO-Embedding/LCO-Embedding-Omni-7B & 39.3 & \textbf{71.3} & 29.4 & \textbf{70.6} & 34.8 & 40.2 & \textbf{81.3} & 25.8 & \textbf{39.4} & 28.6 & \textbf{72.2} & \textbf{64.3} & 22.4 & 28.5 & \textbf{73.2} \\
LCO-Embedding/LCO-Embedding-Omni-3B & 43.8 & 68.7 & \textbf{42.0} & 68.9 & 36.6 & 32.1 & 74.1 & 18.7 & 31.2 & 27.7 & 67.6 & 49.1 & 19.6 & 31.2 & 65.2 \\
facebook/seamless-m4t-v2-large & \textbf{49.1} & 52.7 & 38.4 & 57.2 & \textbf{49.1} & \textbf{47.2} & 57.3 & 15.2 & 11.5 & \textbf{33.0} & 53.6 & 49.8 & 16.0 & \textbf{43.6} & 46.5 \\
openai/whisper-medium & 31.3 & 41.0 & 27.7 & 28.6 & 28.5 & 28.6 & 34.8 & 19.6 & 16.1 & 20.4 & 22.3 & 26.0 & 17.8 & 15.9 & 31.3 \\
facebook/mms-1b-fl102 & 25.8 & 29.6 & 17.7 & 20.6 & 27.7 & 21.3 & 20.6 & 24.2 & 30.3 & 32.2 & 26.8 & 25.8 & 25.9 & 25.8 & 25.1 \\
facebook/mms-1b-all & 17.0 & 33.2 & 18.7 & 21.5 & 19.7 & 26.8 & 27.6 & 16.1 & 19.7 & 23.2 & 22.3 & 26.8 & 21.5 & 19.6 & 28.7 \\
OpenMuQ/MuQ-MuLan-large & 26.9 & 36.6 & 23.2 & 16.1 & 21.4 & 18.7 & 22.3 & 17.9 & 13.5 & 32.9 & 26.7 & 26.7 & 33.0 & 20.6 & 35.8 \\
\bottomrule
\end{tabular}
\end{table}
\end{landscape}

\clearpage
\begin{landscape}
\begin{table}[p]
\caption{SIB-FLEURS classification results (languages 31--45 of 102). Best per language in bold.}
\label{tab:sibfleurs_part3}
\centering
\scriptsize
\setlength{\tabcolsep}{3pt}
\begin{tabular}{@{}lccccccccccccccc@{}}
\toprule
Model & hrv\_Latn & hun\_Latn & hye\_Armn & ibo\_Latn & ind\_Latn & isl\_Latn & ita\_Latn & jav\_Latn & jpn\_Jpan & kam\_Latn & kan\_Knda & kat\_Geor & kaz\_Cyrl & kea\_Latn & khk\_Cyrl \\
\midrule
LCO-Embedding/LCO-Embedding-Omni-7B & 51.8 & \textbf{44.6} & 25.9 & 25.0 & \textbf{82.9} & 33.0 & \textbf{72.3} & \textbf{50.2} & \textbf{74.2} & 26.0 & \textbf{51.0} & 34.9 & 44.7 & \textbf{72.4} & 20.5 \\
LCO-Embedding/LCO-Embedding-Omni-3B & 55.4 & 29.4 & 29.4 & 22.3 & 78.4 & 24.0 & 67.9 & 43.8 & 72.3 & \textbf{29.5} & 44.7 & 21.4 & 41.0 & 65.2 & 24.2 \\
facebook/seamless-m4t-v2-large & \textbf{56.2} & 40.1 & \textbf{48.4} & 20.4 & 54.4 & \textbf{61.6} & 49.1 & 44.7 & 49.4 & 14.3 & 49.1 & \textbf{57.2} & \textbf{48.1} & 36.6 & \textbf{34.7} \\
openai/whisper-medium & 31.2 & 37.4 & 20.5 & 19.6 & 33.8 & 15.1 & 40.9 & 19.6 & 26.8 & 24.1 & 17.8 & 16.1 & 24.0 & 28.5 & 23.1 \\
facebook/mms-1b-fl102 & 28.5 & 27.7 & 22.3 & \textbf{27.6} & 30.3 & 29.5 & 25.9 & 19.6 & 19.8 & 24.2 & 20.5 & 23.1 & 26.8 & 32.1 & 31.3 \\
facebook/mms-1b-all & 27.7 & 26.8 & 23.1 & 22.3 & 27.6 & 25.9 & 31.1 & 22.4 & 19.8 & 18.8 & 23.3 & 24.9 & 24.1 & 25.9 & 18.8 \\
OpenMuQ/MuQ-MuLan-large & 24.2 & 23.2 & 22.4 & 13.5 & 22.4 & 21.3 & 24.9 & 17.0 & 17.9 & 25.0 & 26.8 & 30.5 & 20.5 & 33.9 & 18.8 \\
\bottomrule
\end{tabular}
\end{table}
\end{landscape}

\clearpage
\begin{landscape}
\begin{table}[p]
\caption{SIB-FLEURS classification results (languages 46--60 of 102). Best per language in bold.}
\label{tab:sibfleurs_part4}
\centering
\scriptsize
\setlength{\tabcolsep}{3pt}
\begin{tabular}{@{}lccccccccccccccc@{}}
\toprule
Model & khm\_Khmr & kir\_Cyrl & kor\_Hang & lao\_Laoo & lin\_Latn & lit\_Latn & ltz\_Latn & lug\_Latn & luo\_Latn & lvs\_Latn & mal\_Mlym & mar\_Deva & mkd\_Cyrl & mlt\_Latn & mri\_Latn \\
\midrule
LCO-Embedding/LCO-Embedding-Omni-7B & 23.3 & 43.8 & 68.7 & \textbf{49.1} & 32.9 & 42.7 & \textbf{65.2} & 21.4 & \textbf{36.6} & 39.2 & 38.3 & \textbf{56.2} & 49.8 & \textbf{59.7} & 21.4 \\
LCO-Embedding/LCO-Embedding-Omni-3B & 22.3 & 48.1 & \textbf{70.5} & 42.7 & \textbf{33.0} & 39.1 & 59.9 & 24.2 & 29.4 & 41.9 & 33.0 & 52.8 & 58.9 & 56.2 & 23.3 \\
facebook/seamless-m4t-v2-large & \textbf{30.4} & \textbf{55.1} & 40.4 & 33.2 & 23.2 & \textbf{55.3} & 37.5 & \textbf{35.7} & 21.4 & \textbf{52.8} & \textbf{44.5} & 41.9 & \textbf{64.3} & 43.8 & 25.9 \\
openai/whisper-medium & 26.8 & 20.4 & 27.7 & 24.0 & 30.4 & 32.8 & 32.0 & 20.6 & 19.5 & 34.9 & 25.0 & 34.7 & 32.2 & 21.3 & \textbf{28.6} \\
facebook/mms-1b-fl102 & 27.7 & 24.9 & 19.7 & 30.4 & 26.7 & 32.0 & 26.0 & 27.7 & 25.7 & 35.6 & 21.3 & 27.6 & 25.1 & 24.9 & 24.2 \\
facebook/mms-1b-all & 18.7 & 16.1 & 29.6 & 23.2 & 23.1 & 23.3 & 22.3 & 25.0 & 22.3 & 26.6 & 25.0 & 32.9 & 26.1 & 28.7 & 22.4 \\
OpenMuQ/MuQ-MuLan-large & 15.2 & 14.4 & 18.7 & 25.0 & 25.1 & 27.7 & 23.2 & 11.6 & 25.8 & 27.7 & 16.0 & 15.1 & 28.7 & 17.9 & 19.8 \\
\bottomrule
\end{tabular}
\end{table}
\end{landscape}

\clearpage
\begin{landscape}
\begin{table}[p]
\caption{SIB-FLEURS classification results (languages 61--75 of 102). Best per language in bold.}
\label{tab:sibfleurs_part5}
\centering
\scriptsize
\setlength{\tabcolsep}{3pt}
\begin{tabular}{@{}lccccccccccccccc@{}}
\toprule
Model & mya\_Mymr & nld\_Latn & nob\_Latn & npi\_Deva & nso\_Latn & nya\_Latn & oci\_Latn & ory\_Orya & pan\_Guru & pbt\_Arab & pes\_Arab & pol\_Latn & por\_Latn & ron\_Latn & rus\_Cyrl \\
\midrule
LCO-Embedding/LCO-Embedding-Omni-7B & 15.3 & \textbf{71.5} & 42.0 & \textbf{55.3} & 23.3 & 23.1 & \textbf{56.3} & \textbf{49.0} & \textbf{53.7} & 29.3 & 38.5 & 66.2 & 71.4 & \textbf{58.0} & 69.7 \\
LCO-Embedding/LCO-Embedding-Omni-3B & 20.5 & 66.0 & \textbf{50.9} & 46.5 & 23.3 & 26.8 & 55.2 & 37.7 & 41.1 & \textbf{29.4} & 27.5 & \textbf{71.3} & \textbf{75.9} & 57.2 & \textbf{75.0} \\
facebook/seamless-m4t-v2-large & \textbf{34.0} & 52.9 & 46.5 & 42.0 & 20.5 & \textbf{45.6} & 30.4 & 40.2 & 50.0 & 25.9 & \textbf{55.3} & 43.8 & 50.9 & 44.9 & 52.7 \\
openai/whisper-medium & 8.1 & 31.3 & 38.3 & 26.7 & 17.7 & 17.8 & 26.7 & 24.9 & 33.0 & 21.4 & 24.8 & 36.6 & 42.8 & 33.2 & 33.0 \\
facebook/mms-1b-fl102 & 27.7 & 32.2 & 30.4 & 24.2 & 30.4 & 33.0 & 26.8 & 26.8 & 25.0 & 25.0 & 21.3 & 22.4 & 33.8 & 18.0 & 29.5 \\
facebook/mms-1b-all & 25.0 & 30.4 & 27.6 & 20.6 & 26.8 & 25.0 & 29.4 & 15.2 & 27.6 & 20.6 & 24.8 & 30.3 & 30.3 & 18.8 & 21.5 \\
OpenMuQ/MuQ-MuLan-large & 22.4 & 28.6 & 20.6 & 18.9 & 14.3 & 33.1 & 20.6 & 29.4 & 23.1 & 17.9 & 31.1 & 9.8 & 18.7 & 20.8 & 23.2 \\
\bottomrule
\end{tabular}
\end{table}
\end{landscape}

\clearpage
\begin{landscape}
\begin{table}[p]
\caption{SIB-FLEURS classification results (languages 76--90 of 102). Best per language in bold.}
\label{tab:sibfleurs_part6}
\centering
\scriptsize
\setlength{\tabcolsep}{3pt}
\begin{tabular}{@{}lccccccccccccccc@{}}
\toprule
Model & slk\_Latn & slv\_Latn & sna\_Latn & snd\_Arab & som\_Latn & spa\_Latn & srp\_Cyrl & swe\_Latn & swh\_Latn & tam\_Taml & tel\_Telu & tgk\_Cyrl & tgl\_Latn & tha\_Thai & tur\_Latn \\
\midrule
LCO-Embedding/LCO-Embedding-Omni-7B & \textbf{51.8} & \textbf{55.3} & 32.3 & \textbf{38.4} & 22.2 & \textbf{75.1} & \textbf{59.8} & 50.9 & 29.5 & 24.0 & \textbf{50.8} & 34.7 & \textbf{50.2} & \textbf{72.2} & 43.9 \\
LCO-Embedding/LCO-Embedding-Omni-3B & 49.0 & 44.5 & \textbf{40.2} & 37.5 & 25.8 & 69.6 & 51.8 & \textbf{52.7} & 29.5 & 31.3 & 50.0 & 28.5 & 43.0 & 65.8 & \textbf{49.1} \\
facebook/seamless-m4t-v2-large & 50.1 & 53.6 & 24.2 & 20.6 & \textbf{30.3} & 53.4 & 56.2 & 45.6 & \textbf{41.8} & \textbf{41.1} & 48.2 & \textbf{42.6} & 37.7 & 44.5 & 49.1 \\
openai/whisper-medium & 33.1 & 29.3 & 28.6 & 23.2 & 15.2 & 30.3 & 33.0 & 27.7 & 26.8 & 27.7 & 29.4 & 28.5 & 26.6 & 34.7 & 33.0 \\
facebook/mms-1b-fl102 & 37.5 & 30.3 & 24.2 & 26.6 & 21.4 & 19.6 & 23.2 & 22.2 & 37.5 & 26.7 & 23.2 & 26.0 & 20.6 & 32.8 & 28.7 \\
facebook/mms-1b-all & 22.3 & 24.1 & 27.8 & 25.8 & 28.6 & 21.4 & 25.0 & 18.7 & 33.1 & 24.0 & 24.9 & 24.1 & 15.1 & 24.0 & 24.3 \\
OpenMuQ/MuQ-MuLan-large & 32.3 & 31.2 & 33.1 & 17.1 & 15.2 & 22.3 & 14.3 & 27.8 & 24.1 & 34.0 & 26.8 & 21.6 & 17.8 & 22.5 & 28.7 \\
\bottomrule
\end{tabular}
\end{table}
\end{landscape}

\clearpage
\begin{landscape}
\begin{table}[p]
\caption{SIB-FLEURS classification results (languages 91--102 of 102). Best per language in bold.}
\label{tab:sibfleurs_part7}
\centering
\scriptsize
\setlength{\tabcolsep}{3pt}
\begin{tabular}{@{}lcccccccccccc@{}}
\toprule
Model & ukr\_Cyrl & umb\_Latn & urd\_Arab & uzn\_Latn & vie\_Latn & wol\_Latn & xho\_Latn & yor\_Latn & zho\_Hans & zho\_Hant & zsm\_Latn & zul\_Latn \\
\midrule
LCO-Embedding/LCO-Embedding-Omni-7B & 68.6 & 17.8 & 55.5 & 51.1 & \textbf{69.6} & \textbf{34.0} & 31.1 & 27.8 & \textbf{67.7} & \textbf{64.2} & \textbf{63.4} & 26.0 \\
LCO-Embedding/LCO-Embedding-Omni-3B & \textbf{71.3} & 20.6 & \textbf{57.2} & 40.2 & 66.0 & 33.0 & 25.0 & 32.1 & 65.9 & 62.5 & 55.5 & 26.7 \\
facebook/seamless-m4t-v2-large & 54.4 & 15.3 & 43.8 & \textbf{58.9} & 50.8 & 25.1 & 19.7 & \textbf{33.8} & 49.0 & 33.1 & 51.8 & 16.1 \\
openai/whisper-medium & 32.1 & 17.8 & 21.5 & 23.0 & 30.4 & 14.2 & 15.2 & 16.9 & 22.3 & 29.5 & 30.5 & 16.0 \\
facebook/mms-1b-fl102 & 32.0 & 27.7 & 24.9 & 31.4 & 22.4 & 28.3 & 22.4 & 26.7 & 26.0 & 24.2 & 26.7 & 26.8 \\
facebook/mms-1b-all & 27.8 & 25.8 & 23.1 & 31.2 & 22.4 & 18.7 & 24.0 & 29.4 & 22.3 & 19.8 & 27.7 & \textbf{28.7} \\
OpenMuQ/MuQ-MuLan-large & 23.3 & 8.9 & 21.5 & 25.1 & 25.1 & 23.2 & \textbf{33.0} & 24.1 & 10.8 & 25.1 & 28.6 & 25.0 \\
\bottomrule
\end{tabular}
\end{table}
\end{landscape}

\clearpage


\begin{table}[htbp]
\caption{VoxPopuli classification results. GenderID = gender classification, LanguageID = language identification. Both tasks are evaluated on multilingual audio samples containing English, French, Spanish, Polish, and German. Best result per task in bold.}
\label{tab:voxpopuli_results}
\centering
\footnotesize
\setlength{\tabcolsep}{3pt}
\begin{tabular}{@{}lcc@{}}
\toprule
Model & GenderID & LanguageID \\
\midrule
laion/larger\_clap\_general & 84.6 & 84.6 \\
Qwen/Qwen2-Audio-7B & 68.2 & 99.0 \\
laion/larger\_clap\_music\_and\_speech & 84.4 & 81.4 \\
openai/whisper-tiny & 66.6 & 96.6 \\
openai/whisper-small & 62.8 & 99.2 \\
openai/whisper-base & 63.6 & 98.0 \\
openai/whisper-medium & 59.2 & \textbf{99.4} \\
\bottomrule
\end{tabular}
\end{table}

\clearpage

\begin{landscape}
\begin{table}[h]
\caption{English clustering results.}
\label{tab:english_clustering_results}
\centering
\scriptsize
\setlength{\tabcolsep}{2pt}
\begin{tabular}{@{}l@{\hspace{4pt}}c@{\hspace{4pt}}c@{\hspace{4pt}}c@{\hspace{4pt}}c@{\hspace{4pt}}c@{\hspace{4pt}}c@{\hspace{4pt}}c@{\hspace{4pt}}c@{\hspace{4pt}}c@{}}
\toprule
Model & AmbientAcoustic & CREMA-D & ESC50 & GTZANGenre & MusicGenre & VehicleSound & VoiceGender & VoxCeleb & VoxPopuliAccent \\
\midrule
laion/larger\_clap\_music\_and\_speech & 43.51 & 14.43 & 94.46 & 63.82 & 45.30 & 4.72 & \textbf{71.33} & 0.70 & 10.61 \\
laion/clap-htsat-unfused & 43.68 & 12.78 & 94.76 & 55.70 & 40.07 & 2.66 & 68.11 & 1.06 & 9.35 \\
laion/larger\_clap\_general & 43.78 & 13.17 & 94.10 & 65.09 & 43.38 & 3.37 & 26.63 & 0.68 & \textbf{11.20} \\
Qwen/Qwen2-Audio-7B & 39.92 & \textbf{32.37} & 88.69 & 73.94 & 41.84 & 5.52 & 13.43 & 0.18 & 4.25 \\
microsoft/msclap-2023 & 44.12 & 10.73 & \textbf{95.33} & 61.91 & 36.57 & 2.86 & 31.43 & 0.78 & 9.58 \\
MIT/ast-finetuned-audioset-10-10-0.4593 & \textbf{46.76} & 7.20 & 91.36 & 61.60 & 47.20 & \textbf{13.37} & 14.07 & 1.01 & 7.55 \\
laion/clap-htsat-fused & 40.43 & 10.82 & 93.30 & 47.10 & 37.00 & 4.71 & 40.80 & 0.81 & 9.34 \\
\bottomrule
\end{tabular}
\end{table}
\end{landscape}

\begin{table}[h]
\caption{VoxPopuli gender clustering results. Task clusters audio samples by speaker gender across multilingual audio samples containing German, English, French, Spanish, and Polish. Best result in bold.}
\label{tab:multilingual_clustering_results}
\centering
\scriptsize
\setlength{\tabcolsep}{2pt}
\begin{tabular}{@{}l@{\hspace{4pt}}c@{}}
\toprule
Model & VoxPopuliGender \\
\midrule
laion/clap-htsat-fused & \textbf{52.68} \\
microsoft/msclap-2022 & 46.37 \\
microsoft/msclap-2023 & 32.09 \\
laion/clap-htsat-unfused & 22.20 \\
google/vggish & 8.20 \\
speechbrain/cnn14-esc50 & 7.72 \\
laion/larger\_clap\_music\_and\_speech & 3.80 \\
\bottomrule
\end{tabular}
\end{table}

\begin{table}[htbp]
\caption{Pair classification results. Models are evaluated on audio pair similarity tasks. Best result per task in bold.}
\label{tab:pair_classification_results}
\centering
\footnotesize
\setlength{\tabcolsep}{2pt}
\resizebox{\textwidth}{!}{%
\begin{tabular}{@{}l@{\hspace{15pt}}c@{\hspace{15pt}}c@{\hspace{15pt}}c@{\hspace{15pt}}c@{\hspace{15pt}}c@{{}}}
\toprule
Model & CREMA-D & ESC50 & NMSQA & VocalSound & VoxPopuliAccent \\
\midrule
LCO-Embedding/LCO-Embedding-Omni-7B & 53.56 & 99.44 & \textbf{97.60} & 94.47 & 50.73 \\
LCO-Embedding/LCO-Embedding-Omni-3B & 52.40 & \textbf{99.54} & 96.92 & \textbf{94.65} & 50.67 \\
microsoft/msclap-2023 & 57.65 & 99.22 & 51.09 & 84.22 & 52.05 \\
laion/clap-htsat-unfused & 56.91 & 99.31 & 46.98 & 81.58 & 53.96 \\
laion/larger\_clap\_music\_and\_speech & 56.66 & 99.35 & 46.14 & 80.36 & 53.49 \\
laion/larger\_clap\_general & 54.71 & 99.14 & 47.42 & 81.09 & 53.55 \\
Qwen/Qwen2-Audio-7B & \textbf{68.87} & 92.54 & 48.85 & 71.99 & 52.93 \\
\bottomrule
\end{tabular}}
\end{table}

\begin{table}[htbp]
\caption{Multilabel classification results. Models are evaluated on audio tagging tasks where each sample can have multiple labels. Best result per task in bold.}
\label{tab:multilabel_classification_results}
\centering
\scalebox{0.9}{%
\begin{tabular}{@{}l@{\hspace{15pt}}c@{\hspace{15pt}}c@{\hspace{15pt}}c@{\hspace{15pt}}c@{{}}}
\toprule
Model & AudioSetMini & BirdSet & FSD2019Kaggle & FSD50K \\
\midrule
MIT/ast-finetuned-audioset-10-10-0.4593 & \textbf{55.10} & \textbf{6.55} & 36.90 & 2.79 \\
LCO-Embedding/LCO-Embedding-Omni-7B & 43.60 & 3.03 & \textbf{44.19} & 4.90 \\
LCO-Embedding/LCO-Embedding-Omni-3B & 42.75 & 3.05 & 42.97 & 4.22 \\
google/yamnet & 46.46 & 2.51 & 18.95 & 1.34 \\
Qwen/Qwen2-Audio-7B & 46.28 & 0.81 & 7.05 & \textbf{13.78} \\
openai/whisper-medium & 38.15 & 2.77 & 17.84 & 3.25 \\
openai/whisper-small & 36.19 & 5.35 & 14.03 & 1.94 \\
\bottomrule
\end{tabular}}
\end{table}

\begin{table}[htbp]
\caption{Zero-shot classification results. Models classify audio using text descriptions without task-specific training. Best result per task in bold.}
\label{tab:zeroshot_classification_results}
\centering
\footnotesize
\setlength{\tabcolsep}{2pt}
\resizebox{\textwidth}{!}{%
\begin{tabular}{@{}l@{\hspace{15pt}}c@{\hspace{15pt}}c@{\hspace{15pt}}c@{\hspace{15pt}}c@{\hspace{15pt}}c@{{}}}
\toprule
Model & ESC50 & Ravdess & SpeechCmd v0.01 & SpeechCmd v0.02 & UrbanSound8k \\
\midrule
LCO-Embedding/LCO-Embedding-Omni-7B & 87.80 & \textbf{31.67} & \textbf{96.96} & \textbf{97.42} & 67.38 \\
LCO-Embedding/LCO-Embedding-Omni-3B & 87.65 & 26.94 & 96.61 & 97.40 & 68.80 \\
laion/larger\_clap\_general & \textbf{90.50} & 17.29 & 12.39 & 12.44 & 79.64 \\
microsoft/msclap-2023 & 89.85 & 15.21 & 9.78 & 10.01 & \textbf{83.01} \\
laion/larger\_clap\_music\_and\_speech & 82.85 & 17.29 & 9.23 & 9.03 & 76.86 \\
microsoft/msclap-2022 & 80.25 & 13.61 & 10.09 & 10.51 & 76.17 \\
laion/clap-htsat-unfused & 81.70 & 13.26 & 9.58 & 9.35 & 74.41 \\
laion/clap-htsat-fused & 74.00 & 14.65 & 11.06 & 11.76 & 60.16 \\
lyrebird/wav2clip & 40.75 & 11.81 & 9.74 & 9.72 & 34.18 \\
microsoft/speecht5\_multimodal & 1.35 & 12.99 & 19.21 & 18.80 & 8.30 \\
OpenMuQ/MuQ-MuLan-large & 2.70 & 14.72 & 10.95 & 10.48 & 15.33 \\
Qwen/Qwen2-Audio-7B & 1.00 & 14.37 & 9.89 & 10.38 & 11.82 \\
laion/larger\_clap\_music & 2.00 & 13.33 & 10.09 & 9.72 & 10.35 \\
\bottomrule
\end{tabular}}
\end{table}

\begin{table}[htbp]
\caption{Reranking results. Models are evaluated on audio-to-audio reranking tasks. Best result per task in bold.}
\label{tab:reranking_results}
\centering
\footnotesize
\setlength{\tabcolsep}{2pt}
\resizebox{\textwidth}{!}{%
\begin{tabular}{@{}l@{\hspace{15pt}}c@{\hspace{15pt}}c@{\hspace{15pt}}c@{\hspace{15pt}}c@{\hspace{15pt}}c@{{}}}
\toprule
Model & ESC50 & FSDnoisy18k & GTZAN & UrbanSound8K & VocalSound \\
\midrule
LCO-Embedding/LCO-Embedding-Omni-7B & 97.58 & 84.61 & 78.71 & \textbf{79.27} & \textbf{89.94} \\
LCO-Embedding/LCO-Embedding-Omni-3B & 97.16 & 83.19 & 75.37 & 77.93 & 89.28 \\
microsoft/msclap-2023 & \textbf{97.98} & \textbf{85.34} & 75.43 & 76.50 & 72.12 \\
Qwen/Qwen2-Audio-7B & 94.34 & 68.44 & 80.85 & 70.66 & 80.29 \\
MIT/ast-finetuned-audioset-10-10-0.4593 & 96.89 & 77.35 & 77.65 & 75.91 & 65.56 \\
microsoft/msclap-2022 & 96.28 & 79.11 & 62.91 & 77.96 & 59.69 \\
laion/clap-htsat-unfused & 89.47 & 71.21 & 66.45 & 69.14 & 52.02 \\
\bottomrule
\end{tabular}}
\end{table}

\begin{landscape}

\begin{table}[!ht]
\caption{English retrieval results (datasets 1--9 of 27).}
\label{tab:english_retrieval_part1}
\centering
\scriptsize
\setlength{\tabcolsep}{2pt}
\begin{tabular}{@{}l@{\hspace{4pt}}c@{\hspace{4pt}}c@{\hspace{4pt}}c@{\hspace{4pt}}c@{\hspace{4pt}}c@{\hspace{4pt}}c@{\hspace{4pt}}c@{\hspace{4pt}}c@{\hspace{4pt}}c@{}}
\toprule
Model & AudioCaps A2T & AudioCaps T2A & AudioSetStrong A2T & AudioSetStrong T2A & CMU Arctic A2T & CMU Arctic T2A & Clotho A2T & Clotho T2A & EmoVDB A2T \\
\midrule
LCO-Embedding/LCO-Embedding-Omni-7B & 56.74 & 53.82 & 44.14 & 42.19 & \textbf{99.92} & 99.54 & 37.32 & 33.89 & 97.37 \\
LCO-Embedding/LCO-Embedding-Omni-3B & 54.25 & 46.95 & 41.21 & 34.77 & 99.70 & \textbf{99.70} & 34.74 & 32.70 & \textbf{98.34} \\
laion/larger\_clap\_general & \textbf{76.22} & 57.06 & \textbf{87.30} & \textbf{85.55} & 21.23 & 34.12 & 44.69 & 33.72 & 1.80 \\
laion/larger\_clap\_music\_and\_speech & 74.18 & \textbf{63.14} & 46.29 & 47.07 & 22.21 & 44.20 & 38.95 & 32.39 & 3.94 \\
microsoft/msclap-2023 & 53.11 & 49.94 & 53.71 & 53.71 & 1.82 & 1.44 & 54.64 & 41.93 & 2.63 \\
laion/clap-htsat-unfused & 75.20 & 62.30 & 46.29 & 46.29 & 1.67 & 0.61 & 44.69 & 34.18 & 1.93 \\
microsoft/msclap-2022 & 64.10 & 52.73 & 33.01 & 25.39 & 0.91 & 0.99 & \textbf{71.77} & \textbf{58.73} & 2.00 \\
laion/clap-htsat-fused & 66.03 & 52.73 & 39.65 & 33.59 & 0.61 & 0.91 & 40.57 & 33.57 & 1.93 \\
OpenMuQ/MuQ-MuLan-large & 1.70 & 2.67 & 6.25 & 5.66 & 0.38 & 0.61 & 2.30 & 2.04 & 0.55 \\
lyrebird/wav2clip & 13.14 & 2.52 & 14.65 & 1.76 & 0.46 & 0.61 & 5.55 & 1.15 & 0.41 \\
microsoft/speecht5\_multimodal & 0.91 & 0.66 & 0.98 & 1.17 & 1.36 & 1.44 & 0.48 & 0.48 & 1.11 \\
Qwen/Qwen2-Audio-7B & 0.45 & 0.64 & 1.56 & 0.78 & 0.83 & 1.21 & 0.77 & 0.45 & 0.48 \\
laion/larger\_clap\_music & 0.34 & 0.57 & 1.37 & 0.98 & 0.38 & 0.38 & 0.57 & 0.47 & 0.28 \\
\bottomrule
\end{tabular}
\end{table}

\vspace{0.5em}
\begin{table}[!ht]
\caption{English retrieval results (datasets 10--18 of 27).}
\label{tab:english_retrieval_part2}
\centering
\scriptsize
\setlength{\tabcolsep}{2pt}
\begin{tabular}{@{}l@{\hspace{4pt}}c@{\hspace{4pt}}c@{\hspace{4pt}}c@{\hspace{4pt}}c@{\hspace{4pt}}c@{\hspace{4pt}}c@{\hspace{4pt}}c@{\hspace{4pt}}c@{\hspace{4pt}}c@{}}
\toprule
Model & EmoVDB T2A & GigaSpeech A2T & GigaSpeech T2A & HiFiTTS A2T & HiFiTTS T2A & JLCorpus A2T & JLCorpus T2A & LibriTTS A2T & LibriTTS T2A \\
\midrule
LCO-Embedding/LCO-Embedding-Omni-7B & 97.24 & \textbf{83.11} & \textbf{83.29} & \textbf{99.67} & \textbf{100.00} & 69.83 & 65.33 & \textbf{99.96} & 99.91 \\
LCO-Embedding/LCO-Embedding-Omni-3B & \textbf{98.69} & 82.96 & 83.13 & \textbf{99.67} & \textbf{100.00} & \textbf{70.50} & \textbf{70.67} & \textbf{99.96} & \textbf{99.94} \\
laion/larger\_clap\_general & 2.42 & 0.13 & 0.16 & 4.33 & 5.33 & 4.46 & 6.67 & 0.21 & 0.64 \\
laion/larger\_clap\_music\_and\_speech & 4.98 & 0.15 & 0.21 & 3.00 & 3.67 & 3.96 & 9.33 & 0.13 & 0.70 \\
microsoft/msclap-2023 & 2.42 & 0.19 & 0.10 & 3.00 & 3.33 & 6.42 & 6.67 & 0.36 & 0.36 \\
laion/clap-htsat-unfused & 2.35 & 0.09 & 0.36 & 4.00 & 5.33 & 4.00 & 6.00 & 0.45 & 0.43 \\
microsoft/msclap-2022 & 1.18 & 0.19 & 0.16 & 2.67 & 3.67 & 3.71 & 3.33 & 0.15 & 0.21 \\
laion/clap-htsat-fused & 1.93 & 0.15 & 0.18 & 1.33 & 5.00 & 3.21 & 3.33 & 0.43 & 0.45 \\
OpenMuQ/MuQ-MuLan-large & 0.62 & 0.12 & 0.01 & 2.00 & 2.33 & 3.54 & 3.33 & 0.08 & 0.06 \\
lyrebird/wav2clip & 0.35 & 0.07 & 0.09 & 0.67 & 2.33 & 2.04 & 2.67 & 0.13 & 0.17 \\
microsoft/speecht5\_multimodal & 0.62 & 0.21 & 0.39 & 2.67 & 11.67 & 6.96 & 4.67 & 0.17 & 1.00 \\
Qwen/Qwen2-Audio-7B & 0.41 & 0.06 & 0.09 & 3.00 & 2.33 & 7.21 & 10.00 & 0.21 & 0.47 \\
laion/larger\_clap\_music & 0.35 & 0.07 & 0.07 & 1.67 & 1.67 & 3.33 & 2.67 & 0.11 & 0.11 \\
\bottomrule
\end{tabular}
\end{table}

\vspace{0.5em}
\begin{table}[!ht]
\caption{English retrieval results (datasets 19--27 of 27).}
\label{tab:english_retrieval_part3}
\centering
\scriptsize
\setlength{\tabcolsep}{2pt}
\begin{tabular}{@{}l@{\hspace{4pt}}c@{\hspace{4pt}}c@{\hspace{4pt}}c@{\hspace{4pt}}c@{\hspace{4pt}}c@{\hspace{4pt}}c@{\hspace{4pt}}c@{\hspace{4pt}}c@{\hspace{4pt}}c@{}}
\toprule
Model & MACS A2T & MACS T2A & MusicCaps A2T & MusicCaps T2A & SoundDescs A2T & SoundDescs T2A & SpokenSQuAD T2A & UrbanSound8K A2T & UrbanSound8K T2A \\
\midrule
LCO-Embedding/LCO-Embedding-Omni-7B & 16.03 & 29.77 & \textbf{20.93} & \textbf{24.83} & 21.63 & 32.79 & \textbf{74.00} & 0.79 & 0.92 \\
LCO-Embedding/LCO-Embedding-Omni-3B & 18.32 & 22.65 & 19.69 & 18.83 & 18.82 & 29.09 & 72.00 & 0.67 & 0.88 \\
laion/larger\_clap\_general & 30.03 & 33.08 & 17.36 & 16.49 & 23.45 & 22.96 & 2.00 & 0.94 & 0.96 \\
laion/larger\_clap\_music\_and\_speech & 28.24 & 30.53 & 13.36 & 14.02 & 24.07 & 24.44 & 2.00 & \textbf{1.02} & 0.94 \\
microsoft/msclap-2023 & 15.52 & 27.23 & 18.48 & 19.22 & \textbf{38.14} & \textbf{37.28} & 0.00 & 0.90 & 0.94 \\
laion/clap-htsat-unfused & 25.19 & 27.99 & 11.70 & 11.89 & 20.52 & 21.33 & 1.33 & 0.98 & \textbf{0.98} \\
microsoft/msclap-2022 & \textbf{40.46} & \textbf{41.73} & 4.51 & 3.27 & 6.71 & 6.89 & 2.33 & 0.90 & 0.88 \\
laion/clap-htsat-fused & 29.52 & 29.26 & 8.48 & 7.54 & 15.99 & 17.81 & 2.00 & 0.59 & 0.94 \\
OpenMuQ/MuQ-MuLan-large & 1.02 & 1.78 & 12.38 & 11.19 & 0.34 & 0.48 & 1.00 & 0.26 & 0.27 \\
lyrebird/wav2clip & 1.78 & 1.78 & 3.88 & 0.61 & 2.85 & 1.44 & 0.00 & 0.37 & 0.18 \\
microsoft/speecht5\_multimodal & 1.53 & 1.53 & 0.12 & 0.16 & 0.08 & 0.00 & 3.33 & 0.14 & 0.18 \\
Qwen/Qwen2-Audio-7B & 1.27 & 1.53 & 0.14 & 0.21 & 0.18 & 0.18 & 1.00 & 0.14 & 0.22 \\
laion/larger\_clap\_music & 1.27 & 1.27 & 0.12 & 0.12 & 0.08 & 0.12 & 4.33 & 0.10 & 0.10 \\
\bottomrule
\end{tabular}
\end{table}

\end{landscape}



\begin{landscape}

\begin{table}[!ht]
\caption{FLEURS A2T retrieval results (languages 1--26 of 102). Best per language in bold.}
\label{tab:fleurs_a2t_part1}
\centering
\scriptsize
\setlength{\tabcolsep}{2pt}
\begin{tabular}{@{}lcccccccccccccccccccccccccc@{}}
\toprule
Model & af & am & ar & as & ast & az & be & bg & bn & bs & ca & ceb & ckb & cmn & cs & cy & da & de & el & en & es & et & fa & ff & fi & fil \\
\midrule
LCO-Embedding/LCO-Embedding-Omni-3B & 88.3 & \textbf{41.9} & \textbf{98.8} & \textbf{43.1} & \textbf{100.0} & \textbf{82.2} & \textbf{98.3} & 91.2 & 59.8 & \textbf{87.4} & \textbf{99.9} & \textbf{80.8} & \textbf{47.6} & \textbf{100.0} & \textbf{89.3} & \textbf{35.5} & 65.3 & \textbf{100.0} & \textbf{61.8} & \textbf{100.0} & \textbf{100.0} & \textbf{57.9} & 37.3 & 62.0 & \textbf{50.4} & \textbf{80.5} \\
LCO-Embedding/LCO-Embedding-Omni-7B & \textbf{90.5} & 37.0 & \textbf{98.8} & 40.5 & \textbf{100.0} & 75.9 & 97.9 & \textbf{92.4} & \textbf{60.8} & 84.2 & 99.6 & 79.1 & 39.3 & \textbf{100.0} & 78.1 & 33.3 & \textbf{68.4} & 99.9 & 39.5 & \textbf{100.0} & \textbf{100.0} & 54.1 & \textbf{40.5} & \textbf{65.2} & 36.5 & 77.6 \\
Qwen/Qwen2-Audio-7B & 1.9 & 1.7 & 2.8 & 1.2 & 1.9 & 1.1 & 0.8 & 0.9 & 1.2 & 1.3 & 1.7 & 1.3 & 1.1 & 1.7 & 1.4 & 0.5 & 1.2 & 0.9 & 2.0 & 3.4 & 3.6 & 1.6 & 0.9 & 1.4 & 1.4 & 0.8 \\
microsoft/speecht5\_multimodal & 2.3 & 1.0 & 2.1 & 1.1 & 0.4 & 0.7 & 0.6 & 1.7 & 0.5 & 0.4 & 1.1 & 1.3 & 1.1 & 1.0 & 1.7 & 1.1 & 1.4 & 0.9 & 1.1 & 1.4 & 0.7 & 0.7 & 0.5 & 1.1 & 1.5 & 0.6 \\
laion/larger\_clap\_music\_and\_speech & 2.7 & 1.4 & 1.4 & 0.2 & 1.9 & 0.8 & 0.4 & 1.5 & 0.5 & 0.9 & 0.7 & 1.7 & 0.5 & 0.5 & 1.0 & 0.8 & 1.2 & 0.3 & 0.6 & 0.9 & 1.2 & 0.6 & 0.6 & 1.1 & 0.5 & 0.3 \\
laion/clap-htsat-fused & 2.7 & 0.4 & 0.9 & 0.6 & 0.8 & 0.3 & 0.4 & 0.5 & 0.4 & 0.4 & 0.4 & 1.1 & 0.8 & 1.1 & 1.1 & 0.6 & 0.8 & 0.6 & 0.2 & 0.9 & 0.4 & 0.8 & 0.6 & 0.6 & 0.8 & 0.3 \\
laion/larger\_clap\_general & 1.9 & 0.8 & 0.9 & 0.7 & 1.2 & 0.8 & 0.6 & 1.2 & 0.5 & 0.5 & 0.7 & 0.6 & 0.5 & 0.3 & 1.4 & 0.5 & 1.1 & 0.7 & 0.9 & 1.5 & 0.9 & 0.3 & 0.9 & 1.1 & 0.8 & 0.6 \\
\bottomrule
\end{tabular}
\end{table}

\vspace{1em}
\begin{table}[!ht]
\caption{FLEURS A2T retrieval results (languages 27--52 of 102). Best per language in bold.}
\label{tab:fleurs_a2t_part2}
\centering
\scriptsize
\setlength{\tabcolsep}{2pt}
\begin{tabular}{@{}lcccccccccccccccccccccccccc@{}}
\toprule
Model & fr & ga & gl & gu & ha & he & hi & hr & hu & hy & id & ig & is & it & ja & jv & ka & kam & kea & kk & km & kn & ko & ky & lb & lg \\
\midrule
LCO-Embedding/LCO-Embedding-Omni-3B & \textbf{100.0} & \textbf{29.2} & \textbf{100.0} & 91.8 & \textbf{40.4} & \textbf{38.3} & \textbf{99.0} & \textbf{95.1} & \textbf{35.5} & \textbf{36.6} & \textbf{99.9} & \textbf{35.5} & \textbf{73.9} & \textbf{100.0} & \textbf{100.0} & \textbf{89.3} & \textbf{52.6} & 33.9 & 98.0 & \textbf{61.4} & 16.0 & 64.0 & \textbf{100.0} & \textbf{75.3} & 92.7 & \textbf{44.5} \\
LCO-Embedding/LCO-Embedding-Omni-7B & \textbf{100.0} & 27.6 & \textbf{100.0} & \textbf{92.2} & 31.1 & 37.2 & 98.8 & 87.7 & 31.7 & 27.5 & \textbf{99.9} & 32.7 & 54.3 & \textbf{100.0} & \textbf{100.0} & \textbf{89.3} & 47.4 & \textbf{38.0} & \textbf{98.6} & 59.3 & \textbf{16.5} & \textbf{64.9} & \textbf{100.0} & 69.7 & \textbf{93.3} & 42.3 \\
Qwen/Qwen2-Audio-7B & 3.3 & 1.2 & 2.9 & 1.2 & 0.8 & 1.0 & 1.9 & 1.5 & 1.2 & 1.3 & 2.5 & 0.8 & 21.7 & 3.1 & 2.9 & 0.8 & 1.3 & 1.0 & 1.5 & 1.6 & 0.4 & 0.6 & 2.6 & 1.4 & 1.2 & 1.0 \\
microsoft/speecht5\_multimodal & 0.9 & 1.1 & 1.1 & 1.1 & 1.1 & 1.0 & 2.2 & 0.3 & 0.9 & 0.9 & 1.6 & 0.7 & 15.2 & 1.2 & 0.9 & 1.1 & 0.9 & 1.0 & 0.6 & 0.6 & 0.6 & 0.7 & 1.3 & 0.7 & 1.2 & 0.7 \\
laion/larger\_clap\_music\_and\_speech & 0.9 & 0.8 & 0.9 & 0.3 & 1.0 & 1.1 & 0.7 & 0.7 & 0.7 & 0.4 & 1.2 & 0.5 & 10.9 & 1.0 & 1.1 & 0.5 & 0.4 & 1.0 & 0.8 & 0.8 & 0.4 & 0.7 & 2.1 & 0.4 & 0.6 & 1.0 \\
laion/clap-htsat-fused & 0.4 & 0.6 & 0.5 & 0.6 & 1.3 & 0.9 & 0.7 & 0.5 & 1.1 & 0.8 & 1.2 & 0.7 & 13.0 & 0.8 & 0.9 & 0.5 & 0.3 & 0.7 & 0.5 & 0.5 & 1.0 & 0.2 & 1.3 & 0.5 & 0.9 & 0.6 \\
laion/larger\_clap\_general & 1.0 & 0.9 & 0.5 & 0.5 & 1.0 & 0.6 & 1.2 & 0.9 & 1.4 & 0.6 & 0.9 & 0.6 & 15.2 & 0.7 & 0.9 & 1.0 & 0.6 & 0.8 & 0.7 & 0.8 & 0.6 & 1.1 & 0.8 & 0.5 & 0.5 & 1.0 \\
\bottomrule
\end{tabular}
\end{table}

\vspace{1em}
\begin{table}[!ht]
\caption{FLEURS A2T retrieval results (languages 53--78 of 102). Best per language in bold.}
\label{tab:fleurs_a2t_part3}
\centering
\scriptsize
\setlength{\tabcolsep}{2pt}
\begin{tabular}{@{}lcccccccccccccccccccccccccc@{}}
\toprule
Model & ln & lo & lt & luo & lv & mi & mk & ml & mn & mr & ms & mt & my & nb & ne & nl & nso & ny & oc & om & or & pa & pl & ps & pt & ro \\
\midrule
LCO-Embedding/LCO-Embedding-Omni-3B & 43.1 & 54.3 & 51.8 & 66.0 & \textbf{57.9} & \textbf{23.8} & \textbf{95.3} & 55.8 & \textbf{31.6} & 80.6 & 99.6 & \textbf{89.5} & 6.2 & \textbf{92.2} & 80.0 & \textbf{100.0} & \textbf{43.4} & \textbf{43.5} & 91.7 & \textbf{97.6} & 64.9 & 85.0 & \textbf{98.3} & 59.8 & \textbf{100.0} & \textbf{96.5} \\
LCO-Embedding/LCO-Embedding-Omni-7B & \textbf{47.5} & \textbf{74.6} & \textbf{55.2} & \textbf{68.0} & 56.5 & 22.4 & 92.9 & \textbf{56.5} & 22.1 & \textbf{82.6} & \textbf{99.9} & 89.0 & \textbf{6.7} & 83.8 & \textbf{80.9} & 99.7 & 43.3 & 41.5 & \textbf{93.1} & 90.2 & \textbf{65.3} & \textbf{89.9} & 86.3 & \textbf{60.0} & \textbf{100.0} & 93.2 \\
Qwen/Qwen2-Audio-7B & 0.4 & 4.0 & 0.6 & 1.6 & 2.1 & 0.8 & 1.5 & 0.8 & 0.9 & 0.9 & 1.6 & 1.9 & 0.7 & 4.5 & 1.1 & 2.2 & 0.6 & 0.8 & 1.0 & 14.6 & 1.5 & 1.2 & 1.1 & 1.0 & 3.7 & 1.2 \\
microsoft/speecht5\_multimodal & 1.7 & 1.7 & 0.5 & 3.1 & 1.4 & 0.7 & 0.8 & 0.7 & 0.6 & 0.9 & 1.3 & 0.8 & 0.6 & 3.1 & 0.6 & 1.9 & 1.0 & 1.2 & 0.7 & 12.2 & 0.9 & 1.2 & 0.5 & 2.0 & 1.0 & 1.0 \\
laion/larger\_clap\_music\_and\_speech & 1.0 & 1.7 & 0.9 & 3.1 & 0.7 & 1.2 & 0.7 & 0.5 & 0.6 & 0.1 & 0.8 & 0.5 & 0.6 & 2.5 & 0.7 & 2.7 & 1.1 & 0.5 & 0.7 & 19.5 & 0.5 & 0.3 & 1.2 & 1.6 & 0.8 & 0.8 \\
laion/clap-htsat-fused & 1.5 & 1.0 & 0.4 & 2.0 & 0.7 & 0.4 & 0.2 & 0.7 & 0.4 & 0.2 & 0.7 & 0.5 & 0.1 & 1.4 & 0.4 & 1.4 & 1.3 & 0.8 & 0.8 & 19.5 & 0.6 & 0.7 & 0.5 & 1.0 & 0.4 & 0.9 \\
laion/larger\_clap\_general & 1.0 & 1.5 & 0.7 & 2.0 & 0.5 & 0.7 & 0.4 & 0.6 & 0.1 & 0.3 & 0.3 & 0.5 & 0.3 & 1.7 & 0.4 & 1.9 & 1.1 & 0.8 & 0.6 & 7.3 & 0.6 & 1.4 & 1.1 & 0.2 & 0.7 & 0.7 \\
\bottomrule
\end{tabular}
\end{table}

\vspace{1em}
\begin{table}[!ht]
\caption{FLEURS A2T retrieval results (languages 79--102 of 102). Best per language in bold.}
\label{tab:fleurs_a2t_part4}
\centering
\scriptsize
\setlength{\tabcolsep}{2pt}
\begin{tabular}{@{}lcccccccccccccccccccccccc@{}}
\toprule
Model & ru & sd & sk & sl & sn & so & sr & sv & sw & ta & te & tg & th & tr & uk & umb & ur & uz & vi & wo & xh & yo & yue & zu \\
\midrule
LCO-Embedding/LCO-Embedding-Omni-3B & \textbf{100.0} & \textbf{77.3} & \textbf{96.6} & 78.1 & \textbf{45.1} & \textbf{37.8} & \textbf{89.7} & \textbf{86.6} & 44.1 & \textbf{32.8} & \textbf{85.8} & 66.5 & \textbf{99.7} & \textbf{99.1} & 98.3 & \textbf{35.1} & \textbf{99.0} & 52.9 & \textbf{99.2} & \textbf{66.8} & \textbf{43.1} & 29.5 & \textbf{100.0} & \textbf{33.0} \\
LCO-Embedding/LCO-Embedding-Omni-7B & \textbf{100.0} & 76.6 & 92.2 & \textbf{83.8} & 43.7 & 33.3 & 89.6 & 78.1 & \textbf{48.9} & 26.2 & 72.0 & \textbf{69.3} & \textbf{99.7} & 97.7 & \textbf{98.9} & 30.1 & \textbf{99.0} & \textbf{54.5} & 99.1 & \textbf{66.8} & 39.4 & \textbf{31.5} & \textbf{100.0} & 27.9 \\
Qwen/Qwen2-Audio-7B & 2.2 & 0.7 & 1.8 & 1.0 & 1.0 & 1.0 & 1.7 & 1.6 & 1.8 & 0.5 & 1.9 & 1.3 & 0.7 & 1.5 & 1.5 & 0.8 & 3.0 & 0.5 & 1.9 & 2.2 & 0.9 & 0.5 & 1.3 & 1.3 \\
microsoft/speecht5\_multimodal & 1.3 & 0.9 & 0.6 & 0.5 & 0.8 & 1.2 & 1.0 & 1.3 & 1.8 & 1.4 & 1.5 & 1.2 & 0.6 & 0.7 & 1.2 & 1.1 & 2.0 & 1.3 & 0.6 & 1.6 & 0.7 & 1.1 & 1.5 & 0.7 \\
laion/larger\_clap\_music\_and\_speech & 0.8 & 0.9 & 1.3 & 1.6 & 0.8 & 0.5 & 0.9 & 1.1 & 0.8 & 1.0 & 1.3 & 1.5 & 0.3 & 0.5 & 0.7 & 0.5 & 2.0 & 0.5 & 1.2 & 1.6 & 0.8 & 0.5 & 0.6 & 0.8 \\
laion/clap-htsat-fused & 0.8 & 0.5 & 0.8 & 0.5 & 0.9 & 0.4 & 0.9 & 0.9 & 1.0 & 1.4 & 1.1 & 0.7 & 0.2 & 0.9 & 1.2 & 0.5 & 1.3 & 0.5 & 0.4 & 0.8 & 0.3 & 0.2 & 0.9 & 0.7 \\
laion/larger\_clap\_general & 0.5 & 0.7 & 1.0 & 0.8 & 0.4 & 0.6 & 0.6 & 0.7 & 1.0 & 1.0 & 1.7 & 0.8 & 0.6 & 0.9 & 0.3 & 1.6 & 0.7 & 0.3 & 1.3 & 1.1 & 0.6 & 0.2 & 0.5 & 1.1 \\
\bottomrule
\end{tabular}
\end{table}

\vspace{1em}
\begin{table}[!ht]
\caption{FLEURS T2A retrieval results (languages 1--26 of 102). Best per language in bold.}
\label{tab:fleurs_t2a_part1}
\centering
\scriptsize
\setlength{\tabcolsep}{2pt}
\begin{tabular}{@{}lcccccccccccccccccccccccccc@{}}
\toprule
Model & af & am & ar & as & ast & az & be & bg & bn & bs & ca & ceb & ckb & cmn & cs & cy & da & de & el & en & es & et & fa & ff & fi & fil \\
\midrule
LCO-Embedding/LCO-Embedding-Omni-3B & \textbf{92.4} & \textbf{26.9} & 97.9 & \textbf{47.9} & \textbf{100.0} & \textbf{82.8} & \textbf{98.1} & 91.9 & \textbf{64.5} & \textbf{87.1} & \textbf{99.4} & \textbf{88.4} & \textbf{43.2} & 99.9 & \textbf{90.2} & 26.1 & \textbf{73.7} & \textbf{99.9} & \textbf{69.2} & \textbf{100.0} & \textbf{100.0} & \textbf{66.9} & \textbf{45.0} & \textbf{65.3} & \textbf{58.9} & \textbf{86.5} \\
LCO-Embedding/LCO-Embedding-Omni-7B & 90.2 & 25.6 & \textbf{98.6} & 42.5 & 99.8 & 76.4 & 97.5 & \textbf{93.0} & 62.9 & 82.2 & 98.7 & 81.9 & 38.0 & \textbf{100.0} & 79.5 & \textbf{26.3} & 68.5 & \textbf{99.9} & 44.5 & \textbf{100.0} & \textbf{100.0} & 55.2 & 43.7 & 54.7 & 39.8 & 80.8 \\
Qwen/Qwen2-Audio-7B & 3.4 & 2.3 & 1.4 & 0.6 & 2.5 & 0.5 & 1.2 & 1.5 & 0.8 & 1.0 & 1.3 & 1.3 & 1.0 & 1.7 & 1.0 & 0.4 & 0.5 & 1.2 & 1.7 & 4.3 & 2.4 & 0.9 & 0.9 & 1.2 & 0.5 & 0.4 \\
microsoft/speecht5\_multimodal & 1.5 & 1.4 & 1.2 & 1.4 & 0.4 & 1.1 & 0.7 & 1.5 & 0.9 & 0.6 & 0.5 & 1.1 & 0.7 & 1.0 & 1.4 & 0.7 & 1.2 & 1.2 & 1.1 & 3.4 & 0.6 & 0.9 & 0.9 & 1.1 & 1.0 & 0.5 \\
laion/larger\_clap\_music\_and\_speech & 3.0 & 1.0 & 1.4 & 0.5 & 1.9 & 0.8 & 0.6 & 0.6 & 0.5 & 1.7 & 1.2 & 0.9 & 0.7 & 0.6 & 1.9 & 0.8 & 0.6 & 1.0 & 0.9 & 1.4 & 1.0 & 1.1 & 0.8 & 0.8 & 0.9 & 1.0 \\
laion/larger\_clap\_general & 4.2 & 1.0 & 0.7 & 0.5 & 1.5 & 0.7 & 0.4 & 0.8 & 0.5 & 1.1 & 0.5 & 0.9 & 0.4 & 0.4 & 1.2 & 0.5 & 1.0 & 1.3 & 0.6 & 1.9 & 0.8 & 0.8 & 0.6 & 1.8 & 0.9 & 0.8 \\
lyrebird/wav2clip & 1.5 & 1.0 & 0.5 & 0.6 & 0.7 & 0.4 & 0.6 & 0.6 & 0.4 & 0.3 & 0.4 & 1.3 & 0.9 & 0.4 & 0.6 & 0.3 & 0.8 & 1.2 & 0.8 & 1.1 & 0.2 & 0.4 & 0.6 & 0.5 & 0.8 & 0.5 \\
\bottomrule
\end{tabular}
\end{table}

\vspace{1em}
\begin{table}[!ht]
\caption{FLEURS T2A retrieval results (languages 27--52 of 102). Best per language in bold.}
\label{tab:fleurs_t2a_part2}
\centering
\scriptsize
\setlength{\tabcolsep}{2pt}
\begin{tabular}{@{}lcccccccccccccccccccccccccc@{}}
\toprule
Model & fr & ga & gl & gu & ha & he & hi & hr & hu & hy & id & ig & is & it & ja & jv & ka & kam & kea & kk & km & kn & ko & ky & lb & lg \\
\midrule
LCO-Embedding/LCO-Embedding-Omni-3B & 99.9 & \textbf{29.8} & 99.9 & 91.8 & \textbf{32.5} & \textbf{43.8} & 98.3 & \textbf{94.6} & \textbf{43.1} & \textbf{39.4} & \textbf{99.9} & \textbf{31.9} & 71.7 & \textbf{100.0} & 99.7 & 91.5 & \textbf{59.4} & 30.8 & 97.5 & \textbf{67.8} & 15.4 & 72.1 & \textbf{100.0} & \textbf{75.1} & \textbf{92.9} & \textbf{38.3} \\
LCO-Embedding/LCO-Embedding-Omni-7B & \textbf{100.0} & 27.2 & \textbf{100.0} & \textbf{92.1} & 31.4 & 42.8 & \textbf{98.6} & 87.3 & 36.2 & 32.2 & \textbf{99.9} & 23.3 & \textbf{76.1} & \textbf{100.0} & \textbf{99.8} & \textbf{93.4} & 52.8 & \textbf{33.6} & \textbf{97.9} & 59.8 & \textbf{15.7} & \textbf{73.5} & \textbf{100.0} & 70.4 & 92.2 & 36.9 \\
Qwen/Qwen2-Audio-7B & 2.5 & 0.5 & 2.3 & 1.3 & 0.5 & 1.3 & 1.2 & 1.1 & 0.4 & 0.8 & 1.0 & 0.5 & 19.6 & 2.7 & 0.9 & 0.8 & 0.8 & 0.6 & 0.7 & 0.7 & 0.9 & 1.0 & 2.4 & 0.5 & 0.6 & 1.1 \\
microsoft/speecht5\_multimodal & 1.0 & 0.7 & 0.9 & 1.2 & 1.0 & 1.0 & 1.9 & 0.9 & 1.0 & 1.0 & 1.6 & 1.3 & 13.0 & 0.7 & 0.8 & 1.2 & 0.6 & 1.1 & 0.9 & 1.1 & 0.9 & 0.7 & 1.6 & 0.5 & 0.7 & 1.1 \\
laion/larger\_clap\_music\_and\_speech & 2.5 & 1.2 & 1.3 & 0.5 & 0.5 & 0.5 & 1.2 & 1.3 & 0.8 & 0.4 & 1.0 & 0.4 & 10.9 & 2.0 & 0.5 & 0.8 & 0.6 & 1.0 & 1.4 & 0.6 & 0.6 & 0.6 & 1.8 & 0.6 & 0.9 & 1.0 \\
laion/larger\_clap\_general & 1.9 & 1.4 & 0.4 & 0.4 & 0.6 & 0.6 & 1.4 & 1.1 & 1.2 & 0.5 & 0.9 & 0.8 & 17.4 & 1.7 & 1.2 & 1.4 & 0.4 & 1.0 & 0.8 & 0.6 & 0.6 & 0.1 & 1.3 & 0.1 & 1.0 & 1.5 \\
lyrebird/wav2clip & 0.6 & 0.6 & 0.9 & 0.5 & 0.8 & 0.6 & 1.7 & 0.3 & 0.6 & 0.4 & 0.6 & 0.4 & 15.2 & 0.3 & 0.6 & 0.7 & 0.4 & 0.7 & 0.7 & 0.5 & 0.5 & 0.7 & 1.6 & 0.5 & 0.5 & 0.4 \\
\bottomrule
\end{tabular}
\end{table}

\vspace{1em}
\begin{table}[!ht]
\caption{FLEURS T2A retrieval results (languages 53--78 of 102). Best per language in bold.}
\label{tab:fleurs_t2a_part3}
\centering
\scriptsize
\setlength{\tabcolsep}{2pt}
\begin{tabular}{@{}lcccccccccccccccccccccccccc@{}}
\toprule
Model & ln & lo & lt & luo & lv & mi & mk & ml & mn & mr & ms & mt & my & nb & ne & nl & nso & ny & oc & om & or & pa & pl & ps & pt & ro \\
\midrule
LCO-Embedding/LCO-Embedding-Omni-3B & \textbf{45.6} & 40.7 & \textbf{63.2} & \textbf{73.0} & \textbf{66.4} & \textbf{23.2} & \textbf{95.0} & \textbf{65.2} & \textbf{24.1} & \textbf{82.5} & \textbf{99.6} & \textbf{89.1} & \textbf{5.2} & \textbf{93.8} & \textbf{82.0} & \textbf{99.7} & \textbf{40.0} & \textbf{38.2} & \textbf{91.0} & \textbf{73.2} & 62.2 & 83.4 & \textbf{97.0} & \textbf{63.5} & \textbf{100.0} & \textbf{95.7} \\
LCO-Embedding/LCO-Embedding-Omni-7B & 43.7 & \textbf{64.0} & 58.2 & 65.2 & 58.6 & 21.0 & 92.2 & 64.2 & 21.3 & 80.5 & \textbf{99.6} & 86.5 & 4.3 & 84.9 & 80.2 & \textbf{99.7} & 31.9 & 35.5 & 90.4 & 70.7 & \textbf{62.5} & \textbf{87.3} & 84.2 & 59.0 & \textbf{100.0} & 91.4 \\
Qwen/Qwen2-Audio-7B & 1.0 & 2.0 & 0.6 & 2.3 & 0.9 & 0.8 & 0.3 & 1.1 & 0.7 & 0.3 & 1.2 & 0.5 & 0.9 & 3.1 & 1.0 & 1.4 & 0.5 & 0.1 & 0.5 & 17.1 & 1.5 & 1.0 & 0.9 & 1.2 & 2.5 & 1.2 \\
microsoft/speecht5\_multimodal & 1.5 & 1.5 & 0.8 & 3.9 & 0.5 & 0.5 & 0.8 & 0.5 & 0.4 & 0.6 & 0.7 & 0.6 & 0.8 & 2.5 & 1.5 & 2.2 & 1.0 & 1.2 & 1.0 & 19.5 & 0.7 & 0.9 & 1.3 & 1.2 & 1.2 & 0.9 \\
laion/larger\_clap\_music\_and\_speech & 2.1 & 1.5 & 1.1 & 2.3 & 1.3 & 1.0 & 0.6 & 0.6 & 0.4 & 0.6 & 0.8 & 1.1 & 0.5 & 1.7 & 1.0 & 2.5 & 0.9 & 0.7 & 0.6 & 12.2 & 0.7 & 0.5 & 2.4 & 1.2 & 2.6 & 1.5 \\
laion/larger\_clap\_general & 1.9 & 1.5 & 1.3 & 3.1 & 0.5 & 1.0 & 0.5 & 0.3 & 0.3 & 0.3 & 0.8 & 1.3 & 0.7 & 2.0 & 0.4 & 2.2 & 1.0 & 1.1 & 0.7 & 7.3 & 0.3 & 0.5 & 1.2 & 0.8 & 1.3 & 1.4 \\
lyrebird/wav2clip & 0.8 & 1.2 & 0.3 & 2.3 & 0.5 & 0.3 & 0.5 & 0.6 & 0.6 & 0.4 & 0.4 & 0.4 & 0.5 & 1.7 & 0.7 & 2.2 & 0.8 & 0.7 & 0.6 & 17.1 & 0.6 & 1.0 & 0.7 & 1.0 & 0.3 & 0.7 \\
\bottomrule
\end{tabular}
\end{table}

\vspace{1em}
\begin{table}[!ht]
\caption{FLEURS T2A retrieval results (languages 79--102 of 102). Best per language in bold.}
\label{tab:fleurs_t2a_part4}
\centering
\scriptsize
\setlength{\tabcolsep}{2pt}
\begin{tabular}{@{}lcccccccccccccccccccccccc@{}}
\toprule
Model & ru & sd & sk & sl & sn & so & sr & sv & sw & ta & te & tg & th & tr & uk & umb & ur & uz & vi & wo & xh & yo & yue & zu \\
\midrule
LCO-Embedding/LCO-Embedding-Omni-3B & \textbf{100.0} & \textbf{75.8} & \textbf{95.6} & \textbf{83.2} & \textbf{35.5} & \textbf{32.1} & \textbf{91.1} & \textbf{83.7} & 45.0 & \textbf{36.2} & \textbf{85.2} & 66.3 & 99.3 & \textbf{99.2} & \textbf{98.3} & \textbf{36.7} & \textbf{100.0} & \textbf{56.8} & 99.0 & \textbf{58.0} & \textbf{33.6} & \textbf{33.8} & 99.9 & \textbf{20.8} \\
LCO-Embedding/LCO-Embedding-Omni-7B & \textbf{100.0} & 74.8 & 91.5 & \textbf{83.2} & 32.3 & 21.6 & 91.0 & 75.4 & \textbf{47.4} & 33.3 & 74.8 & \textbf{68.0} & \textbf{99.7} & 98.0 & 97.5 & 29.8 & 99.0 & 53.8 & \textbf{99.2} & 56.1 & 28.8 & 31.9 & \textbf{100.0} & 19.3 \\
Qwen/Qwen2-Audio-7B & 1.4 & 0.4 & 0.9 & 1.0 & 0.5 & 0.7 & 1.3 & 0.5 & 2.1 & 0.8 & 1.7 & 1.2 & 0.7 & 1.1 & 1.6 & 1.6 & 2.0 & 0.5 & 1.3 & 1.9 & 0.8 & 1.1 & 0.7 & 0.8 \\
microsoft/speecht5\_multimodal & 1.2 & 0.6 & 1.0 & 1.2 & 0.6 & 0.4 & 1.7 & 1.3 & 1.6 & 1.4 & 1.3 & 1.2 & 0.5 & 0.5 & 0.8 & 1.8 & 2.3 & 0.8 & 0.5 & 1.3 & 1.0 & 1.0 & 1.1 & 1.2 \\
laion/larger\_clap\_music\_and\_speech & 0.8 & 0.5 & 1.3 & 1.0 & 0.4 & 0.8 & 2.4 & 1.7 & 1.2 & 0.7 & 1.1 & 0.8 & 0.5 & 0.9 & 1.2 & 1.3 & 1.7 & 0.5 & 0.8 & 1.3 & 0.8 & 0.7 & 0.6 & 0.9 \\
laion/larger\_clap\_general & 0.8 & 0.5 & 1.6 & 0.8 & 0.8 & 0.5 & 2.0 & 1.3 & 0.8 & 1.0 & 0.8 & 0.8 & 0.4 & 1.8 & 0.8 & 2.1 & 1.3 & 0.8 & 0.9 & 1.6 & 0.8 & 0.5 & 0.7 & 1.1 \\
lyrebird/wav2clip & 0.6 & 0.5 & 0.6 & 0.6 & 0.5 & 0.4 & 0.7 & 0.4 & 1.6 & 0.8 & 1.1 & 0.8 & 0.7 & 0.5 & 0.9 & 1.1 & 2.0 & 0.6 & 0.6 & 1.6 & 0.5 & 0.6 & 0.6 & 0.7 \\
\bottomrule
\end{tabular}
\end{table}

\end{landscape}

\clearpage


\begin{table}[!ht]
\caption{JamAlt A2T retrieval results (languages 1--4 of 4). Best per language in bold.}
\label{tab:jamalt_a2t_part1}
\centering
\footnotesize
\setlength{\tabcolsep}{2pt}
\begin{tabular}{@{}lcccc@{}}
\toprule
Model & de & en & es & fr \\
\midrule
LCO-Embedding/LCO-Embedding-Omni-3B & \textbf{73.5} & \textbf{76.4} & \textbf{81.4} & \textbf{72.5} \\
LCO-Embedding/LCO-Embedding-Omni-7B & 66.1 & 71.1 & 77.2 & 71.8 \\
microsoft/msclap-2023 & 0.9 & 0.7 & 1.8 & 1.0 \\
Qwen/Qwen2-Audio-7B & 0.7 & 0.8 & 1.1 & 1.5 \\
microsoft/speecht5\_multimodal & 0.7 & 1.7 & 0.8 & 0.8 \\
laion/larger\_clap\_music\_and\_speech & 1.0 & 0.5 & 0.8 & 1.6 \\
laion/larger\_clap\_music & 1.7 & 0.7 & 0.6 & 0.8 \\
laion/larger\_clap\_general & 0.9 & 0.7 & 0.9 & 1.1 \\
laion/clap-htsat-fused & 0.9 & 1.0 & 0.4 & 1.1 \\
laion/clap-htsat-unfused & 0.6 & 0.6 & 1.1 & 0.9 \\
microsoft/msclap-2022 & 0.5 & 0.9 & 0.5 & 0.8 \\
lyrebird/wav2clip & 0.4 & 0.5 & 0.4 & 1.0 \\
OpenMuQ/MuQ-MuLan-large & 0.7 & 0.8 & 0.4 & 0.2 \\
\bottomrule
\end{tabular}
\end{table}

\begin{table}[!ht]
\caption{JamAlt T2A retrieval results (languages 1--4 of 4). Best per language in bold.}
\label{tab:jamalt_t2a_part1}
\centering
\footnotesize
\setlength{\tabcolsep}{2pt}
\begin{tabular}{@{}lcccc@{}}
\toprule
Model & de & en & es & fr \\
\midrule
LCO-Embedding/LCO-Embedding-Omni-3B & \textbf{75.6} & \textbf{79.6} & \textbf{81.7} & 75.4 \\
LCO-Embedding/LCO-Embedding-Omni-7B & 72.4 & 78.6 & 80.8 & \textbf{76.0} \\
Qwen/Qwen2-Audio-7B & 1.2 & 2.0 & 1.7 & 1.8 \\
microsoft/speecht5\_multimodal & 1.1 & 3.0 & 1.0 & 0.9 \\
laion/larger\_clap\_music\_and\_speech & 1.5 & 1.3 & 0.6 & 1.0 \\
laion/clap-htsat-unfused & 0.8 & 1.5 & 0.9 & 0.7 \\
laion/larger\_clap\_general & 0.8 & 1.2 & 0.8 & 1.0 \\
microsoft/msclap-2023 & 0.8 & 1.1 & 0.6 & 0.7 \\
laion/clap-htsat-fused & 0.5 & 0.8 & 0.7 & 1.0 \\
lyrebird/wav2clip & 0.8 & 0.5 & 0.7 & 0.8 \\
microsoft/msclap-2022 & 0.5 & 0.5 & 0.8 & 0.9 \\
laion/larger\_clap\_music & 0.7 & 0.6 & 0.7 & 0.7 \\
OpenMuQ/MuQ-MuLan-large & 0.6 & 0.7 & 0.3 & 0.3 \\
\bottomrule
\end{tabular}
\end{table}

\begin{table}[!ht]
\caption{JamAlt A2A retrieval results (languages 1--4 of 4). Best per language in bold.}
\label{tab:jamalt_a2a_part1}
\centering
\footnotesize
\setlength{\tabcolsep}{2pt}
\begin{tabular}{@{}lcccc@{}}
\toprule
Model & de & en & es & fr \\
\midrule
laion/larger\_clap\_music\_and\_speech & \textbf{99.1} & 94.3 & \textbf{97.9} & \textbf{96.3} \\
OpenMuQ/MuQ-MuLan-large & 97.6 & \textbf{95.2} & 97.6 & 94.7 \\
laion/larger\_clap\_general & 97.9 & 93.2 & 96.8 & 94.6 \\
MIT/ast-finetuned-audioset-10-10-0.4593 & 97.7 & 90.2 & 96.2 & 94.4 \\
laion/clap-htsat-unfused & 97.0 & 88.8 & 94.2 & 91.8 \\
microsoft/msclap-2023 & 96.5 & 87.3 & 94.9 & 92.5 \\
google/vggish & 92.4 & 83.8 & 90.5 & 89.2 \\
\bottomrule
\end{tabular}
\end{table}

\clearpage


\begin{landscape}

\begin{table}[!ht]
\caption{CommonVoiceMini17 A2T retrieval results (languages 1--13 of 50). Best per language in bold.}
\label{tab:commonvoicemini17_a2t_part1}
\centering
\footnotesize
\setlength{\tabcolsep}{2pt}
\begin{tabular}{@{}lccccccccccccc@{}}
\toprule
Model & ar & ast & be & bg & bn & br & cs & cy & da & de & el & en & es \\
\midrule
LCO-Embedding/LCO-Embedding-Omni-3B & 94.4 & 98.3 & \textbf{99.4} & \textbf{69.4} & \textbf{58.0} & \textbf{67.4} & \textbf{86.6} & \textbf{38.4} & \textbf{56.6} & \textbf{100.0} & \textbf{46.4} & \textbf{99.8} & \textbf{100.0} \\
LCO-Embedding/LCO-Embedding-Omni-7B & \textbf{94.8} & \textbf{99.2} & \textbf{99.4} & 66.2 & 56.6 & 56.6 & 70.2 & 27.2 & 49.0 & 99.8 & 21.6 & 99.6 & \textbf{100.0} \\
Qwen/Qwen2-Audio-7B & 2.6 & 8.4 & 2.8 & 2.8 & 2.0 & 1.4 & 2.6 & 1.4 & 2.2 & 4.4 & 2.6 & 4.2 & 6.2 \\
microsoft/speecht5\_multimodal & 1.2 & 8.4 & 0.2 & 1.0 & 0.8 & 1.6 & 1.8 & 1.0 & 1.6 & 1.2 & 0.6 & 1.0 & 2.0 \\
laion/clap-htsat-unfused & 1.0 & 3.4 & 1.4 & 1.4 & 1.6 & 2.0 & 2.2 & 1.2 & 1.2 & 1.2 & 1.4 & 1.4 & 0.4 \\
microsoft/msclap-2023 & 0.8 & 4.2 & 1.4 & 1.0 & 0.8 & 0.8 & 1.8 & 1.2 & 1.6 & 0.8 & 1.6 & 1.2 & 1.6 \\
microsoft/msclap-2022 & 1.2 & 7.6 & 1.4 & 1.2 & 1.2 & 1.2 & 0.8 & 1.4 & 1.2 & 1.0 & 1.2 & 1.2 & 1.4 \\
laion/clap-htsat-fused & 1.8 & 5.9 & 0.6 & 0.4 & 0.8 & 1.4 & 1.0 & 1.4 & 0.8 & 1.0 & 1.2 & 1.4 & 1.0 \\
OpenMuQ/MuQ-MuLan-large & 1.0 & 6.7 & 1.0 & 1.0 & 1.2 & 0.6 & 1.6 & 1.2 & 0.8 & 1.4 & 1.0 & 0.4 & 1.0 \\
laion/larger\_clap\_general & 0.4 & 6.7 & 1.6 & 1.4 & 1.4 & 1.0 & 0.4 & 1.6 & 2.0 & 0.8 & 1.0 & 1.2 & 0.2 \\
lyrebird/wav2clip & 1.2 & 3.4 & 1.0 & 1.4 & 1.2 & 1.0 & 0.6 & 1.2 & 1.4 & 0.8 & 1.4 & 1.4 & 1.0 \\
laion/larger\_clap\_music\_and\_speech & 0.8 & 5.9 & 0.4 & 1.0 & 1.4 & 1.4 & 1.8 & 1.4 & 1.2 & 1.8 & 0.8 & 1.6 & 0.4 \\
laion/larger\_clap\_music & 1.0 & 4.2 & 1.0 & 1.0 & 0.8 & 1.0 & 1.2 & 1.0 & 0.8 & 1.0 & 1.0 & 1.0 & 1.2 \\
\bottomrule
\end{tabular}
\end{table}

\vspace{1em}
\begin{table}[!ht]
\caption{CommonVoiceMini17 A2T retrieval results (languages 14--26 of 50). Best per language in bold.}
\label{tab:commonvoicemini17_a2t_part2}
\centering
\footnotesize
\setlength{\tabcolsep}{2pt}
\begin{tabular}{@{}lccccccccccccc@{}}
\toprule
Model & et & fa & fi & fr & frold & gl & ha & hi & hu & it & ja & ka & ko \\
\midrule
LCO-Embedding/LCO-Embedding-Omni-3B & \textbf{46.8} & \textbf{25.2} & \textbf{39.0} & \textbf{99.8} & \textbf{99.8} & \textbf{100.0} & \textbf{61.6} & 94.0 & \textbf{50.0} & \textbf{99.6} & 95.2 & \textbf{61.2} & \textbf{99.1} \\
LCO-Embedding/LCO-Embedding-Omni-7B & 39.4 & 21.4 & 24.2 & 99.4 & 99.4 & \textbf{100.0} & 51.2 & \textbf{94.4} & 42.4 & \textbf{99.6} & \textbf{95.8} & 54.4 & 98.8 \\
Qwen/Qwen2-Audio-7B & 2.6 & 2.8 & 2.0 & 4.6 & 4.6 & 3.6 & 2.4 & 1.4 & 2.0 & 6.0 & 3.4 & 2.8 & 4.4 \\
microsoft/speecht5\_multimodal & 1.4 & 1.4 & 3.2 & 2.4 & 2.4 & 1.8 & 1.8 & 0.8 & 1.8 & 2.0 & 1.2 & 1.2 & 1.2 \\
laion/clap-htsat-unfused & 1.0 & 1.0 & 1.2 & 0.6 & 0.6 & 1.4 & 2.1 & 1.4 & 1.2 & 0.8 & 1.4 & 1.2 & 2.4 \\
microsoft/msclap-2023 & 1.0 & 1.2 & 2.0 & 1.6 & 1.6 & 1.0 & 1.5 & 1.0 & 1.0 & 1.4 & 1.0 & 1.6 & 2.1 \\
microsoft/msclap-2022 & 1.2 & 1.6 & 1.2 & 0.4 & 0.4 & 0.8 & 1.2 & 0.2 & 0.8 & 1.0 & 1.0 & 1.2 & 1.8 \\
laion/clap-htsat-fused & 1.0 & 1.4 & 0.6 & 0.8 & 0.8 & 1.8 & 2.4 & 1.6 & 1.8 & 0.6 & 2.2 & 0.8 & 1.2 \\
OpenMuQ/MuQ-MuLan-large & 1.4 & 0.2 & 1.0 & 0.6 & 0.6 & 0.8 & 1.5 & 0.8 & 1.2 & 0.6 & 1.2 & 0.6 & 1.8 \\
laion/larger\_clap\_general & 0.6 & 0.4 & 0.6 & 1.2 & 1.2 & 1.2 & 1.8 & 0.6 & 0.8 & 1.2 & 1.0 & 1.0 & 1.5 \\
lyrebird/wav2clip & 1.0 & 0.8 & 1.0 & 0.6 & 0.6 & 1.2 & 1.8 & 0.8 & 1.2 & 1.2 & 1.0 & 0.8 & 2.1 \\
laion/larger\_clap\_music\_and\_speech & 1.0 & 0.4 & 1.2 & 0.8 & 0.8 & 0.8 & 2.4 & 1.0 & 1.6 & 1.8 & 1.0 & 1.0 & 0.6 \\
laion/larger\_clap\_music & 1.0 & 1.0 & 0.8 & 1.0 & 1.0 & 0.8 & 1.5 & 0.8 & 1.0 & 1.0 & 0.8 & 1.0 & 1.5 \\
\bottomrule
\end{tabular}
\end{table}

\vspace{1em}
\begin{table}[!ht]
\caption{CommonVoiceMini17 A2T retrieval results (languages 27--39 of 50). Best per language in bold.}
\label{tab:commonvoicemini17_a2t_part3}
\centering
\footnotesize
\setlength{\tabcolsep}{2pt}
\begin{tabular}{@{}lccccccccccccc@{}}
\toprule
Model & lt & lv & mk & ml & mn & mr & nl & oc & pl & pt & ro & ru & sk \\
\midrule
LCO-Embedding/LCO-Embedding-Omni-3B & \textbf{62.8} & \textbf{43.4} & \textbf{76.2} & \textbf{57.8} & \textbf{20.2} & 78.6 & \textbf{99.0} & \textbf{90.9} & \textbf{88.4} & \textbf{99.0} & \textbf{86.0} & \textbf{99.2} & \textbf{82.2} \\
LCO-Embedding/LCO-Embedding-Omni-7B & 59.6 & 38.8 & 72.9 & 40.8 & 11.8 & \textbf{81.2} & \textbf{99.0} & 90.2 & 66.2 & 98.8 & 78.8 & \textbf{99.2} & 80.2 \\
Qwen/Qwen2-Audio-7B & 1.8 & 2.2 & 3.3 & 2.2 & 2.0 & 2.0 & 2.8 & 5.1 & 2.6 & 2.4 & 2.6 & 3.0 & 2.4 \\
microsoft/speecht5\_multimodal & 2.4 & 2.0 & 1.5 & 0.8 & 1.0 & 2.4 & 2.2 & 2.0 & 1.8 & 1.6 & 1.0 & 0.6 & 1.8 \\
laion/clap-htsat-unfused & 0.8 & 1.4 & 1.5 & 1.6 & 1.2 & 1.4 & 0.8 & 2.4 & 1.8 & 0.6 & 1.0 & 1.6 & 1.0 \\
microsoft/msclap-2023 & 1.4 & 0.4 & 1.5 & 1.4 & 0.8 & 1.2 & 1.0 & 2.4 & 1.8 & 1.2 & 1.4 & 1.4 & 0.8 \\
microsoft/msclap-2022 & 1.2 & 1.2 & 0.9 & 1.0 & 1.2 & 1.4 & 0.6 & 2.0 & 1.0 & 1.6 & 1.2 & 0.6 & 0.8 \\
laion/clap-htsat-fused & 0.8 & 1.2 & 2.1 & 0.8 & 0.6 & 1.6 & 1.4 & 3.5 & 1.4 & 1.4 & 0.8 & 1.6 & 2.0 \\
OpenMuQ/MuQ-MuLan-large & 1.0 & 1.0 & 1.5 & 1.0 & 1.2 & 1.2 & 1.2 & 2.4 & 1.2 & 0.8 & 1.2 & 0.8 & 1.0 \\
laion/larger\_clap\_general & 0.4 & 1.8 & 2.4 & 1.2 & 1.0 & 0.8 & 1.0 & 3.1 & 2.0 & 1.8 & 1.2 & 0.6 & 1.8 \\
lyrebird/wav2clip & 1.6 & 1.2 & 1.5 & 1.6 & 1.0 & 1.4 & 1.2 & 1.6 & 1.4 & 1.4 & 1.0 & 1.2 & 0.8 \\
laion/larger\_clap\_music\_and\_speech & 1.4 & 0.8 & 3.0 & 0.4 & 0.8 & 1.8 & 0.6 & 3.1 & 1.0 & 0.6 & 0.8 & 1.2 & 0.6 \\
laion/larger\_clap\_music & 1.0 & 0.8 & 1.2 & 0.8 & 1.0 & 1.0 & 1.0 & 2.0 & 1.0 & 1.0 & 1.0 & 1.0 & 1.0 \\
\bottomrule
\end{tabular}
\end{table}

\vspace{1em}
\begin{table}[!ht]
\caption{CommonVoiceMini17 A2T retrieval results (languages 40--50 of 50). Best per language in bold.}
\label{tab:commonvoicemini17_a2t_part4}
\centering
\footnotesize
\setlength{\tabcolsep}{2pt}
\begin{tabular}{@{}lccccccccccc@{}}
\toprule
Model & sl & sr & sv & sw & ta & te & th & tr & uk & ur & vi \\
\midrule
LCO-Embedding/LCO-Embedding-Omni-3B & 59.8 & \textbf{58.6} & \textbf{68.2} & \textbf{53.0} & \textbf{42.4} & \textbf{91.8} & \textbf{92.0} & \textbf{72.2} & \textbf{86.6} & 82.2 & 84.4 \\
LCO-Embedding/LCO-Embedding-Omni-7B & \textbf{63.0} & 54.0 & 46.8 & 49.4 & 21.6 & 59.2 & \textbf{92.0} & 63.4 & \textbf{86.6} & \textbf{84.6} & \textbf{86.6} \\
Qwen/Qwen2-Audio-7B & 1.2 & 1.8 & 2.2 & 2.0 & 2.0 & 16.3 & 1.4 & 1.2 & 2.2 & 1.2 & 1.6 \\
microsoft/speecht5\_multimodal & 1.0 & 0.4 & 1.8 & 2.4 & 0.6 & 6.1 & 0.8 & 0.8 & 1.6 & 1.0 & 1.0 \\
laion/clap-htsat-unfused & 0.8 & 1.2 & 1.0 & 1.4 & 1.0 & 12.2 & 1.2 & 2.0 & 1.2 & 1.2 & 1.4 \\
microsoft/msclap-2023 & 2.0 & 2.0 & 0.4 & 1.4 & 1.4 & 10.2 & 1.4 & 1.2 & 1.2 & 1.2 & 0.8 \\
microsoft/msclap-2022 & 0.6 & 1.8 & 1.0 & 1.4 & 1.2 & 14.3 & 1.2 & 1.6 & 1.8 & 1.2 & 0.6 \\
laion/clap-htsat-fused & 0.8 & 1.0 & 1.2 & 1.6 & 0.8 & 10.2 & 0.4 & 1.2 & 1.4 & 1.0 & 1.4 \\
OpenMuQ/MuQ-MuLan-large & 1.2 & 0.8 & 0.2 & 1.0 & 2.0 & 16.3 & 1.8 & 1.4 & 1.2 & 0.6 & 1.6 \\
laion/larger\_clap\_general & 1.8 & 0.6 & 1.6 & 0.6 & 1.0 & 10.2 & 1.8 & 1.6 & 1.6 & 0.6 & 0.6 \\
lyrebird/wav2clip & 1.0 & 1.4 & 1.0 & 1.0 & 1.0 & 10.2 & 0.8 & 0.8 & 1.0 & 1.2 & 0.6 \\
laion/larger\_clap\_music\_and\_speech & 0.6 & 0.6 & 1.8 & 1.6 & 0.2 & 4.1 & 1.2 & 1.2 & 0.8 & 1.2 & 1.2 \\
laion/larger\_clap\_music & 1.2 & 0.8 & 1.0 & 1.0 & 1.0 & 10.2 & 1.0 & 1.0 & 1.0 & 1.0 & 1.2 \\
\bottomrule
\end{tabular}
\end{table}

\vspace{1em}
\begin{table}[!ht]
\caption{CommonVoiceMini17 T2A retrieval results (languages 1--13 of 50). Best per language in bold.}
\label{tab:commonvoicemini17_t2a_part1}
\centering
\footnotesize
\setlength{\tabcolsep}{2pt}
\begin{tabular}{@{}lccccccccccccc@{}}
\toprule
Model & ar & ast & be & bg & bn & br & cs & cy & da & de & el & en & es \\
\midrule
LCO-Embedding/LCO-Embedding-Omni-3B & 94.2 & \textbf{99.2} & 98.8 & \textbf{73.2} & \textbf{58.2} & \textbf{51.0} & \textbf{87.4} & \textbf{20.0} & \textbf{63.2} & \textbf{99.8} & \textbf{41.2} & 99.6 & \textbf{100.0} \\
LCO-Embedding/LCO-Embedding-Omni-7B & \textbf{95.0} & 97.5 & \textbf{99.0} & 65.6 & 49.4 & 47.0 & 72.6 & 16.2 & 56.0 & 99.6 & 19.4 & \textbf{99.8} & \textbf{100.0} \\
Qwen/Qwen2-Audio-7B & 1.6 & 8.4 & 3.0 & 1.0 & 2.2 & 1.8 & 2.0 & 2.8 & 2.2 & 3.0 & 1.0 & 5.2 & 5.6 \\
microsoft/speecht5\_multimodal & 1.0 & 9.2 & 1.0 & 1.4 & 1.4 & 2.0 & 1.8 & 1.6 & 1.4 & 1.6 & 0.8 & 2.0 & 1.4 \\
microsoft/msclap-2023 & 0.8 & 10.1 & 1.0 & 1.6 & 1.0 & 1.6 & 1.2 & 1.8 & 1.6 & 1.6 & 1.0 & 0.8 & 1.6 \\
laion/clap-htsat-fused & 0.6 & 4.2 & 1.6 & 1.8 & 0.6 & 1.2 & 1.6 & 1.6 & 1.6 & 1.8 & 1.2 & 1.2 & 0.8 \\
laion/clap-htsat-unfused & 1.0 & 6.7 & 0.8 & 1.2 & 1.2 & 1.4 & 1.4 & 1.4 & 1.4 & 1.2 & 1.4 & 1.0 & 0.2 \\
microsoft/msclap-2022 & 1.8 & 6.7 & 1.2 & 1.0 & 1.0 & 1.0 & 0.8 & 0.8 & 0.6 & 0.8 & 0.8 & 1.6 & 1.6 \\
laion/larger\_clap\_general & 0.8 & 7.6 & 1.0 & 1.0 & 1.2 & 1.6 & 1.2 & 0.8 & 1.4 & 0.6 & 1.6 & 1.4 & 0.6 \\
laion/larger\_clap\_music\_and\_speech & 1.0 & 4.2 & 1.2 & 1.0 & 0.8 & 0.8 & 1.6 & 0.8 & 1.6 & 2.6 & 0.8 & 0.8 & 0.8 \\
OpenMuQ/MuQ-MuLan-large & 1.6 & 3.4 & 1.4 & 0.8 & 0.4 & 1.0 & 1.2 & 1.4 & 1.0 & 0.8 & 1.8 & 0.4 & 0.6 \\
lyrebird/wav2clip & 1.4 & 5.0 & 0.8 & 0.8 & 0.8 & 1.2 & 1.0 & 0.8 & 1.2 & 0.8 & 1.2 & 0.8 & 0.6 \\
laion/larger\_clap\_music & 1.0 & 4.2 & 1.0 & 1.0 & 1.0 & 1.0 & 1.0 & 1.0 & 1.0 & 1.2 & 0.8 & 1.0 & 1.2 \\
\bottomrule
\end{tabular}
\end{table}

\vspace{1em}
\begin{table}[!ht]
\caption{CommonVoiceMini17 T2A retrieval results (languages 14--26 of 50). Best per language in bold.}
\label{tab:commonvoicemini17_t2a_part2}
\centering
\footnotesize
\setlength{\tabcolsep}{2pt}
\begin{tabular}{@{}lccccccccccccc@{}}
\toprule
Model & et & fa & fi & fr & frold & gl & ha & hi & hu & it & ja & ka & ko \\
\midrule
LCO-Embedding/LCO-Embedding-Omni-3B & \textbf{46.6} & 25.6 & \textbf{41.6} & \textbf{99.6} & \textbf{99.6} & \textbf{99.8} & \textbf{49.7} & \textbf{92.8} & \textbf{48.8} & 99.6 & 95.0 & \textbf{58.8} & \textbf{98.8} \\
LCO-Embedding/LCO-Embedding-Omni-7B & 41.8 & \textbf{26.0} & 22.0 & 99.4 & 99.4 & \textbf{99.8} & 40.9 & 91.8 & 42.2 & \textbf{99.8} & \textbf{96.2} & 50.6 & 98.5 \\
Qwen/Qwen2-Audio-7B & 2.8 & 1.4 & 0.8 & 4.2 & 4.2 & 4.4 & 2.4 & 1.6 & 1.2 & 6.4 & 1.4 & 2.4 & 4.4 \\
microsoft/speecht5\_multimodal & 2.0 & 1.0 & 1.8 & 1.0 & 1.0 & 1.4 & 1.8 & 1.2 & 1.4 & 1.2 & 1.0 & 0.8 & 0.9 \\
microsoft/msclap-2023 & 1.8 & 1.2 & 1.4 & 1.2 & 1.2 & 0.8 & 0.6 & 1.2 & 1.4 & 1.2 & 0.8 & 0.8 & 2.1 \\
laion/clap-htsat-fused & 1.2 & 1.0 & 1.8 & 1.4 & 1.4 & 1.2 & 1.5 & 1.0 & 1.2 & 0.4 & 1.2 & 1.0 & 1.8 \\
laion/clap-htsat-unfused & 1.4 & 0.6 & 1.4 & 1.2 & 1.2 & 1.6 & 1.5 & 1.4 & 0.8 & 1.2 & 2.6 & 1.0 & 2.1 \\
microsoft/msclap-2022 & 1.4 & 1.0 & 0.6 & 1.4 & 1.4 & 0.6 & 3.7 & 0.8 & 0.6 & 1.2 & 1.2 & 0.6 & 1.2 \\
laion/larger\_clap\_general & 0.6 & 1.0 & 0.6 & 1.0 & 1.0 & 0.4 & 1.2 & 1.0 & 1.2 & 1.2 & 0.8 & 1.2 & 1.2 \\
laion/larger\_clap\_music\_and\_speech & 0.4 & 1.4 & 1.8 & 2.4 & 2.4 & 1.0 & 1.2 & 1.0 & 0.6 & 0.8 & 1.0 & 1.0 & 1.2 \\
OpenMuQ/MuQ-MuLan-large & 1.2 & 1.2 & 1.6 & 1.0 & 1.0 & 0.8 & 0.9 & 1.0 & 0.6 & 1.0 & 1.0 & 1.0 & 2.1 \\
lyrebird/wav2clip & 1.0 & 1.4 & 1.0 & 1.0 & 1.0 & 0.8 & 1.8 & 1.0 & 1.4 & 1.2 & 1.2 & 1.0 & 0.9 \\
laion/larger\_clap\_music & 1.2 & 1.0 & 1.0 & 1.0 & 1.0 & 1.0 & 1.5 & 1.0 & 1.2 & 1.0 & 1.0 & 0.8 & 1.8 \\
\bottomrule
\end{tabular}
\end{table}

\vspace{1em}
\begin{table}[!ht]
\caption{CommonVoiceMini17 T2A retrieval results (languages 27--39 of 50). Best per language in bold.}
\label{tab:commonvoicemini17_t2a_part3}
\centering
\footnotesize
\setlength{\tabcolsep}{2pt}
%
\end{table}

\end{landscape}

\clearpage


\begin{landscape}

\begin{table}[!ht]
\caption{CommonVoiceMini21 A2T retrieval results (languages 1--29 of 114). Best per language in bold.}
\label{tab:commonvoicemini21_a2t_part1}
\centering
\scriptsize
\setlength{\tabcolsep}{2pt}
%
\end{table}

\vspace{1em}
\begin{table}[!ht]
\caption{CommonVoiceMini21 A2T retrieval results (languages 30--58 of 114). Best per language in bold.}
\label{tab:commonvoicemini21_a2t_part2}
\centering
\scriptsize
\setlength{\tabcolsep}{2pt}
%
\end{table}

\vspace{1em}
\begin{table}[!ht]
\caption{CommonVoiceMini21 A2T retrieval results (languages 59--87 of 114). Best per language in bold.}
\label{tab:commonvoicemini21_a2t_part3}
\centering
\scriptsize
\setlength{\tabcolsep}{2pt}
%
\end{table}

\vspace{1em}
\begin{table}[!ht]
\caption{CommonVoiceMini21 A2T retrieval results (languages 88--114 of 114). Best per language in bold.}
\label{tab:commonvoicemini21_a2t_part4}
\centering
\scriptsize
\setlength{\tabcolsep}{2pt}
%
\end{table}

\vspace{1em}
\begin{table}[!ht]
\caption{CommonVoiceMini21 T2A retrieval results (languages 1--29 of 114). Best per language in bold.}
\label{tab:commonvoicemini21_t2a_part1}
\centering
\scriptsize
\setlength{\tabcolsep}{2pt}
%
\end{table}

\vspace{1em}
\begin{table}[!ht]
\caption{CommonVoiceMini21 T2A retrieval results (languages 30--58 of 114). Best per language in bold.}
\label{tab:commonvoicemini21_t2a_part2}
\centering
\scriptsize
\setlength{\tabcolsep}{2pt}
%
\end{table}

\vspace{1em}
\begin{table}[!ht]
\caption{CommonVoiceMini21 T2A retrieval results (languages 59--87 of 114). Best per language in bold.}
\label{tab:commonvoicemini21_t2a_part3}
\centering
\scriptsize
\setlength{\tabcolsep}{2pt}
%
\end{table}

\end{landscape}

\clearpage


\end{document}